\documentclass[a4paper,11pt]{article}
\pdfoutput=1
\usepackage{jheppub}
\usepackage[utf8]{inputenc}
\usepackage{url}

\usepackage{pdflscape}
\usepackage{tikz}
\usetikzlibrary{positioning}
\usetikzlibrary{arrows}
\usetikzlibrary{decorations.pathreplacing}
\usetikzlibrary{shapes.geometric}
\usetikzlibrary{calc}
\usepackage{refcount}
\usepackage{multirow}

\usetikzlibrary{shapes.misc}
\tikzset{gauge1/.style={draw=none,minimum size=0.6cm,fill=white,circle, draw}}
\tikzset{gauge3/.style={draw=none,minimum size=0.4cm,fill=white,circle, draw}}
\tikzset{crosses/.style={cross out, draw=black, minimum size=0.3cm, inner sep=0pt, outer sep=0pt},
cross/.default={1pt}}
\tikzset{blank/.style={draw=none,minimum size=0.4cm,fill=none,circle, draw}}
\tikzset{flavour2/.style={draw=none,minimum size=0.5cm,fill=white,regular polygon sides=4,draw}}
\tikzset{flavourBlue/.style={draw=none,minimum size=0.4cm,fill=blue,regular polygon sides=4,draw}}
\tikzset{flavourRed/.style={draw=none,minimum size=0.4cm,fill=red,regular polygon sides=4,draw}}
\tikzset{none/.style={draw=none}}
\tikzset{redgauge/.style={draw=none,minimum size=0.4cm,fill=red,circle, draw}}
\tikzset{miniU/.style={draw=none,minimum size=0.1cm,fill=red,circle, draw}}
\tikzset{smallgauge1/.style={draw=none,minimum size=0.1cm,fill=white,circle, draw}}
\tikzset{miniBlue/.style={draw=none,minimum size=0.1cm,fill=blue,circle, draw}}
\tikzset{gauge2/.style={draw=none,minimum size=0.35mm,fill=red,circle, draw}}
\tikzset{bluegauge/.style={draw=none,minimum size=0.4cm,fill=blue,circle, draw}}
\tikzset{flavour1/.style={draw=none,minimum size=0.35mm,fill=blue, regular polygon,regular polygon sides=4,draw}}
\tikzset{flavour0/.style={draw=none,minimum size=0.35mm,fill=white, regular polygon,regular polygon sides=4,draw}}
\tikzset{smalldot/.style={draw=none,minimum size=0.1mm,fill=black, circle,draw}}
\tikzset{dotsize/.style={circle,fill,inner sep=1.5pt,draw}}
\tikzset{doubleguys/.style={double, double distance = 3pt}}
\tikzset{tripleguys/.style={triple}}
\tikzset{new edge style 1/.style={dashed}}
\tikzset{thickline/.style={line width=0.06cm}}
\tikzset{darke/.style={line width=0.3mm,black}}
\tikzset{brace/.style={decorate,decoration={brace,amplitude=10pt}}}
\pgfdeclarelayer{edgelayer}
\pgfdeclarelayer{nodelayer}
\usepackage[noabbrev,capitalise]{cleveref}
\usetikzlibrary{decorations.pathreplacing}
\pgfsetlayers{edgelayer,nodelayer,main}

\usepackage{booktabs}
\usepackage{todonotes}

\usepackage{enumitem}

\usepackage{adjustbox}
\tikzset{hasse/.style={circle, fill,inner sep=2pt}}
\tikzset{gauge/.style={inner sep=1mm,draw=none,fill=white,minimum size=2mm,circle, draw}}
\tikzset{flavour/.style={draw=none,minimum size=0.3mm,fill=white, regular polygon,regular polygon sides=4,draw}}
\tikzset{bd/.style={circle, draw=black, inner sep=0pt, fill=black, minimum size=2mm}}
\tikzset{gd/.style={circle, draw=green, inner sep=0pt, fill=green, minimum size=2mm}}
\tikzset{gauge3/.style={draw=none,minimum size=0.35mm,fill=white,circle, draw}}
\tikzset{none/.style={draw=none}}
\tikzset{flavor1/.style={draw=none,minimum size=0.35mm,fill=white, regular polygon,regular polygon sides=4,draw}}
\tikzset{flavour2/.style={draw=none,minimum size=0.35mm,fill=white, regular polygon,regular polygon sides=4,draw}}
\tikzset{blankflavor/.style={draw=none,minimum size=0.5mm,fill=none, regular polygon,regular polygon sides=4,draw}}
\pgfdeclarelayer{edgelayer}
\pgfdeclarelayer{nodelayer}
\pgfsetlayers{edgelayer,nodelayer,main}
\tikzset{brace/.style={decorate,decoration={brace,amplitude=10pt}}}
\DeclareFontFamily{U}{rcjhbltx}{}
\DeclareFontShape{U}{rcjhbltx}{m}{n}{<->rcjhbltx}{}
\DeclareSymbolFont{hebrewletters}{U}{rcjhbltx}{m}{n}

\let\aleph\relax\let\beth\relax
\let\gimel\relax\let\daleth\relax

\DeclareMathSymbol{\aleph}{\mathord}{hebrewletters}{39}
\DeclareMathSymbol{\beth}{\mathord}{hebrewletters}{98}
\DeclareMathSymbol{\gimel}{\mathord}{hebrewletters}{103}
\DeclareMathSymbol{\daleth}{\mathord}{hebrewletters}{100}

\DeclareMathSymbol{\lamed}{\mathord}{hebrewletters}{108}
\DeclareMathSymbol{\mem}{\mathord}{hebrewletters}{109}
\DeclareMathSymbol{\ayin}{\mathord}{hebrewletters}{96}
\DeclareMathSymbol{\tsadi}{\mathord}{hebrewletters}{118}
\DeclareMathSymbol{\qof}{\mathord}{hebrewletters}{114}
\DeclareMathSymbol{\shin}{\mathord}{hebrewletters}{152}
\DeclareMathSymbol{\thet}{\mathord}{hebrewletters}{84}

\preprint{Imperial/TP/20/AH/06}

\title{Magnetic quivers for rank 1 theories}

\author[\lamed]{Antoine Bourget}
\author[\lamed]{, Julius F. Grimminger}
\author[\lamed]{, Amihay Hanany}
\author[\beth]{, Marcus Sperling}
\author[\thet,\mem]{, Gabi Zafrir}
\author[\lamed]{and Zhenghao Zhong}
\affiliation[\lamed]{Theoretical Physics Group, The Blackett Laboratory, Imperial College London, Prince Consort Road
London, SW7 2AZ, UK}
\affiliation[\beth]{Yau Mathematical Sciences Center, Tsinghua University, Haidian District, Beijing, 100084, China}
\affiliation[\thet]{Kavli Institute for the Physics and Mathematics of the Universe (WPI), University of Tokyo
Kashiwa, Chiba 277-8583, Japan}
\affiliation[\mem]{Dipartimento di Fisica, Universit`a di Milano-Bicocca \& INFN,
Sezione di Milano-Bicocca, I-20126 Milano, Italy}
\emailAdd{a.bourget@imperial.ac.uk}
\emailAdd{julius.grimminger17@imperial.ac.uk}
\emailAdd{a.hanany@imperial.ac.uk}
\emailAdd{marcus.sperling@univie.ac.at}
\emailAdd{gabi.zafrir@unimib.it}
\emailAdd{zhenghao.zhong14@imperial.ac.uk}

\abstract{Magnetic quivers and Hasse diagrams for Higgs branches of rank 1 $4d$ $\mathcal{N}=2$ SCFTs are provided. These rank 1 theories fit naturally into families of higher rank theories, originating from higher dimensions, which are addressed.}

\begin{document}

\maketitle

\section{Introduction}
Coulomb branches of $4d$ $\mathcal{N}=2$ gauge theories and superconformal field theories (SCFTs) are well studied objects. Their geometrical properties were first studied in \cite{Seiberg:1994rs,Seiberg:1994aj}. Amongst the $4d$ $\mathcal{N}=2$ theories, SCFTs hold a special place. A crude estimate of their complexity is provided by the complex dimension of the Coulomb branch, called the \emph{rank} of the SCFT. The classification of all $4d$ $\mathcal{N}=2$ SCFTs based on all possible Coulomb branch geometries is a long standing problem. It has been addressed systematically for rank 1 theories in \cite{Argyres:2015ffa,Argyres:2015gha,Argyres:2016xua,Argyres:2016xmc}. A nice feature of some of these SCFTs is the existence of an \emph{enhanced} Coulomb branch. In this case the pure Coulomb branch of the theory is contained as a subvariety in a mixed branch, which is called the enhanced Coulomb branch. This implies that there is a partial Higgsing from a rank 1 SCFT with an enhanced Coulomb branch, to a rank 1 SCFT where the Coulomb branch is not enhanced \cite{Apruzzi:2020pmv}. 

(Classical) Higgs branches of Lagrangian $4d$ $\mathcal{N}=2$ theories are constructed through the hyper-K\"ahler quotient \cite{Hitchin:1986ea,Argyres:1996eh,Antoniadis:1996ra}. However, when the theory is non-Lagrangian, different methods are needed. A $3d$ $\mathcal{N}=4$ theory may have a 3d mirror theory \cite{Intriligator:1996ex}. In this case computing the Higgs branch of one theory corresponds to computing the Coulomb branch of the other theory and vice versa. A powerful tool to compute the Higgs branch of a theory with 8 supercharges in any dimension\footnote{Note that dimensional reduction to 3 dimensions is not necessary to obtain a magnetic quiver \cite{Ferlito:2017xdq}. Furthermore some 3d theories do not have a 3d mirror, but they have several magnetic quivers \cite{Ferlito:2016grh,Bourget:2019rtl}.}, is through the use of \emph{magnetic quivers} \cite{Hanany:1996ie,DelZotto:2014kka,Cremonesi:2015lsa,Ferlito:2017xdq,Hanany:2018vph,Cabrera:2018jxt,Cabrera:2019izd,Cabrera:2019dob,Bourget:2020gzi}. A crucial ingredient in deriving such quivers is the usage of non-simply laced quivers, whose Coulomb branches were first computed in \cite{Cremonesi:2014xha}. This note is dedicated to the study of magnetic quivers for rank 1 4d theories, and some sequences they naturally fall into. In this case the non-simply laced magnetic quivers can be obtained from the folding of simply laced magnetic quivers \cite{Dey:2014tka,Dey:2016qqp,Kimura:2017hez,Nakajima:2019olw,Bourget:2020bxh}, which arise in the context of five dimensional theories. This can be deduced from a brane web picture. The Higgs branches of some of the theories were previously studied through different methods in \cite{Xie:2014pua,Zafrir:2016wkk,Ohmori:2018ona}.

Any proposed magnetic quiver is constrained by conditions on global symmetry, rank of gauge groups, scaling dimension of the operators in the Coulomb branch, as well as the identification of the $\mathbb{Z}_k$ S-folds with $k$-folded non-simply laced edges. It turns out, these conditions are restrictive enough to determine a candidate magnetic quiver. The Coulomb branch Hilbert series and Hasse diagrams of the magnetic quivers are consistent with previous computations. 
We also provide the Hasse diagrams for the full moduli space of these theories in 3 dimensions. Hasse diagrams for the full moduli space of $4d$ $\mathcal{N}=2$ theories are discussed in \cite{Mario,MarioPhilip} and for $4d$ $\mathcal{N}=3$ theories in \cite{Argyres:2019yyb}.

\section{Rank 1 \texorpdfstring{$\mathcal{N}=2$}{Nis2} SCFTs}
\label{Nis2}
We begin with a brief review of the classification of these theories \cite{Argyres:2015ffa,Argyres:2015gha,Argyres:2016xua,Argyres:2016xmc,Argyres:2020nrr}. It is a geometric classification, based on the geometry of the Coulomb branch $\mathcal{C}$, which by definition is a complex singular space of dimension 1. In the case where there is no enhanced Coulomb branch, at a generic point on $\mathcal{C}$ the theory is a free $U(1)$ gauge theory, and interesting physics emerges at singular points. Scale invariance indicates that the singular locus is reduced to a single point, which we take to be the origin $O$ of the Coulomb branch. On the non simply connected regular locus $\mathcal{C}-\{O\}$, the locally trivial physics undergoes non-trivial topological effects, incarnated by a non-trivial monodromy in the electromagnetic duality group $\mathrm{SL}(2,\mathbb{Z})$. Scale invariance constrains the geometry of the Coulomb branch to be one of those listed in the left part of Table \ref{rank1table}. These geometries can be characterized by their Kodaira type \cite{Dasgupta:1996ij}
\begin{equation}
    K \in \{II^\ast ,III^\ast ,IV^\ast , I_0^\ast , IV, III, II\} \, . 
\end{equation}
When there is an enhanced Coulomb branch, on a generic point of the Coulomb branch there are also $d>0$ hypermultiplets which can take vacuum expectation values, parametrizing a geometry which can be globally non-trivial (in the rank 1 case, studied in this paper, these take the form of orbifolds $h_{d,k}$, see Appendix \ref{orbifoldappendix}). 

The geometry of the scale-invariant geometry $\mathcal{C}$ is not sufficient to fully characterize the SCFT: one also needs to understand possible $\mathcal{N}=2$ preserving mass deformations. These deformations break conformal invariance, so the geometry after deformation does not need to contain a single singular point. Rather, it contains a finite number of singularities which can be characterized by a finite set of Kodaira classes. The deformation parameters are invariant under (the Weyl group of) a flavor symmetry $F$. There can also be chiral deformation parameters. 
Barring the issue of discrete gauging \cite{Argyres:2016yzz}, a pair $[K , F]$ entirely identifies a 4d $\mathcal{N}=2$ rank-1 SCFT. For instance, the $\mathfrak{su}(2)$ gauge theory with $N_f=4$ has $K = I_0^\ast$ (which can be deformed to $\{I_1^6\}$), $F=D_4$ and has an exactly marginal deformation parameter. 

The classification gives a list of 17 rank 1 $\mathcal{N}=2$ SCFTs (not counting IR-free theories). It turns out they can be uniquely identified by the flavor symmetry $F$, and in the following we use this flavor symmetry to label the theories. The list of 17 SCFTs is presented in the right part of Table \ref{rank1table}. Alternatively, the theories can be obtained from $\mathbb{Z}_k$ $\mathcal{N}=2$ S-fold constructions \cite{Apruzzi:2020pmv}, from the compactification of $(1,0)$ $6d$ SCFTs with non-trivial global symmetry background \cite{Ohmori:2018ona}, or from $\mathbb{Z}_k$ twisted compactification from 5d $\mathcal{N}=1$ SCFTs, as detailed in Section \ref{generalfamily}. This is the reason for the organization into rows and columns: the row gives the geometry of the Coulomb branch in the singular limit, and the column gives the $\mathbb{Z}_k$ group. 

\begin{table}[t]
    \centering
    \begin{tabular}{cc||ccccc} \toprule
        Geometry $K$  & $[u]$ & $\mathbb{Z}_1$ & $\mathbb{Z}_2$ & $\mathbb{Z}_3$ & $\mathbb{Z}_4$ & $\mathbb{Z}_6$ \\ \midrule 
        $II^\ast$  & $6$ & $E_8$ & $C_5$ & $A_3 \rtimes \mathbb{Z}_2$ & $A_2 \rtimes \mathbb{Z}_2$ &  \\ 
        $III^\ast$& $4$ & $E_7$ & $C_3A_1$ & $A_1U_1 \rtimes \mathbb{Z}_2$ & \textcolor{green}{$U_1 \rtimes \mathbb{Z}_2$} &  \\ 
        $IV^\ast$  & $3$  & $E_6$ & $C_2U_1$ & \textcolor{green}{$U_1$} & $\emptyset$ &  \\ 
        $I_0^\ast$  & $2$ & $D_4 \chi_0$ & \textcolor{blue}{$C_1\chi_0$} & & &  \\ 
        $IV$   & $\frac{3}{2}$    & $A_2\chi_{\frac{1}{2}}$ & & & &  \\ 
        $III$ & $\frac{4}{3}$     & $A_1\chi_{\frac{2}{3}}$ & & & &  \\ 
        $II$  & $\frac{6}{5}$     & $\chi_{\frac{4}{5}}$ & & & &  \\ 
         \bottomrule
    \end{tabular}
    \caption{\texttt{Left:} List of the seven singular Coulomb branch geometries at rank 1. These are freely generated, and $[u]$ is the scaling dimension of the generator.  \texttt{Right:} List of 4d $\mathcal{N}=2$ rank 1 SCFTs (IR-free theories are omitted). Each entry represents one theory, labeled by its flavor symmetry. In the rest of the paper, for conciseness we ignore the discrete $\mathbb{Z}_2$ in the naming of the theories.  The symbol $\chi_\delta$ signals the existence of a chiral deformation parameter of scaling dimension $\delta$. The magnetic quivers for the theories of the $\mathbb{Z}_k$ column involve $k$-laced edges. The theory in blue is $\mathcal{N}=4$ SYM with gauge group $SU(2)$. Theories in green are $\mathcal{N}=3$ S-fold theories \cite{Aharony:2016kai}.  }
    \label{rank1table}
\end{table}

\section{Magnetic quivers}\label{magnetic quiver}
In this section, the magnetic quivers $\mathsf{Q}$ of the $4d$ $\mathcal{N}=2$ rank 1 SCFTs are presented. The $3d$ $\mathcal{N}=4$ Coulomb branches of the magnetic quivers are equal to the Higgs branches of the rank 1 SCFTs. As given in Table \ref{resulttable}, the magnetic quivers are non-simply laced. Hence the framing, called \emph{ungauging scheme} in \cite{Hanany:2020jzl}, has to be prescribed. It is indicated by a \emph{squircle} (gauge node with a box around it). When the non-simply laced quiver is ungauged on a long node in the sense of a long root in a Dynkin diagram, it does not matter which of the long nodes is chosen. This is a remnant of the `Crawley-Boevey trick' \cite{Crawley-Boevey2001} for simply laced quivers, i.e.\ the freedom to change between framed and unframed simply laced quivers. In the case of simply laced quivers there is no need to denote an ungauged node.

\begin{table}[t]
\small
\centering
\begin{adjustbox}{center}
	\begin{tabular}{cc}
\toprule
		Rank 1 SCFT & Magnetic quiver \\ 
\midrule
       $C_5$ &		
    \begin{tabular}{c}
    \scalebox{0.70}{\begin{tikzpicture}
	\begin{pgfonlayer}{nodelayer}
		\node [style=gauge3] (2) at (-1, 0) {};
		\node [style=none] (6) at (-1, -0.5) {4};
		\node [style=gauge3] (12) at (-2, 0) {};
		\node [style=none] (15) at (-2, -0.5) {3};
		\node [style=gauge3] (24) at (0.2, 0) {};
		\node [style=none] (29) at (0.2, -0.5) {5};
		\node [style=none] (30) at (-0.85, 0.075) {};
		\node [style=none] (31) at (0.2, 0.075) {};
		\node [style=none] (32) at (-0.85, -0.075) {};
		\node [style=none] (33) at (0.2, -0.075) {};
		\node [style=none] (34) at (-0.55, 0) {};
		\node [style=none] (35) at (-0.175, 0.375) {};
		\node [style=none] (36) at (-0.175, -0.375) {};
		\node [style=gauge3] (37) at (0.95, 0) {};
		\node [style=none] (38) at (0.95, -0.5) {2};
		\node [style=gauge3] (39) at (-3, 0) {};
		\node [style=gauge3] (40) at (-4, 0) {};
		\node [style=none] (41) at (-4, -0.5) {1};
		\node [style=none] (42) at (-3, -0.5) {2};
		\node [style=blankflavor] (43) at (0.2, 0) {};
	\end{pgfonlayer}
	\begin{pgfonlayer}{edgelayer}
		\draw (30.center) to (31.center);
		\draw (33.center) to (32.center);
		\draw (35.center) to (34.center);
		\draw (34.center) to (36.center);
		\draw (37) to (24);
		\draw (12) to (2);
		\draw (40) to (39);
		\draw (39) to (12);
	\end{pgfonlayer}
\end{tikzpicture}}
    \end{tabular}	\\ 
          $C_3 \times A_1$ &		
    \begin{tabular}{c}
    \scalebox{0.70}{\begin{tikzpicture}
	\begin{pgfonlayer}{nodelayer}
		\node [style=gauge3] (2) at (-1, 0) {};
		\node [style=none] (6) at (-1, -0.5) {2};
		\node [style=gauge3] (12) at (-2, 0) {};
		\node [style=none] (15) at (-2, -0.5) {1};
		\node [style=gauge3] (24) at (0.25, 0) {};
		\node [style=none] (29) at (0.25, -0.5) {3};
		\node [style=none] (30) at (-1, 0.075) {};
		\node [style=none] (31) at (0.25, 0.075) {};
		\node [style=none] (32) at (-1, -0.075) {};
		\node [style=none] (33) at (0.25, -0.075) {};
		\node [style=none] (34) at (-0.5, 0) {};
		\node [style=none] (35) at (-0.125, 0.375) {};
		\node [style=none] (36) at (-0.125, -0.375) {};
		\node [style=gauge3] (37) at (1, 0) {};
		\node [style=none] (38) at (1, -0.5) {2};
		\node [style=gauge3] (39) at (2.25, 0) {};
		\node [style=none] (40) at (1, 0.075) {};
		\node [style=none] (41) at (2.25, 0.075) {};
		\node [style=none] (42) at (1, -0.075) {};
		\node [style=none] (43) at (2.25, -0.075) {};
		\node [style=none] (44) at (1.75, 0) {};
		\node [style=none] (45) at (1.375, 0.375) {};
		\node [style=none] (46) at (1.375, -0.375) {};
		\node [style=none] (49) at (2.25, -0.5) {1};
		\node [style=blankflavor] (50) at (0.25, 0) {};
	\end{pgfonlayer}
	\begin{pgfonlayer}{edgelayer}
		\draw (30.center) to (31.center);
		\draw (33.center) to (32.center);
		\draw (35.center) to (34.center);
		\draw (34.center) to (36.center);
		\draw (37) to (24);
		\draw (40.center) to (41.center);
		\draw (43.center) to (42.center);
		\draw (45.center) to (44.center);
		\draw (44.center) to (46.center);
		\draw (12) to (2);
	\end{pgfonlayer}
\end{tikzpicture}
}
    \end{tabular}	\\ 
      $C_2 \times U_1$ &		
    \begin{tabular}{c}
    \scalebox{0.70}{\begin{tikzpicture}
	\begin{pgfonlayer}{nodelayer}
		\node [style=gauge3] (2) at (-1, 0) {};
		\node [style=none] (6) at (-1, -0.5) {1};
		\node [style=gauge3] (24) at (0.25, 0) {};
		\node [style=none] (29) at (0.25, -0.5) {2};
		\node [style=none] (30) at (-1, 0.075) {};
		\node [style=none] (31) at (0.25, 0.075) {};
		\node [style=none] (32) at (-1, -0.075) {};
		\node [style=none] (33) at (0.25, -0.075) {};
		\node [style=none] (34) at (-0.5, 0) {};
		\node [style=none] (35) at (-0.125, 0.375) {};
		\node [style=none] (36) at (-0.125, -0.375) {};
		\node [style=gauge3] (37) at (1.25, 0.5) {};
		\node [style=gauge3] (38) at (1.25, -0.5) {};
		\node [style=none] (39) at (1.75, 0.5) {1};
		\node [style=none] (40) at (1.75, -0.5) {1};
		\node [style=blankflavor] (41) at (0.25, 0) {};
	\end{pgfonlayer}
	\begin{pgfonlayer}{edgelayer}
		\draw (30.center) to (31.center);
		\draw (33.center) to (32.center);
		\draw (35.center) to (34.center);
		\draw (34.center) to (36.center);
		\draw (37) to (33.center);
		\draw (38) to (31.center);
		\draw (37) to (38);
	\end{pgfonlayer}
\end{tikzpicture}}
    \end{tabular}	 \\ 
    $A_3$ &		
    \begin{tabular}{c}
    \scalebox{0.70}{
\begin{tikzpicture}
	\begin{pgfonlayer}{nodelayer}
		\node [style=gauge3] (2) at (-1, 0) {};
		\node [style=none] (6) at (-1, -0.5) {3};
		\node [style=gauge3] (24) at (0.5, 0) {};
		\node [style=none] (29) at (0.5, -0.5) {4};
		\node [style=none] (30) at (-1, 0.15) {};
		\node [style=none] (31) at (0.5, 0.15) {};
		\node [style=none] (32) at (-1, -0.15) {};
		\node [style=none] (33) at (0.5, -0.15) {};
		\node [style=none] (34) at (-0.5, 0) {};
		\node [style=none] (35) at (0, 0.5) {};
		\node [style=none] (36) at (0, -0.5) {};
		\node [style=gauge3] (37) at (-2, 0) {};
		\node [style=gauge3] (38) at (-3, 0) {};
		\node [style=none] (39) at (-2, -0.5) {2};
		\node [style=none] (40) at (-3, -0.5) {1};
		\node [style=blankflavor] (41) at (0.5, 0) {};
	\end{pgfonlayer}
	\begin{pgfonlayer}{edgelayer}
		\draw (2) to (24);
		\draw (30.center) to (31.center);
		\draw (33.center) to (32.center);
		\draw (35.center) to (34.center);
		\draw (34.center) to (36.center);
		\draw (37) to (2);
		\draw (38) to (37);
	\end{pgfonlayer}
\end{tikzpicture}
}
    \end{tabular}	\\ 

      $A_1 \times U_1$ &		
    \begin{tabular}{c}
    \scalebox{0.70}{\begin{tikzpicture}
	\begin{pgfonlayer}{nodelayer}
		\node [style=gauge3] (2) at (-1, 0) {};
		\node [style=none] (6) at (-1, -0.5) {1};
		\node [style=gauge3] (24) at (0.5, 0) {};
		\node [style=none] (29) at (0.5, -0.5) {2};
		\node [style=none] (30) at (-1, 0.15) {};
		\node [style=none] (31) at (0.5, 0.15) {};
		\node [style=none] (32) at (-1, -0.15) {};
		\node [style=none] (33) at (0.5, -0.15) {};
		\node [style=none] (34) at (-0.5, 0) {};
		\node [style=none] (35) at (0, 0.5) {};
		\node [style=none] (36) at (0, -0.5) {};
		\node [style=gauge3] (37) at (1.5, 0) {};
		\node [style=none] (38) at (0.5, 0.075) {};
		\node [style=none] (39) at (0.5, -0.1) {};
		\node [style=none] (40) at (1.5, 0.075) {};
		\node [style=none] (41) at (1.5, -0.1) {};
		\node [style=none] (42) at (1.5, -0.5) {1};
		\node [style=blankflavor] (43) at (0.5, 0) {};
	\end{pgfonlayer}
	\begin{pgfonlayer}{edgelayer}
		\draw (2) to (24);
		\draw (30.center) to (31.center);
		\draw (33.center) to (32.center);
		\draw (35.center) to (34.center);
		\draw (34.center) to (36.center);
		\draw (40.center) to (38.center);
		\draw (39.center) to (41.center);
	\end{pgfonlayer}
\end{tikzpicture}}
    \end{tabular} \\ 
        $A_2$ &		
    \begin{tabular}{c}
    \scalebox{0.70}{
\begin{tikzpicture}
	\begin{pgfonlayer}{nodelayer}
		\node [style=gauge3] (2) at (-1, 0) {};
		\node [style=none] (6) at (-1, -0.5) {2};
		\node [style=gauge3] (24) at (0.5, 0) {};
		\node [style=none] (29) at (0.5, -0.5) {3};
		\node [style=none] (30) at (-1, 0.05) {};
		\node [style=none] (31) at (0.5, 0.05) {};
		\node [style=none] (32) at (-1, -0.075) {};
		\node [style=none] (33) at (0.5, -0.075) {};
		\node [style=none] (34) at (-0.5, 0) {};
		\node [style=none] (35) at (0, 0.5) {};
		\node [style=none] (36) at (0, -0.5) {};
		\node [style=none] (37) at (-1, 0.15) {};
		\node [style=none] (38) at (0.5, 0.15) {};
		\node [style=none] (39) at (-1, -0.175) {};
		\node [style=none] (40) at (0.5, -0.175) {};
		\node [style=gauge3] (41) at (-2, 0) {};
		\node [style=none] (42) at (-2, -0.5) {1};
		\node [style=blankflavor] (43) at (0.5, 0) {};
	\end{pgfonlayer}
	\begin{pgfonlayer}{edgelayer}
		\draw (30.center) to (31.center);
		\draw (33.center) to (32.center);
		\draw (35.center) to (34.center);
		\draw (34.center) to (36.center);
		\draw (37.center) to (38.center);
		\draw (39.center) to (40.center);
		\draw (41) to (2);
	\end{pgfonlayer}
\end{tikzpicture}}
    \end{tabular}  	\\ 
    \bottomrule
	\end{tabular}
\end{adjustbox}
\caption{The magnetic quivers of $4d$ $\mathcal{N}=2$ rank 1 theories with enhanced Coulomb branches (labeled by their global symmetry).}
\label{resulttable}
\end{table}

\begin{table}[t]
\small
\centering
\scalebox{0.7}{
\begin{adjustbox}{center}
	\begin{tabular}{c|c|l}
\toprule
	\begin{tabular}{c}	Rank 1\\SCFT \end{tabular} &Hilbert series  &Refined PL[HS] \\ 
\midrule
    $C_5$	& 
  $\dfrac
	{\scriptsize \left(\begin{array}{c}1 + 2 t + 40 t^2 + 194 t^3 + 1007 t^4 + 4704 t^5 + 18683 t^6 + 
 67030 t^7 + 220700 t^8 + 657352 t^9 + 1796735 t^{10}\\ + 4540442 t^{11} + 
 10610604 t^{12} + 23011366 t^{13} + 46535540 t^{14} + 87887734 t^{15} + 
 155277056 t^{16}\\ + 257288236 t^{17} + 400453203 t^{18} + 585971786 t^{19} + 
 807195575 t^{20} + 1047954388 t^{21}\\ + 1282842123 t^{22} + 
 1481462886 t^{23} + 1615002952 t^{24} + 1662191888 t^{25} + \cdots\text{palindrome}\cdots+ t^{50}\end{array}\right)} {(-1 + t)^{32} (1 + t)^{18} (1 + t + t^2)^{16}} $ &
    \begin{tabular}{c}
        \parbox{6cm}{$t^2: [20000]\\
t^3:[00001]\\
t^4: -[01000]\\
t^5: -[10010] \\
t^6: -[00200]-[20000]+[01000]$ }
    \end{tabular}\\ 
    \midrule
         $C_3 \times A_1$	& 
$\dfrac
	{\scriptsize \left(\begin{array}{c}1 + 2 t + 17 t^2 + 66 t^3 + 205 t^4 + 572 t^5 + 1415 t^6 + 2914 t^7 + 
 5368 t^8 + 8874 t^9 + 12992 t^{10}\\ + 16856 t^{11} + 19865 t^{12} + 
 21032 t^{13} + 19865 t^{14} + 16856 t^{15} + 12992 t^{16} + 8874 t^{17} + 
 5368 t^{18}\\ + 2914 t^{19} + 1415 t^{20} + 572 t^{21} + 205 t^{22} + 66 t^{23} + 
 17 t^{24} + 2 t^{25} + t^{26} \end{array}\right)} {(-1 + t)^{16} (1 + t)^{10}(1 + t + t^2)^8} $  &
    \begin{tabular}{c}
        \parbox{6cm}{$t^2: [000;2] + [200;0]\\
t^3: [001;2] \\
t^4: -[000;0]-[010;0] \\ 
t^5: -[001;2]-[001;0] -[110;2] \\
t^6: -[000;2]-[002;0]-[020;2]\\-[200;4]-[200;0]+[010;0]$}
    \end{tabular}\\ 
    \midrule
        $C_2 \times U_1$ 	& 
  $\dfrac
	{\scriptsize \left(\begin{array}{c}1 + 2 t + 8 t^2 + 20 t^3 + 41 t^4 + 62 t^5 + 87 t^6 + 96 t^7 + 
 87 t^8 + 62 t^9 + 41 t^{10} + 20 t^{11} + 8 t^{12} + 2 t^{13} + t^{14} \end{array}\right)} {(-1 + t)^8 (1 + t)^6 (1 + t + t^2)^4} $ &
    \begin{tabular}{c}
        \parbox{6cm}{$t^2: [00] + [20]\\
t^3: (1/q + q ) [01]\\
t^4: -[00] -[01]\\
t^5: -(q+1/q)([01]+[20])\\
t^6: -(1+q^2+1/q^2)[00] + [01]\\ - [02] - [20]$}
    \end{tabular}\\ 
    \midrule
    $A_3$ 	& $\dfrac
	{\scriptsize \left(\begin{array}{c}1 - t + 10 t^2 + 23 t^3 + 67 t^4 + 190 t^5 + 525 t^6 + 1053 t^7 + 
 2292 t^8 + 4167 t^9 + 7299 t^{10}\\ + 11494 t^{11} + 17114 t^{12} + 
 23080 t^{13} + 29925 t^{14} + 35107 t^{15}\\ + 39221 t^{16} + 40320 t^{17} +\cdots\text{palindrome}+\cdots+ t^{34}\end{array}\right)} {(-1 + t)^{18} (1 + t)^{10} (1 + t^2)^5 (1 + t + t^2)^7} $ 
&
    \begin{tabular}{c}
        \parbox{6cm}{$t^2:[101]\\ t^3:[003]+[300]\\ t^4:[030] \\t^5:-[011]-[110]\\t^6:-[000]-[022]-[101]-[111]-[202]\\-[220]\\t^7: [001]-[003]+[100]-[112]-[122]\\-[211]-[221]-[300] \\t^8:-[000]+[002]+[012]-[020]+[022]\\-[040]+2[101]+[103]+2[111]+[200]\\+[210]+[220]-[222]+[301]$}
    \end{tabular}\\ 
    \midrule
    
     $A_1 \times U_1$	& 
$\dfrac
	{\scriptsize \begin{array}{c}(1 - t + t^2) (1 + 2 t^2 + 4 t^3 + 4 t^4 + 4 t^5 + 2 t^6 + t^8) \end{array}} {(-1 + t)^6 (1 + t)^2 (1 + t^2) (1 + t + t^2)^3} $  &
    \begin{tabular}{c}
        \parbox{6cm}{$t^2: [2] + [0]\\
t^3: (q + 1/q) [3]\\
t^4: (q^2 + 1/q^2)[0]\\
t^5: - (q + 1/q)[1]\\
t^6: - [0] - (1 + 1/q + q) [2] - [4]\\
t^7: -(q+\frac{1}{q})[3]\\
t^8: -[0]+(2+q^2+\frac{1}{q^2})[2]+[4]$
}
    \end{tabular}\\ 
    \midrule
    
      $A_2$	& 
    $\dfrac
	{\scriptsize \left(\begin{array}{c}1 + 3 t^2 + 31 t^4 + 55 t^6 + 156 t^8 + 132 t^{10} + 156 t^{12} + 
 55 t^{14} + 31 t^{16} + 3 t^{18} + t^{20}\end{array}\right)} {(-1 + t)^{10} (1 + t)^{10} (1 + t^2)^5} $  &
    \begin{tabular}{c}
        \parbox{6cm}{$t^2:[11]\\t^4:[04]+[40]\\t^6: -[12]-[21]\\t^8: -[00]+[01]-[04]+[10]-[11]\\-[22]-[24]-[33]-[40]-[42]$}
    \end{tabular}\\ 
    \bottomrule
  	\end{tabular}
\end{adjustbox}}
\caption{Higgs branch Hilbert series of the $4d$ $\mathcal{N}=2$ SCFTs (labeled by their global symmetry) as well as the refined plethystic logarithm (PL).}
\label{resultHS}
\end{table}
The magnetic quivers of the $4d$ $\mathcal{N}=2$ theories in question can be obtained from magnetic quivers of $5d$ $\mathcal{N}=1$ theories compactified with a $\mathbb{Z}_k$ twist. In this case one starts with the magnetic quiver $\mathsf{Q}'$ of the 5d theory, which contains $k$ identical simply-laced legs, and obtains the magnetic quiver $\mathsf{Q}$ of the $4d$ $\mathcal{N}=2$ theory by folding the $k$ legs of $\mathsf{Q}'$: 
\begin{equation}
    \mathcal{F}_k: Q' \mapsto  Q
\end{equation}
where $\mathcal{F}_k$ is the action of folding $k$ identical legs. Remarkably, most of the $5d$ theories in question are all well known, simple theories.

\paragraph{Hilbert series and chiral ring.}
The Coulomb branch Hilbert series and refined plethystic logarithm (PL) of $\mathsf{Q}$ are given in Table \ref{resultHS}. The refined PL encodes information on the generators of the chiral ring and their relations \cite{Benvenuti:2006qr}. The first few positive terms are representations of the generators whereas the first few negative terms are the representations of the relations. Higher order terms are often higher syzygies. The terms at order $t^2$ transform in the adjoint representations of the global symmetry group \cite{1992math......4227B}. 

The $C_5$, $C_3 \times A_1$, and $C_2\times U_1$ magnetic quivers can be derived from $5d$ $\mathcal{N}=1$ SQCD magnetic quivers $\mathsf{Q}'$ through folding \cite{Ferlito:2017xdq}, as detailed in Section \ref{generalfamily} below. Therefore, one can expect that the highest order relations exist at order $t^{4\Delta_B}$ where $\Delta_B=3/2$ is the conformal dimension of the baryonic/instanton generators, see for example \cite[Sec.\ 2.2]{Ferlito:2017xdq}.  As a result, there should be no relations beyond $t^6$. This is consistent with the fact that there are no negative terms in the PL at order $t^7$ and $t^8$. At higher orders, negative terms reemerge in the form of higher syzygies, i.e.\ relations between relations \cite{Benvenuti:2006qr}. As a result, for these three families, all the generators and relations can be seen in Table \ref{resultHS}. 

In contrast, the $A_3$, $A_1 \times U_1,$ and $A_2$ rank 1 SCFTs originate from folding more complicated magnetic quivers. For example, the $A_3$ magnetic quiver comes from the folding of the $T_4$ theory which has a diverse set of chiral ring relations \cite{Maruyoshi:2013hja}. Therefore, only some of the relations (up to $t^8$) are listed in Table \ref{resultHS}.

\paragraph{Hasse diagrams.}
The Higgs branch Hasse diagrams \cite{Bourget:2019aer} for the $4d$ $\mathcal{N}=2$ rank 1 theories with enhanced Coulomb branch are provided in \crefrange{tab:Hasse_C5}{tab:Hasse_A2}. The Higgs branches are symplectic singularities and, therefore, their geometric structure can be encoded in a Hasse diagram which details the symplectic leaves and transverse slices. To explore the Higgs branch of the rank 1 theories, one can study the $3d$ $\mathcal{N}=4$ Coulomb branch of their magnetic quivers. For the Hasse diagrams presented below, the quivers to the right are magnetic quivers whose Coulomb branch are closures of the symplectic leaves labeled by the symbol $\bullet$. The transverse slice between neighboring leaves is called an elementary slice and is labeled by a line segment. The elementary slices are also symplectic singularities and can be identified with the Coulomb branch of a quiver listed on the left of the line segment. The Hasse diagrams for the Higgs branches of the theories in the $\mathbb{Z}_1$ column of Table \ref{rank1table} are simple, as the Higgs branches are closures of minimal nilpotent orbits, and are not discussed below.

In \cite{Cabrera:2016vvv,Cabrera:2017njm,Bourget:2019aer}, the elementary slices are taken to be either closures of minimal nilpotent orbits or Kleinian singularities. Quivers that are weighted affine Dynkin diagrams have Coulomb branches that are closures of minimal nilpotent orbits, see for example \cite[Tab.\ (12)]{Bourget:2019aer} for a list. In Table \ref{nilpotent} several new non-simply laced quivers, whose Coulomb branches are also closures of minimal nilpotent orbits, are presented. These quivers are necessary in performing quiver subtraction on the rank 1 magnetic quivers and their families as in Section \ref{generalfamily}. These new quivers arise as the $n=1$ members of the families presented in Section \ref{generalfamily}. Furthermore, new conjectured elementary slices, denoted as $h_{d,k}$, of the form of $\mathbb{H}^d/\mathbb{Z}_k$ orbifolds with $U(d)$ global symmetry (or $\mathrm{Sp}(d)$ for $k=2$), are presented in Appendix \ref{orbifoldappendix}.

The Hasse diagram is constructed by taking the magnetic quiver of the rank 1 theory, at the top of the diagram, and \emph{subtracting} the quiver of the elementary slice. The resulting quiver has a Coulomb branch that is the closure of the \emph{minimal degeneration}, i.e.\ the leaf below. The prescription for \emph{quiver subtraction} is detailed in \cite{Cabrera:2016vvv,Cabrera:2017njm,Bourget:2019aer}. This process is repeated until one reaches the bottom/trivial leaf.  

When performing quiver subtraction on non-simply laced quivers it is important to align the squircles of the magnetic quiver and the slice to be subtracted. For example, one cannot subtract a quiver where the squircle is on a short node from a quiver where the squircle is on a long node and vice-versa.

\cite[Eq.\ (36), Eq.\ (86)]{Apruzzi:2020pmv} makes a prediction of the dimension $d$ of the lower slice in the Hasse diagram. This is in full agreement with the results of this section. Also, \cite{Apruzzi:2020pmv} correctly identifies that away from the origin the moduli space looks flat, which is consistent with the fact that all lower slices are orbifolds of $\mathbb{H}^d$.
Applying \cite[Eq.\ (86)]{Apruzzi:2020pmv} to the green and blue parts of Table \ref{rank1table} gives $d=1$. This is consistent with the observed property that these cases have accidentally enhanced supersymmetry, and that these spaces have Hasse diagrams with only 2 leaves as opposed to 3 for the other cases. Naturally, all these cases fit into infinite families on their own merit and are presented in Table \ref{orbifolds}. Following \cite{JuliusGrimminger} we conjecture the Hasse diagram of the full moduli space of the rank 1 theories, in 3 dimensions. We now turn to the analysis of each individual case.

\subsection{\texorpdfstring{$C_5$}{C5} rank 1 SCFT}
The global symmetry for this theory is $C_5$, the dimension of the Higgs branch is 16, and there is a generator of the chiral ring at $[00001]_{C_5}$ and scaling dimension $3/2$. These facts lead to a guess for a magnetic quiver among the minimally unbalanced quivers of \cite{Cabrera:2018uvz}, where the unbalanced node is attached to the last node. Luckily, the dimension of this moduli space is indeed 16 and the imbalance indeed leads to a scaling dimension $3/2$. The Hilbert series for the $C_5$ theory, see Table \ref{resultHS}, is consistent with the prediction in \cite[Sec.\ 2.1]{Ohmori:2018ona} obtained by compactifying on a torus with a non-trivial flavour background of the $6d$ $\mathcal{N}=(1,0)$ SCFT, which is the UV completion of an $\mathrm{SU}(2)$ gauge theory with 10 flavours.

The Hasse diagram is given in Table \ref{tab:Hasse_C5} for the Higgs branch of the $4d$ $\mathcal{N}=2$ rank 1 theory as well as the full moduli space of the $3d$ $\mathcal{N}=4$ theory. Notice that the pure Coulomb branch is the leftmost point in the right diagram, which is the geometry $\mathbb{H}/E_6$ (Klein singularity of type $E_6$). This is included as a subvariety in the enhanced Coulomb branch, with transverse slice $c_5=h_{5,2}=\mathbb{H}^5/\mathbb{Z}_2$. This inclusion is the geometric translation of the fact that, in the notation of Section \ref{Nis2}, the 4d Coulomb branch which is of Kodaira type $II^{\ast}$ (with symmetry $E_8$) admits deformations to $\{I_1^6 , I_4\}$ and not $\{I_1^{10}\}$ as would be the case for the $E_8$ theory in Table \ref{rank1table}.

\begin{table}[t]
    \centering
    \begin{tabular}{c||c} \toprule 
        Hasse diagram for the  & Hasse diagram for the full \\ 
        Higgs branch of the 4d theory  &  moduli space of the 3d theory \\ 
        \midrule 
       \raisebox{-.5\height}{
    \scalebox{0.70}{\begin{tikzpicture}
	\begin{pgfonlayer}{nodelayer}
		\node [style=none] (41) at (-5, 0.25) {};
		\node [style=none] (42) at (-5, -2.75) {};
		\node [style=none] (43) at (-5, -6) {};
		\node [style=bd] (44) at (-5, 0.25) {};
		\node [style=bd] (45) at (-5, -2.75) {};
		\node [style=bd] (46) at (-5, -6) {};
		\node [style=none] (60) at (-11, -1) {$e_6$};
		\node [style=none] (97) at (-11, -4) {$c_5$};
		\node [style=gauge3] (171) at (0, 0.25) {};
		\node [style=none] (172) at (0, -0.25) {4};
		\node [style=gauge3] (173) at (-1, 0.25) {};
		\node [style=none] (174) at (-1, -0.25) {3};
		\node [style=gauge3] (175) at (1.15, 0.25) {};
		\node [style=none] (176) at (1.15, -0.25) {5};
		\node [style=none] (177) at (0.15, 0.325) {};
		\node [style=none] (178) at (1.15, 0.325) {};
		\node [style=none] (179) at (0.15, 0.175) {};
		\node [style=none] (180) at (1.15, 0.175) {};
		\node [style=none] (181) at (0.4, 0.25) {};
		\node [style=none] (182) at (0.775, 0.625) {};
		\node [style=none] (183) at (0.775, -0.125) {};
		\node [style=gauge3] (184) at (1.9, 0.25) {};
		\node [style=none] (185) at (1.9, -0.25) {2};
		\node [style=gauge3] (186) at (-2, 0.25) {};
		\node [style=gauge3] (187) at (-3, 0.25) {};
		\node [style=none] (188) at (-3, -0.25) {1};
		\node [style=none] (189) at (-2, -0.25) {2};
		\node [style=gauge3] (190) at (-7.75, -1) {};
		\node [style=none] (191) at (-7.75, -1.5) {3};
		\node [style=gauge3] (192) at (-8.75, -1) {};
		\node [style=none] (193) at (-8.75, -1.5) {2};
		\node [style=gauge3] (194) at (-6.575, -1) {};
		\node [style=none] (195) at (-6.575, -1.5) {4};
		\node [style=none] (196) at (-7.575, -0.925) {};
		\node [style=none] (197) at (-6.575, -0.925) {};
		\node [style=none] (198) at (-7.575, -1.075) {};
		\node [style=none] (199) at (-6.575, -1.075) {};
		\node [style=none] (200) at (-7.325, -1) {};
		\node [style=none] (201) at (-6.95, -0.625) {};
		\node [style=none] (202) at (-6.95, -1.375) {};
		\node [style=gauge3] (203) at (-5.825, -1) {};
		\node [style=none] (204) at (-5.825, -1.5) {2};
		\node [style=gauge3] (205) at (-9.75, -1) {};
		\node [style=none] (208) at (-9.75, -1.5) {1};
		\node [style=gauge3] (209) at (0, -3.25) {};
		\node [style=none] (210) at (0, -3.75) {1};
		\node [style=gauge3] (211) at (-1, -3.25) {};
		\node [style=none] (212) at (-1, -3.75) {1};
		\node [style=gauge3] (213) at (1.175, -3.25) {};
		\node [style=none] (214) at (1.175, -3.75) {1};
		\node [style=none] (215) at (0.175, -3.175) {};
		\node [style=none] (216) at (1.175, -3.175) {};
		\node [style=none] (217) at (0.175, -3.325) {};
		\node [style=none] (218) at (1.175, -3.325) {};
		\node [style=none] (219) at (0.425, -3.25) {};
		\node [style=none] (220) at (0.8, -2.875) {};
		\node [style=none] (221) at (0.8, -3.625) {};
		\node [style=gauge3] (224) at (-2, -3.25) {};
		\node [style=gauge3] (225) at (-3, -3.25) {};
		\node [style=none] (226) at (-3, -3.75) {1};
		\node [style=none] (227) at (-2, -3.75) {1};
		\node [style=flavour2] (228) at (-3, -2.25) {};
		\node [style=none] (229) at (-3, -1.75) {1};
		\node [style=gauge3] (230) at (-7, -4.5) {};
		\node [style=none] (231) at (-7, -5) {1};
		\node [style=gauge3] (232) at (-8, -4.5) {};
		\node [style=none] (233) at (-8, -5) {1};
		\node [style=gauge3] (234) at (-5.825, -4.5) {};
		\node [style=none] (235) at (-5.825, -5) {1};
		\node [style=none] (236) at (-6.825, -4.425) {};
		\node [style=none] (237) at (-5.825, -4.425) {};
		\node [style=none] (238) at (-6.825, -4.575) {};
		\node [style=none] (239) at (-5.825, -4.575) {};
		\node [style=none] (240) at (-6.575, -4.5) {};
		\node [style=none] (241) at (-6.2, -4.125) {};
		\node [style=none] (242) at (-6.2, -4.875) {};
		\node [style=gauge3] (243) at (-9, -4.5) {};
		\node [style=gauge3] (244) at (-10, -4.5) {};
		\node [style=none] (245) at (-10, -5) {1};
		\node [style=none] (246) at (-9, -5) {1};
		\node [style=flavour2] (247) at (-10, -3.5) {};
		\node [style=none] (248) at (-10, -3) {1};
		\node [style=blankflavor] (249) at (1.9, 0.25) {};
		\node [style=blankflavor] (250) at (-5.825, -1) {};
		\node [style=gauge3] (251) at (-1.5, -6) {};
		\node [style=none] (252) at (-1.5, -6.5) {1};
	\end{pgfonlayer}
	\begin{pgfonlayer}{edgelayer}
		\draw (44) to (45);
		\draw (45) to (46);
		\draw (177.center) to (178.center);
		\draw (180.center) to (179.center);
		\draw (182.center) to (181.center);
		\draw (181.center) to (183.center);
		\draw (184) to (175);
		\draw (173) to (171);
		\draw (187) to (186);
		\draw (186) to (173);
		\draw (196.center) to (197.center);
		\draw (199.center) to (198.center);
		\draw (201.center) to (200.center);
		\draw (200.center) to (202.center);
		\draw (203) to (194);
		\draw (192) to (190);
		\draw (205) to (192);
		\draw (215.center) to (216.center);
		\draw (218.center) to (217.center);
		\draw (220.center) to (219.center);
		\draw (219.center) to (221.center);
		\draw (211) to (209);
		\draw (225) to (224);
		\draw (224) to (211);
		\draw (228) to (225);
		\draw (236.center) to (237.center);
		\draw (239.center) to (238.center);
		\draw (241.center) to (240.center);
		\draw (240.center) to (242.center);
		\draw (232) to (230);
		\draw (244) to (243);
		\draw (243) to (232);
		\draw (247) to (244);
	\end{pgfonlayer}
\end{tikzpicture}}
}  & \raisebox{-.5\height}{\scalebox{.7}{ \begin{tikzpicture}
	\begin{pgfonlayer}{nodelayer}
		\node [style=bd] (0) at (-1.5, 2) {};
		\node [style=bd] (1) at (1.5, 2) {};
		\node [style=bd] (2) at (0, 0) {};
		\node [style=none] (3) at (1.25, 1) {$c_5$};
		\node [style=none] (4) at (-1.25, 0.75) {$E_6$};
		\node [style=bd] (5) at (0, 4) {};
		\node [style=bd] (6) at (3, 4) {};
		\node [style=none] (7) at (2.75, 2.75) {$e_6$};
		\node [style=none] (8) at (1.25, 3.25) {$E_6$};
		\node [style=none] (9) at (-1.45, 3.25) {$h_{5,2}=c_5$};
	\end{pgfonlayer}
	\begin{pgfonlayer}{edgelayer}
		\draw [style=blue] (2) to (1);
		\draw [style=red] (2) to (0);
		\draw [style=blue] (0) to (5);
		\draw [style=red] (1) to (5);
		\draw [style=blue] (1) to (6);
	\end{pgfonlayer}
\end{tikzpicture}}} \\ \bottomrule
    \end{tabular}
    \caption{$C_5$ theory. \texttt{Left:} Hasse diagram for the Higgs branch of the $4d$ $\mathcal{N}=2$ rank 1 theory. To the right of the Hasse diagram are the magnetic quivers for the closures of the leaves and to the left are magnetic quivers for the slices. This Hasse diagram is already explored in \cite{Bourget:2019aer,JuliusGrimminger}. \texttt{Right:} Hasse diagram for the full moduli space of the theory in $3d$ $\mathcal{N}=4$.  }
    \label{tab:Hasse_C5}
\end{table}

\subsection{\texorpdfstring{$C_3 \times A_1$}{C3 x A1} rank 1 SCFT}
The computation in Table \ref{resultHS} agrees with the Hilbert series computation in \cite[Eq.\ (40)]{Zafrir:2016wkk}. 
The approach is as follows: The Higgs branches of the $5d$ $\mathcal{N}=1$ theories $T_2$ and $T_3$ have a $\mathbb{Z}_2$ symmetry that we can twist by. Both of the resulting Higgs branches, $T_{2}^{\mathbb{Z}_2 \mathrm{twisted}}$ and $T_{3}^{\mathbb{Z}_2 \mathrm{twisted}}$, have an $SU(2)$ isometry subgroup. Hence, the diagonal $SU(2)$ can be gauged such that the resulting Higgs branch, $T_{2}^{\mathbb{Z}_2 \mathrm{twisted}} \times_{SU(2)} T_{3}^{\mathbb{Z}_2 \mathrm{twisted}}$, becomes the Higgs branch of the $C_3 \times A_1$ rank 1 SCFT. Consequently, the Higgs branch is given by the hyper-K\"ahler quotient
\begin{align}
    \mathcal{H}\left(C_3 \times A_1 \text{ SCFT}\right) =    
    \left( \mathcal{H}\left(T_{2}^{\mathbb{Z}_2 \mathrm{twisted}}\right) 
    \times
    \mathcal{H}\left( T_{3}^{\mathbb{Z}_2 \mathrm{twisted}}\right) \right) /// SU(2) \,.
\end{align}
One can equivalently perform this gauging process from the point of view of magnetic quivers. The $\mathcal{F}_2$ folding of the magnetic quivers for the $5d$ $\mathcal{N}=1$ theories $T_2$ and $T_3$ proceeds as follows:
\begin{equation}
\raisebox{-.5\height}{
\begin{tikzpicture}
	\begin{pgfonlayer}{nodelayer}
		\node [style=none] (0) at (0.75, 5) {$T_3$};
		\node [style=gauge3] (1) at (4, 3) {};
		\node [style=gauge3] (2) at (4, 4) {};
		\node [style=gauge3] (3) at (4, 5) {};
		\node [style=gauge3] (4) at (5, 3) {};
		\node [style=gauge3] (5) at (6, 3) {};
		\node [style=gauge3] (6) at (3, 3) {};
		\node [style=gauge3] (7) at (2, 3) {};
		\node [style=none] (8) at (2, 2.5) {1};
		\node [style=none] (9) at (3, 2.5) {2};
		\node [style=none] (10) at (4, 2.5) {3};
		\node [style=none] (11) at (5, 2.5) {2};
		\node [style=none] (12) at (6, 2.5) {1};
		\node [style=none] (13) at (4.5, 4) {2};
		\node [style=none] (14) at (4.5, 5) {1};
		\node [style=none] (15) at (-4.75, 5) {$T_2$};
		\node [style=gauge3] (16) at (-2.75, 3) {};
		\node [style=gauge3] (17) at (-2.75, 4) {};
		\node [style=gauge3] (19) at (-1.75, 3) {};
		\node [style=gauge3] (21) at (-3.75, 3) {};
		\node [style=none] (24) at (-3.75, 2.5) {1};
		\node [style=none] (25) at (-2.75, 2.5) {2};
		\node [style=none] (26) at (-1.75, 2.5) {1};
		\node [style=none] (28) at (-2.25, 4) {1};
		\node [style=none] (29) at (-2.75, 1.5) {};
		\node [style=none] (30) at (-2.75, 0) {};
		\node [style=none] (31) at (4, 1.5) {};
		\node [style=none] (32) at (4, 0) {};
		\node [style=none] (33) at (-2.25, 0.75) {$\mathcal{F}_2$};
		\node [style=none] (34) at (4.5, 0.75) {$\mathcal{F}_2$};
		\node [style=gauge3] (38) at (-3.75, -1) {};
		\node [style=none] (39) at (-3.75, -1.5) {1};
		\node [style=gauge3] (52) at (1.85, -1) {};
		\node [style=none] (53) at (1.85, -1.5) {1};
		\node [style=gauge3] (98) at (-1.85, -1) {};
		\node [style=none] (99) at (-1.85, -1.5) {1};
		\node [style=none] (100) at (-2.85, -0.925) {};
		\node [style=none] (101) at (-1.85, -0.925) {};
		\node [style=none] (102) at (-2.85, -1.075) {};
		\node [style=none] (103) at (-1.85, -1.075) {};
		\node [style=none] (104) at (-2.175, -1) {};
		\node [style=none] (105) at (-2.55, -0.625) {};
		\node [style=none] (106) at (-2.55, -1.375) {};
		\node [style=gauge3] (107) at (-2.75, -1) {};
		\node [style=none] (108) at (-2.75, -1.5) {2};
		\node [style=gauge3] (109) at (3, -1) {};
		\node [style=none] (110) at (3, -1.5) {2};
		\node [style=gauge3] (111) at (4.9, -1) {};
		\node [style=none] (112) at (4.9, -1.5) {2};
		\node [style=none] (113) at (3.9, -0.925) {};
		\node [style=none] (114) at (4.9, -0.925) {};
		\node [style=none] (115) at (3.9, -1.075) {};
		\node [style=none] (116) at (4.9, -1.075) {};
		\node [style=none] (117) at (4.575, -1) {};
		\node [style=none] (118) at (4.2, -0.625) {};
		\node [style=none] (119) at (4.2, -1.375) {};
		\node [style=gauge3] (120) at (4, -1) {};
		\node [style=none] (121) at (4, -1.5) {3};
		\node [style=gauge3] (122) at (5.85, -1) {};
		\node [style=none] (123) at (5.85, -1.5) {1};
		\node [style=none] (124) at (-4.75, -1) {$\mathcal{C}$};
		\node [style=none] (125) at (-4, 0) {};
		\node [style=none] (126) at (-4, -2) {};
		\node [style=none] (127) at (1.5, 0) {};
		\node [style=none] (128) at (1.5, -2) {};
		\node [style=none] (129) at (6.25, 0) {};
		\node [style=none] (130) at (6.25, -2) {};
		\node [style=none] (131) at (-1.75, 0) {};
		\node [style=none] (132) at (-1.75, -2) {};
		\node [style=none] (133) at (0.75, -1) {$\mathcal{C}$};
		\node [style=none] (134) at (-0.75, 3.25) {=\;$\mathbb{H}^4$};
		\node [style=none] (135) at (7.5, -1) {=\;$\overline{\mathcal{O}}^{\mathfrak{f}_4}_{\text{min}}$};
		\node [style=blankflavor] (136) at (-2.75, -1) {};
		\node [style=blankflavor] (137) at (4, -1) {};
		\node [style=none] (138) at (-4.75, 3.25) {$\mathcal{C}$};
		\node [style=none] (139) at (-4, 4.25) {};
		\node [style=none] (140) at (-4, 2.25) {};
		\node [style=none] (141) at (-1.75, 4.25) {};
		\node [style=none] (142) at (-1.75, 2.25) {};
		\node [style=none] (143) at (1.5, 4.25) {};
		\node [style=none] (144) at (1.5, 2.25) {};
		\node [style=none] (145) at (6.25, 4.25) {};
		\node [style=none] (146) at (6.25, 2.25) {};
		\node [style=none] (147) at (0.75, 3.25) {$\mathcal{C}$};
		\node [style=none] (148) at (-0.75, -1) {=\;$\mathbb{H}^3$};
		\node [style=none] (149) at (7.5, 3.25) {=\;$\overline{\mathcal{O}}^{\mathfrak{e}_6}_{\text{min}}$};
	\end{pgfonlayer}
	\begin{pgfonlayer}{edgelayer}
		\draw (3) to (2);
		\draw (2) to (1);
		\draw (1) to (4);
		\draw (4) to (5);
		\draw (6) to (1);
		\draw (7) to (6);
		\draw (17) to (16);
		\draw (16) to (19);
		\draw (21) to (16);
		\draw [style=->] (29.center) to (30.center);
		\draw [style=->] (31.center) to (32.center);
		\draw (100.center) to (101.center);
		\draw (103.center) to (102.center);
		\draw (105.center) to (104.center);
		\draw (104.center) to (106.center);
		\draw (38) to (107);
		\draw (113.center) to (114.center);
		\draw (116.center) to (115.center);
		\draw (118.center) to (117.center);
		\draw (117.center) to (119.center);
		\draw (109) to (120);
		\draw (109) to (52);
		\draw (122) to (111);
		\draw [bend right=45] (125.center) to (126.center);
		\draw [bend right=45] (127.center) to (128.center);
		\draw [bend left=60] (129.center) to (130.center);
		\draw [bend left=60] (131.center) to (132.center);
		\draw [bend right=45] (139.center) to (140.center);
		\draw [bend left=60] (141.center) to (142.center);
		\draw [bend right=45] (143.center) to (144.center);
		\draw [bend left=60] (145.center) to (146.center);
	\end{pgfonlayer}
\end{tikzpicture}
}
\end{equation}
The folded theories, with the $U(1)$ ungauged on a long node, have Coulomb branches $\mathbb{H}^3$ with $Sp(3)$ global symmetry and the closure of the minimal nilpotent orbit of $F_4$ respectively. Following the prescription in \cite{Hanany:2018vph}, the gauging of an $SU(2)$ subgroup of the Coulomb branch global symmetry of the two magnetic quivers is performed as follows:
\begin{equation}
\raisebox{-.5\height}{
\begin{tikzpicture}
	\begin{pgfonlayer}{nodelayer}
		\node [style=gauge3] (38) at (-1.75, -1.25) {};
		\node [style=none] (39) at (-1.75, -1.75) {1};
		\node [style=gauge3] (52) at (1.85, -1.25) {};
		\node [style=none] (53) at (1.85, -1.75) {1};
		\node [style=gauge3] (98) at (-4.1, -1.25) {};
		\node [style=none] (99) at (-4.1, -1.75) {1};
		\node [style=none] (100) at (-3.1, -1.175) {};
		\node [style=none] (101) at (-4.1, -1.175) {};
		\node [style=none] (102) at (-3.1, -1.325) {};
		\node [style=none] (103) at (-4.1, -1.325) {};
		\node [style=none] (104) at (-3.675, -1.25) {};
		\node [style=none] (105) at (-3.3, -0.875) {};
		\node [style=none] (106) at (-3.3, -1.625) {};
		\node [style=gauge3] (107) at (-3, -1.25) {};
		\node [style=none] (108) at (-3, -1.75) {2};
		\node [style=gauge3] (109) at (3, -1.25) {};
		\node [style=none] (110) at (3, -1.75) {2};
		\node [style=gauge3] (111) at (4.9, -1.25) {};
		\node [style=none] (112) at (4.9, -1.75) {2};
		\node [style=none] (113) at (3.9, -1.175) {};
		\node [style=none] (114) at (4.9, -1.175) {};
		\node [style=none] (115) at (3.9, -1.325) {};
		\node [style=none] (116) at (4.9, -1.325) {};
		\node [style=none] (117) at (4.575, -1.25) {};
		\node [style=none] (118) at (4.2, -0.875) {};
		\node [style=none] (119) at (4.2, -1.625) {};
		\node [style=gauge3] (120) at (4, -1.25) {};
		\node [style=none] (121) at (4, -1.75) {3};
		\node [style=gauge3] (122) at (5.85, -1.25) {};
		\node [style=none] (123) at (5.85, -1.75) {1};
		\node [style=none] (124) at (-2.35, -0.5) {};
		\node [style=none] (125) at (-2.35, -2) {};
		\node [style=none] (126) at (-1.1, -2) {};
		\node [style=none] (127) at (-1.1, -0.5) {};
		\node [style=none] (128) at (-1.75, -0.25) {\color{red}{ $\mathrm{SU}(2)$}};
		\node [style=none] (129) at (1.85, -0.25) {\color{red}{ $\mathrm{SU}(2)$}};
		\node [style=none] (130) at (1.25, -0.5) {};
		\node [style=none] (131) at (1.25, -2) {};
		\node [style=none] (132) at (2.5, -2) {};
		\node [style=none] (133) at (2.5, -0.5) {};
		\node [style=none] (134) at (-1.75, -2.25) {};
		\node [style=none] (135) at (2, -2.25) {};
		\node [style=none] (136) at (1.3, -3.5) {Gauge $SU(2)$};
		\node [style=none] (137) at (0.125, -2.8) {};
		\node [style=none] (138) at (0.125, -4.25) {};
		\node [style=gauge3] (143) at (2.15, -5.25) {};
		\node [style=none] (144) at (2.15, -5.75) {2};
		\node [style=none] (145) at (1.15, -5.175) {};
		\node [style=none] (146) at (2.15, -5.175) {};
		\node [style=none] (147) at (1.15, -5.325) {};
		\node [style=none] (148) at (2.15, -5.325) {};
		\node [style=none] (149) at (1.825, -5.25) {};
		\node [style=none] (150) at (1.45, -4.875) {};
		\node [style=none] (151) at (1.45, -5.625) {};
		\node [style=gauge3] (152) at (1.25, -5.25) {};
		\node [style=none] (153) at (1.25, -5.75) {3};
		\node [style=gauge3] (154) at (3.1, -5.25) {};
		\node [style=none] (155) at (3.1, -5.75) {1};
		\node [style=gauge3] (156) at (-1, -5.25) {};
		\node [style=none] (157) at (-1, -5.75) {1};
		\node [style=none] (158) at (0, -5.175) {};
		\node [style=none] (159) at (-1, -5.175) {};
		\node [style=none] (160) at (0, -5.325) {};
		\node [style=none] (161) at (-1, -5.325) {};
		\node [style=none] (162) at (-0.575, -5.25) {};
		\node [style=none] (163) at (-0.2, -4.875) {};
		\node [style=none] (164) at (-0.2, -5.625) {};
		\node [style=gauge3] (165) at (0.1, -5.25) {};
		\node [style=none] (166) at (0.1, -5.75) {2};
		\node [style=none] (168) at (-1.3, -4.25) {};
		\node [style=none] (169) at (-1.3, -6.25) {};
		\node [style=none] (170) at (-2.05, -5.25) {$\mathcal{C}$};
		\node [style=none] (171) at (3.45, -4.25) {};
		\node [style=none] (172) at (3.45, -6.25) {};
		\node [style=none] (174) at (6.35, -5.25) {$=\mathcal{H}(C_3\times A_1 \;\mathrm{SCFT} )$};
		\node [style=blankflavor] (175) at (1.25, -5.25) {};
		\node [style=blankflavor] (176) at (4, -1.25) {};
		\node [style=blankflavor] (177) at (-3, -1.25) {};
	\end{pgfonlayer}
	\begin{pgfonlayer}{edgelayer}
		\draw (100.center) to (101.center);
		\draw (103.center) to (102.center);
		\draw (105.center) to (104.center);
		\draw (104.center) to (106.center);
		\draw (38) to (107);
		\draw (113.center) to (114.center);
		\draw (116.center) to (115.center);
		\draw (118.center) to (117.center);
		\draw (117.center) to (119.center);
		\draw (109) to (120);
		\draw (109) to (52);
		\draw (122) to (111);
		\draw [style=dotted, red] (124.center) to (127.center);
		\draw [style=dotted, draw=red] (127.center) to (126.center);
		\draw [style=dotted, red] (126.center) to (125.center);
		\draw [style=dotted, red] (125.center) to (124.center);
		\draw [style=dotted, red] (130.center) to (133.center);
		\draw [style=dotted, red] (133.center) to (132.center);
		\draw [style=dotted, red] (132.center) to (131.center);
		\draw [style=dotted, red] (131.center) to (130.center);
		\draw [style=<->, bend right] (134.center) to (135.center);
		\draw [style=->] (137.center) to (138.center);
		\draw (145.center) to (146.center);
		\draw (148.center) to (147.center);
		\draw (150.center) to (149.center);
		\draw (149.center) to (151.center);
		\draw (154) to (143);
		\draw (158.center) to (159.center);
		\draw (161.center) to (160.center);
		\draw (163.center) to (162.center);
		\draw (162.center) to (164.center);
		\draw (165) to (152);
		\draw [bend right=45] (168.center) to (169.center);
		\draw [bend left=45] (171.center) to (172.center);
	\end{pgfonlayer}
\end{tikzpicture}
}
\label{gluing}
\end{equation}
which provides the magnetic quiver of the $C_3\times A_1$ theory. 

\begin{table}[t]
    \centering
    \begin{tabular}{c||c} \toprule 
        Hasse diagram for the  & Hasse diagram for the full \\ 
        Higgs branch of the 4d theory  &  moduli space of the 3d theory \\ 
        \midrule 
       \raisebox{-.5\height}{
    \scalebox{0.70}{\begin{tikzpicture}
	\begin{pgfonlayer}{nodelayer}
		\node [style=none] (41) at (-5, 0.25) {};
		\node [style=none] (42) at (-5, -2.75) {};
		\node [style=none] (43) at (-5, -6) {};
		\node [style=bd] (44) at (-5, 0.25) {};
		\node [style=bd] (45) at (-5, -2.75) {};
		\node [style=bd] (46) at (-5, -6) {};
		\node [style=none] (60) at (-11, -1) {$d_4$};
		\node [style=none] (97) at (-10.925, -4.5) {$c_3$};
		\node [style=gauge3] (308) at (-2.35, 0.25) {};
		\node [style=none] (309) at (-2.35, -0.25) {2};
		\node [style=gauge3] (310) at (-3.35, 0.25) {};
		\node [style=none] (311) at (-3.35, -0.25) {1};
		\node [style=gauge3] (312) at (-1.175, 0.25) {};
		\node [style=none] (313) at (-1.175, -0.25) {3};
		\node [style=none] (314) at (-2.175, 0.325) {};
		\node [style=none] (315) at (-1.175, 0.325) {};
		\node [style=none] (316) at (-2.175, 0.175) {};
		\node [style=none] (317) at (-1.175, 0.175) {};
		\node [style=none] (318) at (-1.925, 0.25) {};
		\node [style=none] (319) at (-1.55, 0.625) {};
		\node [style=none] (320) at (-1.55, -0.125) {};
		\node [style=gauge3] (321) at (-0.425, 0.25) {};
		\node [style=none] (322) at (-0.425, -0.25) {2};
		\node [style=gauge3] (323) at (0.775, 0.25) {};
		\node [style=none] (324) at (-0.425, 0.325) {};
		\node [style=none] (325) at (0.775, 0.325) {};
		\node [style=none] (326) at (-0.425, 0.175) {};
		\node [style=none] (327) at (0.775, 0.175) {};
		\node [style=none] (328) at (0.325, 0.25) {};
		\node [style=none] (329) at (-0.05, 0.625) {};
		\node [style=none] (330) at (-0.05, -0.125) {};
		\node [style=none] (331) at (0.775, -0.25) {1};
		\node [style=gauge3] (386) at (-8.975, -1) {};
		\node [style=none] (387) at (-8.975, -1.5) {1};
		\node [style=gauge3] (390) at (-7.8, -1) {};
		\node [style=none] (391) at (-7.8, -1.5) {2};
		\node [style=none] (392) at (-8.8, -0.925) {};
		\node [style=none] (393) at (-7.8, -0.925) {};
		\node [style=none] (394) at (-8.8, -1.075) {};
		\node [style=none] (395) at (-7.8, -1.075) {};
		\node [style=none] (396) at (-8.55, -1) {};
		\node [style=none] (397) at (-8.175, -0.625) {};
		\node [style=none] (398) at (-8.175, -1.375) {};
		\node [style=gauge3] (399) at (-7.05, -1) {};
		\node [style=none] (400) at (-7.05, -1.5) {2};
		\node [style=gauge3] (401) at (-5.85, -1) {};
		\node [style=none] (402) at (-7.05, -0.925) {};
		\node [style=none] (403) at (-5.85, -0.925) {};
		\node [style=none] (404) at (-7.05, -1.075) {};
		\node [style=none] (405) at (-5.85, -1.075) {};
		\node [style=none] (406) at (-6.3, -1) {};
		\node [style=none] (407) at (-6.675, -0.625) {};
		\node [style=none] (408) at (-6.675, -1.375) {};
		\node [style=none] (409) at (-5.85, -1.5) {1};
		\node [style=gauge3] (410) at (-1.35, -2.775) {};
		\node [style=none] (411) at (-1.35, -3.275) {1};
		\node [style=gauge3] (412) at (-0.175, -2.775) {};
		\node [style=none] (413) at (-0.175, -3.275) {1};
		\node [style=none] (414) at (-1.175, -2.7) {};
		\node [style=none] (415) at (-0.175, -2.7) {};
		\node [style=none] (416) at (-1.175, -2.85) {};
		\node [style=none] (417) at (-0.175, -2.85) {};
		\node [style=none] (418) at (-0.925, -2.775) {};
		\node [style=none] (419) at (-0.55, -2.4) {};
		\node [style=none] (420) at (-0.55, -3.15) {};
		\node [style=blankflavor] (434) at (-1.175, 0.25) {};
		\node [style=blankflavor] (435) at (-7.8, -1) {};
		\node [style=gauge3] (454) at (-2.5, -2.775) {};
		\node [style=none] (455) at (-2.5, -3.275) {1};
		\node [style=flavour2] (456) at (-2.5, -1.775) {};
		\node [style=none] (457) at (-2.5, -1.25) {1};
		\node [style=gauge3] (458) at (-7.075, -4.775) {};
		\node [style=none] (459) at (-7.075, -5.275) {1};
		\node [style=gauge3] (460) at (-5.9, -4.775) {};
		\node [style=none] (461) at (-5.9, -5.275) {1};
		\node [style=none] (462) at (-6.9, -4.7) {};
		\node [style=none] (463) at (-5.9, -4.7) {};
		\node [style=none] (464) at (-6.9, -4.85) {};
		\node [style=none] (465) at (-5.9, -4.85) {};
		\node [style=none] (466) at (-6.65, -4.775) {};
		\node [style=none] (467) at (-6.275, -4.4) {};
		\node [style=none] (468) at (-6.275, -5.15) {};
		\node [style=gauge3] (469) at (-8.225, -4.775) {};
		\node [style=none] (470) at (-8.225, -5.275) {1};
		\node [style=flavour2] (471) at (-8.225, -3.775) {};
		\node [style=none] (472) at (-8.225, -3.25) {1};
		\node [style=gauge3] (473) at (-1.75, -6) {};
		\node [style=none] (474) at (-1.75, -6.5) {1};
	\end{pgfonlayer}
	\begin{pgfonlayer}{edgelayer}
		\draw (44) to (45);
		\draw (45) to (46);
		\draw (314.center) to (315.center);
		\draw (317.center) to (316.center);
		\draw (319.center) to (318.center);
		\draw (318.center) to (320.center);
		\draw (321) to (312);
		\draw (324.center) to (325.center);
		\draw (327.center) to (326.center);
		\draw (329.center) to (328.center);
		\draw (328.center) to (330.center);
		\draw (310) to (308);
		\draw (392.center) to (393.center);
		\draw (395.center) to (394.center);
		\draw (397.center) to (396.center);
		\draw (396.center) to (398.center);
		\draw (399) to (390);
		\draw (402.center) to (403.center);
		\draw (405.center) to (404.center);
		\draw (407.center) to (406.center);
		\draw (406.center) to (408.center);
		\draw (414.center) to (415.center);
		\draw (417.center) to (416.center);
		\draw (419.center) to (418.center);
		\draw (418.center) to (420.center);
		\draw (456) to (454);
		\draw (454) to (410);
		\draw (462.center) to (463.center);
		\draw (465.center) to (464.center);
		\draw (467.center) to (466.center);
		\draw (466.center) to (468.center);
		\draw (471) to (469);
		\draw (469) to (458);
	\end{pgfonlayer}
\end{tikzpicture}}
}  & \raisebox{-.37\height}{\scalebox{.7}{ \begin{tikzpicture}
	\begin{pgfonlayer}{nodelayer}
		\node [style=bd] (0) at (-1.5, 2) {};
		\node [style=bd] (1) at (1.5, 2) {};
		\node [style=bd] (2) at (0, 0) {};
		\node [style=none] (3) at (1.25, 1) {$c_3$};
		\node [style=none] (4) at (-1.25, 0.75) {$D_4$};
		\node [style=bd] (5) at (0, 4) {};
		\node [style=bd] (6) at (3, 4) {};
		\node [style=none] (7) at (2.75, 2.75) {$d_4$};
		\node [style=none] (9) at (-1.45, 3.25) {$h_{3,2}=c_3$};
		\node [style=none] (10) at (1.25, 3.25) {$D_4$};
	\end{pgfonlayer}
	\begin{pgfonlayer}{edgelayer}
		\draw [style=blue] (2) to (1);
		\draw [style=red] (2) to (0);
		\draw [style=blue] (0) to (5);
		\draw [style=red] (1) to (5);
		\draw [style=blue] (1) to (6);
	\end{pgfonlayer}
\end{tikzpicture}}} \\ \bottomrule
    \end{tabular}
    \caption{$C_3\times A_1$ theory. \texttt{Left:} Hasse diagram for the Higgs branch of the $4d$ $\mathcal{N}=2$ rank 1 theory. To the right of the Hasse diagram are the magnetic quivers for the closures of the leaves and to the left are magnetic quivers for the slices. \texttt{Right:} Hasse diagram for the full moduli space of the theory in $3d$ $\mathcal{N}=4$. }
    \label{tab:Hasse_C3xA1}
\end{table}
The Hasse diagram for the Higgs branch of this $4d$ $\mathcal{N}=2$ rank 1 theory is given in Table \ref{tab:Hasse_C3xA1}. As detailed in \cite{Bourget:2019aer}, the non-abelian part of the global symmetry of a magnetic quiver can be identified by taking the product of the global symmetry of the slices connected to the trivial leaf $\{1\}$. In this case, however, there is only one such slice which is the minimal nilpotent orbit $c_3$. The absence of an $A_1$ slice connecting to the trivial leaf may indicate additional rules required for quiver subtraction of non-simply laced quivers or that only a subgroup of the global symmetry can be read off from the Hasse diagram.  The full moduli space of the $3d$ $\mathcal{N}=4$ theory is also tabulated in Table \ref{tab:Hasse_C3xA1}.
\subsection{\texorpdfstring{$C_2 \times U_1$}{C2 x U1} rank 1 SCFT}
For this case we note that the global symmetry is $C_2 \times U_1$ and there is a simple way of getting the magnetic quiver by attaching 2 nodes of 1 to the $C_2$ quiver as in Table \ref{resulttable}. Nicely enough this guess verifies that the dimension is 4 and the two generators in the chiral ring are in $[01]_{C_2}$ with scaling dimension $3/2$ and $U(1)$ charges $\pm1$, respectively. The Hilbert series results of Table \ref{resultHS} agree with \cite[Sec.\ 3.3]{Chacaltana:2014nya} which obtain the rank 1 theory as the class $\mathcal{S}$ theory of a sphere with a minimal untwisted $A_2$ puncture and two maximal twisted $A_2$ punctures connected with a cylinder with a $\mathbb{Z}_2$ twist line.  

\begin{table}[t]
    \centering
    \begin{tabular}{c||c} \toprule 
        Hasse diagram for the  & Hasse diagram for the full \\ 
        Higgs branch of the 4d theory  &  moduli space of the 3d theory \\ 
        \midrule 
       \raisebox{-.5\height}{
    \scalebox{0.70}{\begin{tikzpicture}
	\begin{pgfonlayer}{nodelayer}
		\node [style=none] (41) at (-5, 0.25) {};
		\node [style=none] (42) at (-5, -2.75) {};
		\node [style=none] (43) at (-5, -6) {};
		\node [style=bd] (44) at (-5, 0.25) {};
		\node [style=bd] (45) at (-5, -2.75) {};
		\node [style=bd] (46) at (-5, -6) {};
		\node [style=none] (60) at (-11, -1) {$a_2$};
		\node [style=none] (97) at (-11, -4) {$c_2$};
		\node [style=gauge3] (251) at (-3.35, 0.25) {};
		\node [style=none] (252) at (-3.35, -0.25) {1};
		\node [style=gauge3] (253) at (-2.1, 0.25) {};
		\node [style=none] (254) at (-2.1, -0.25) {2};
		\node [style=none] (255) at (-3.35, 0.325) {};
		\node [style=none] (256) at (-2.1, 0.325) {};
		\node [style=none] (257) at (-3.35, 0.175) {};
		\node [style=none] (258) at (-2.1, 0.175) {};
		\node [style=none] (259) at (-2.85, 0.25) {};
		\node [style=none] (260) at (-2.475, 0.625) {};
		\node [style=none] (261) at (-2.475, -0.125) {};
		\node [style=gauge3] (262) at (-1.1, 0.75) {};
		\node [style=gauge3] (263) at (-1.1, -0.25) {};
		\node [style=none] (264) at (-0.6, 0.75) {1};
		\node [style=none] (265) at (-0.6, -0.25) {1};
		\node [style=blankflavor] (266) at (-2.1, 0.25) {};
		\node [style=gauge3] (269) at (-8.075, -1) {};
		\node [style=none] (270) at (-8.075, -1.5) {1};
		\node [style=none] (272) at (-8.075, -0.925) {};
		\node [style=none] (274) at (-8.075, -1.075) {};
		\node [style=gauge3] (278) at (-7.075, -0.5) {};
		\node [style=gauge3] (279) at (-7.075, -1.5) {};
		\node [style=none] (280) at (-6.575, -0.5) {1};
		\node [style=none] (281) at (-6.575, -1.5) {1};
		\node [style=gauge3] (282) at (-3.35, -3.25) {};
		\node [style=none] (283) at (-3.35, -3.75) {1};
		\node [style=gauge3] (284) at (-2.1, -3.25) {};
		\node [style=none] (285) at (-2.1, -3.75) {1};
		\node [style=none] (286) at (-3.35, -3.175) {};
		\node [style=none] (287) at (-2.1, -3.175) {};
		\node [style=none] (288) at (-3.35, -3.325) {};
		\node [style=none] (289) at (-2.1, -3.325) {};
		\node [style=none] (290) at (-2.85, -3.25) {};
		\node [style=none] (291) at (-2.475, -2.875) {};
		\node [style=none] (292) at (-2.475, -3.625) {};
		\node [style=flavour2] (293) at (-3.35, -2.25) {};
		\node [style=none] (294) at (-3.35, -1.75) {1};
		\node [style=gauge3] (295) at (-8.1, -4.75) {};
		\node [style=none] (296) at (-8.1, -5.25) {1};
		\node [style=gauge3] (297) at (-6.85, -4.75) {};
		\node [style=none] (298) at (-6.85, -5.25) {1};
		\node [style=none] (299) at (-8.1, -4.675) {};
		\node [style=none] (300) at (-6.85, -4.675) {};
		\node [style=none] (301) at (-8.1, -4.825) {};
		\node [style=none] (302) at (-6.85, -4.825) {};
		\node [style=none] (303) at (-7.6, -4.75) {};
		\node [style=none] (304) at (-7.225, -4.375) {};
		\node [style=none] (305) at (-7.225, -5.125) {};
		\node [style=flavour2] (306) at (-8.1, -3.75) {};
		\node [style=none] (307) at (-8.1, -3.25) {1};
		\node [style=gauge3] (308) at (-3, -6) {};
		\node [style=none] (309) at (-3, -6.5) {1};
	\end{pgfonlayer}
	\begin{pgfonlayer}{edgelayer}
		\draw (44) to (45);
		\draw (45) to (46);
		\draw (255.center) to (256.center);
		\draw (258.center) to (257.center);
		\draw (260.center) to (259.center);
		\draw (259.center) to (261.center);
		\draw (262) to (258.center);
		\draw (263) to (256.center);
		\draw (262) to (263);
		\draw (278) to (274.center);
		\draw (279) to (272.center);
		\draw (278) to (279);
		\draw (286.center) to (287.center);
		\draw (289.center) to (288.center);
		\draw (291.center) to (290.center);
		\draw (290.center) to (292.center);
		\draw (293) to (288.center);
		\draw (299.center) to (300.center);
		\draw (302.center) to (301.center);
		\draw (304.center) to (303.center);
		\draw (303.center) to (305.center);
		\draw (306) to (301.center);
	\end{pgfonlayer}
\end{tikzpicture}}
}  & \raisebox{-.5\height}{\scalebox{.7}{ \begin{tikzpicture}
		\begin{pgfonlayer}{nodelayer}
		\node [style=bd] (0) at (-1.5, 2) {};
		\node [style=bd] (1) at (1.5, 2) {};
		\node [style=bd] (2) at (0, 0) {};
		\node [style=none] (3) at (1.25, 1) {$c_2$};
		\node [style=none] (4) at (-1.25, 0.75) {$A_2$};
		\node [style=bd] (5) at (0, 4) {};
		\node [style=bd] (6) at (3, 4) {};
		\node [style=none] (7) at (2.75, 2.75) {$a_2$};
		\node [style=none] (9) at (-1.45, 3.25) {$h_{2,2}=c_2$};
		\node [style=none] (10) at (1.25, 3.25) {$A_2$};
	\end{pgfonlayer}
	\begin{pgfonlayer}{edgelayer}
		\draw [style=blue] (2) to (1);
		\draw [style=red] (2) to (0);
		\draw [style=blue] (0) to (5);
		\draw [style=red] (1) to (5);
		\draw [style=blue] (1) to (6);
	\end{pgfonlayer}
\end{tikzpicture}}} \\ \bottomrule
    \end{tabular}
    \caption{$C_2\times U_1$ theory.  \texttt{Left:} Hasse diagram for the Higgs branch of the $4d$ $\mathcal{N}=2$ rank 1 theory. To the right of the Hasse diagram are the magnetic quivers for the closures of the leaves and to the left are magnetic quivers for the slices. \texttt{Right:} Hasse diagram for the full moduli space of the theory in $3d$ $\mathcal{N}=4$. }
    \label{tab:Hasse_C2xU1}
\end{table}
The Hasse diagram is given in Table \ref{tab:Hasse_C2xU1} for the Higgs branch of the $4d$ $\mathcal{N}=2$ rank 1 theory as well as the full moduli space of the $3d$ $\mathcal{N}=4$ theory.

\subsection{\texorpdfstring{$A_3$}{A3} rank 1 SCFT}
For this case, the global symmetry and the $Z_3$ twist lead to a very natural guess of folding a theory that has an $SU(4)^3$ global symmetry. There is a very natural candidate for such a theory. The so called $T_4$ theory. This leads to the guess as in Table \ref{resulttable}. This guess is verified by a set of consistency checks like the dimension of the Higgs branch, the generators in the chiral ring, etc.
The Hilbert series in Table \ref{resultHS} is consistent with the prediction of \cite[Eq.\ 5.7]{Chacaltana:2016shw} obtained by a class $\mathcal{S}$ construction of compactifying a $\mathbb{Z}_3$ twisted $D_4$ theory. Moreover, the result also agrees with the prediction in \cite[Sec.\ 2.2]{Ohmori:2018ona} obtained by compactifying on a torus with a non-trivial flavour background of the $6d$ $\mathcal{N}=(1,0)$ SCFT, which is the UV completion of an $\mathrm{SU}(3)$ gauge theory with 12 flavours.  

The Hasse diagram is given in Table \ref{tab:Hasse_A3} for the Higgs branch of the $4d$ $\mathcal{N}=2$ rank 1 theory as well as the full moduli space of the $3d$ $\mathcal{N}=4$ theory. 

\begin{table}[t]
    \centering
    \begin{tabular}{c||c} \toprule 
        Hasse diagram for the  & Hasse diagram for the full \\ 
        Higgs branch of the 4d theory  &  moduli space of the 3d theory \\ 
        \midrule 
       \raisebox{-.5\height}{
    \scalebox{0.70}{\begin{tikzpicture}
	\begin{pgfonlayer}{nodelayer}
		\node [style=gauge3] (2) at (-1, 0.25) {};
		\node [style=none] (6) at (-1, -0.25) {3};
		\node [style=gauge3] (24) at (0.5, 0.25) {};
		\node [style=none] (29) at (0.5, -0.25) {4};
		\node [style=none] (30) at (-1, 0.4) {};
		\node [style=none] (31) at (0.5, 0.4) {};
		\node [style=none] (32) at (-1, 0.1) {};
		\node [style=none] (33) at (0.5, 0.1) {};
		\node [style=none] (34) at (-0.5, 0.25) {};
		\node [style=none] (35) at (0, 0.75) {};
		\node [style=none] (36) at (0, -0.25) {};
		\node [style=gauge3] (37) at (-2, 0.25) {};
		\node [style=gauge3] (38) at (-3, 0.25) {};
		\node [style=none] (39) at (-2, -0.25) {2};
		\node [style=none] (40) at (-3, -0.25) {1};
		\node [style=none] (41) at (-5, 0.25) {};
		\node [style=none] (42) at (-5, -2.75) {};
		\node [style=none] (43) at (-5, -6) {};
		\node [style=bd] (44) at (-5, 0.25) {};
		\node [style=bd] (45) at (-5, -2.75) {};
		\node [style=bd] (46) at (-5, -6) {};
		\node [style=gauge3] (47) at (-7.25, -1.25) {};
		\node [style=none] (48) at (-7.25, -1.75) {2};
		\node [style=gauge3] (49) at (-5.75, -1.25) {};
		\node [style=none] (50) at (-5.75, -1.75) {3};
		\node [style=none] (51) at (-7.25, -1.1) {};
		\node [style=none] (52) at (-5.75, -1.1) {};
		\node [style=none] (53) at (-7.25, -1.4) {};
		\node [style=none] (54) at (-5.75, -1.4) {};
		\node [style=none] (55) at (-6.75, -1.25) {};
		\node [style=none] (56) at (-6.25, -0.75) {};
		\node [style=none] (57) at (-6.25, -1.75) {};
		\node [style=gauge3] (58) at (-8.25, -1.25) {};
		\node [style=none] (59) at (-8.25, -1.75) {1};
		\node [style=none] (60) at (-10.25, -1.25) {$d_4$};
		\node [style=gauge3] (61) at (-1, -3.25) {};
		\node [style=none] (62) at (-1, -3.75) {1};
		\node [style=gauge3] (63) at (0.5, -3.25) {};
		\node [style=none] (64) at (0.5, -3.75) {1};
		\node [style=none] (65) at (-1, -3.1) {};
		\node [style=none] (66) at (0.5, -3.1) {};
		\node [style=none] (67) at (-1, -3.4) {};
		\node [style=none] (68) at (0.5, -3.4) {};
		\node [style=none] (69) at (-0.5, -3.25) {};
		\node [style=none] (70) at (0, -2.75) {};
		\node [style=none] (71) at (0, -3.75) {};
		\node [style=gauge3] (72) at (-2, -3.25) {};
		\node [style=gauge3] (73) at (-3, -3.25) {};
		\node [style=none] (74) at (-2, -3.75) {1};
		\node [style=none] (75) at (-3, -3.75) {1};
		\node [style=blankflavor] (76) at (0.5, 0.25) {};
		\node [style=blankflavor] (77) at (-5.75, -1.25) {};
		\node [style=flavour2] (78) at (-3, -2.25) {};
		\node [style=none] (79) at (-3, -1.75) {1};
		\node [style=gauge3] (80) at (-7.25, -4.75) {};
		\node [style=none] (81) at (-7.25, -5.25) {1};
		\node [style=gauge3] (82) at (-5.75, -4.75) {};
		\node [style=none] (83) at (-5.75, -5.25) {1};
		\node [style=none] (84) at (-7.25, -4.6) {};
		\node [style=none] (85) at (-5.75, -4.6) {};
		\node [style=none] (86) at (-7.25, -4.9) {};
		\node [style=none] (87) at (-5.75, -4.9) {};
		\node [style=none] (88) at (-6.75, -4.75) {};
		\node [style=none] (89) at (-6.25, -4.25) {};
		\node [style=none] (90) at (-6.25, -5.25) {};
		\node [style=gauge3] (91) at (-8.25, -4.75) {};
		\node [style=gauge3] (92) at (-9.25, -4.75) {};
		\node [style=none] (93) at (-8.25, -5.25) {1};
		\node [style=none] (94) at (-9.25, -5.25) {1};
		\node [style=flavour2] (95) at (-9.25, -3.75) {};
		\node [style=none] (96) at (-9.25, -3.25) {1};
		\node [style=none] (97) at (-10.75, -4.75) {$h_{4,3}$};
		\node [style=gauge3] (98) at (-1.25, -6) {};
		\node [style=none] (99) at (-1.25, -6.5) {1};
	\end{pgfonlayer}
	\begin{pgfonlayer}{edgelayer}
		\draw (2) to (24);
		\draw (30.center) to (31.center);
		\draw (33.center) to (32.center);
		\draw (35.center) to (34.center);
		\draw (34.center) to (36.center);
		\draw (37) to (2);
		\draw (38) to (37);
		\draw (41.center) to (42.center);
		\draw (42.center) to (43.center);
		\draw (47) to (49);
		\draw (51.center) to (52.center);
		\draw (54.center) to (53.center);
		\draw (56.center) to (55.center);
		\draw (55.center) to (57.center);
		\draw (58) to (47);
		\draw (61) to (63);
		\draw (65.center) to (66.center);
		\draw (68.center) to (67.center);
		\draw (70.center) to (69.center);
		\draw (69.center) to (71.center);
		\draw (72) to (61);
		\draw (73) to (72);
		\draw (78) to (73);
		\draw (80) to (82);
		\draw (84.center) to (85.center);
		\draw (87.center) to (86.center);
		\draw (89.center) to (88.center);
		\draw (88.center) to (90.center);
		\draw (91) to (80);
		\draw (92) to (91);
		\draw (95) to (92);
	\end{pgfonlayer}
\end{tikzpicture}}
}  & \raisebox{-.37\height}{\scalebox{.8}{ \begin{tikzpicture}
		\begin{pgfonlayer}{nodelayer}
		\node [style=bd] (0) at (-1.5, 2) {};
		\node [style=bd] (1) at (1.5, 2) {};
		\node [style=bd] (2) at (0, 0) {};
		\node [style=none] (3) at (1.25, 1) {$h_{4,3}$};
		\node [style=none] (4) at (-1.25, 0.75) {$D_4$};
		\node [style=bd] (5) at (0, 4) {};
		\node [style=bd] (6) at (3, 4) {};
		\node [style=none] (7) at (2.75, 2.75) {$d_4$};
		\node [style=none] (9) at (-1.25, 3.25) {$h_{4,3}$};
		\node [style=none] (10) at (1.25, 3.25) {$D_4$};
	\end{pgfonlayer}
	\begin{pgfonlayer}{edgelayer}
		\draw [style=blue] (2) to (1);
		\draw [style=red] (2) to (0);
		\draw [style=blue] (0) to (5);
		\draw [style=red] (1) to (5);
		\draw [style=blue] (1) to (6);
	\end{pgfonlayer}
\end{tikzpicture}}} \\ \bottomrule
    \end{tabular}
    \caption{$A_3$ theory.  \texttt{Left:} Hasse diagram for the Higgs branch of the $4d$ $\mathcal{N}=2$ rank 1 theory. To the right of the Hasse diagram are the magnetic quivers for the closures of the leaves and to the left are magnetic quivers for the slices. \texttt{Right:} Hasse diagram for the full moduli space of the theory in $3d$ $\mathcal{N}=4$. }
    \label{tab:Hasse_A3}
\end{table}
\subsection{\texorpdfstring{$A_1\times U_1$}{A1 x U1} rank 1 SCFT}
The global symmetry of the Coulomb branch of a quiver can be read from the balance of its nodes. The low rank of the global symmetry for this SCFT places a very strong restriction on the form of the quiver. The twist is $\mathbb{Z}_3$, implying that there should be a triply-laced edge. These conditions make the magnetic quiver in Table \ref{resulttable} a very natural guess. 

In addition, we also conjecture an explicit construction of the Higgs branch using a hyper-K\"ahler quotient. We check that these two independent descriptions of the Higgs branch are consistent, by computing the Hilbert series in both cases and find perfect agreement.

The Higgs branch of the $A_1 \times U_1$ SCFT is conjectured to be given by the hyper-K\"ahler quotient:
\begin{align}
    \mathcal{H}\left(A_1 \times U_1 \text{ SCFT}\right) =    
    \left( \overline{\mathcal{O}}^{\text{min}}_{G_2} \times\mathbb{H}^2
    \times
    \mathbb{C}^2/\mathbb{Z}_2  \right) /// SU(2) \,.
\end{align}
Explicitly, we first take the refined Hilbert series $\mathrm{HS}_{\overline{\mathcal{O}}^{\text{min}}_{G_2} }(x_1,x_2,t)$, where $x_1,x_2$ are the fundamental weight fugacities of $G_2$ and  decompose it with respect to the branching $G_2 \rightarrow SU(2)\times SU(2) $. The resulting Hilbert series is   $\mathrm{HS}_{\overline{\mathcal{O}}^{\text{min}}_{G_2} }(y,z,t)$ where $y$ and $z$ are the weight fugacities of $SU(2)\times SU(2)$ and is given in \cite[Tab.\ 10]{Hanany:2015hxa}. The $\mathbb{H}^2$ theory can be described by the Higgs branch of $[SU(2)]-[O2]$ with Hilbert series $ \mathrm{HS}_{\mathbb{H}^2 }(z,q,t) $, where $z$ is the fugacity for $SU(2)$ and $q$ is the fugacity for $O(2)$.
The hyper-K\"ahler quotient then takes the form:
\begin{align}
\begin{aligned}
    \mathrm{HS}_{A_1\times U_1}(y,q,t)  =  \oint_{\mathrm{SU}(2)} d\mu_{\mathrm{SU}(2)} &\mathrm{HS}_{\overline{\mathcal{O}}^{\text{min}}_{G_2} }(y,z,t) \mathrm{HS}_{\mathbb{H}^2 }(z,q,t) \mathrm{HS}_{\mathbb{C}^2/\mathbb{Z}_2}(z,t)  \\ &  \qquad  \cdot (1-z^2t^2)\left(1-\frac{1}{z^2}t^2 \right)(1-t^2)
    \end{aligned}
\end{align}
which is consistent with the result in Table \ref{resultHS}. Following an analagous gluing process in \eqref{gluing}, one would expect:
\begin{equation}
\raisebox{-.5\height}{
\begin{tikzpicture}
	\begin{pgfonlayer}{nodelayer}
		\node [style=none] (13) at (1, 1) {$\mathbb{H}^2$};
		\node [style=gauge3] (15) at (2.25, -4.25) {};
		\node [style=none] (16) at (1, -4.125) {};
		\node [style=none] (17) at (2.25, -4.125) {};
		\node [style=none] (18) at (1, -4.375) {};
		\node [style=none] (19) at (2.25, -4.375) {};
		\node [style=none] (20) at (4.75, 1) {$\mathbb{C}^2/\mathbb{Z}_2$};
		\node [style=none] (21) at (-2.5, -1.5) {};
		\node [style=none] (22) at (4.75, -1.5) {};
		\node [style=none] (23) at (1, -3.5) {};
		\node [style=none] (24) at (1, -2.1) {};
		\node [style=none] (45) at (2.25, -4.75) {1};
		\node [style=none] (47) at (2.5, -2.75) {Gauge $SU(2)$};
		\node [style=none] (59) at (-4.5, -1) {1};
		\node [style=none] (60) at (-3, -1) {2};
		\node [style=none] (61) at (-3.5, 1) {$\overline{\mathcal{O}}^{\text{min}}_{G_2}$};
		\node [style=none] (63) at (-1.75, -1) {1};
		\node [style=gauge3] (64) at (-4.5, -0.5) {};
		\node [style=gauge3] (66) at (-3, -0.5) {};
		\node [style=none] (68) at (-4.5, -0.35) {};
		\node [style=none] (69) at (-3, -0.35) {};
		\node [style=none] (70) at (-4.5, -0.65) {};
		\node [style=none] (71) at (-3, -0.65) {};
		\node [style=none] (72) at (-4, -0.5) {};
		\node [style=none] (73) at (-3.5, 0) {};
		\node [style=none] (74) at (-3.5, -1) {};
		\node [style=blankflavor] (75) at (-3, -0.5) {};
		\node [style=gauge3] (76) at (-1.75, -0.5) {};
		\node [style=none] (77) at (0.25, -1) {1};
		\node [style=none] (78) at (1.75, -1) {2};
		\node [style=gauge3] (80) at (0.25, -0.5) {};
		\node [style=gauge3] (81) at (1.75, -0.5) {};
		\node [style=none] (82) at (0.25, -0.35) {};
		\node [style=none] (83) at (1.75, -0.35) {};
		\node [style=none] (84) at (0.25, -0.65) {};
		\node [style=none] (85) at (1.75, -0.65) {};
		\node [style=none] (86) at (0.75, -0.5) {};
		\node [style=none] (87) at (1.25, 0) {};
		\node [style=none] (88) at (1.25, -1) {};
		\node [style=blankflavor] (89) at (1.75, -0.5) {};
		\node [style=none] (90) at (-0.5, -4.75) {1};
		\node [style=none] (91) at (1, -4.75) {2};
		\node [style=gauge3] (92) at (-0.5, -4.25) {};
		\node [style=gauge3] (93) at (1, -4.25) {};
		\node [style=none] (94) at (-0.5, -4.1) {};
		\node [style=none] (95) at (1, -4.1) {};
		\node [style=none] (96) at (-0.5, -4.4) {};
		\node [style=none] (97) at (1, -4.4) {};
		\node [style=none] (98) at (0, -4.25) {};
		\node [style=none] (99) at (0.5, -3.75) {};
		\node [style=none] (100) at (0.5, -4.75) {};
		\node [style=blankflavor] (101) at (1, -4.25) {};
		\node [style=gauge3] (102) at (5.75, -0.5) {};
		\node [style=none] (103) at (4.5, -0.375) {};
		\node [style=none] (104) at (5.75, -0.375) {};
		\node [style=none] (105) at (4.5, -0.625) {};
		\node [style=none] (106) at (5.75, -0.625) {};
		\node [style=none] (107) at (5.75, -1) {1};
		\node [style=none] (108) at (4.5, -1) {1};
		\node [style=gauge3] (109) at (4.5, -0.5) {};
		\node [style=none] (110) at (4.5, -0.35) {};
		\node [style=none] (111) at (4.5, -0.65) {};
	\end{pgfonlayer}
	\begin{pgfonlayer}{edgelayer}
		\draw (17.center) to (16.center);
		\draw (18.center) to (19.center);
		\draw [bend right=15] (21.center) to (22.center);
		\draw [style=->] (24.center) to (23.center);
		\draw (64) to (66);
		\draw (68.center) to (69.center);
		\draw (71.center) to (70.center);
		\draw (73.center) to (72.center);
		\draw (72.center) to (74.center);
		\draw (76) to (75);
		\draw (80) to (81);
		\draw (82.center) to (83.center);
		\draw (85.center) to (84.center);
		\draw (87.center) to (86.center);
		\draw (86.center) to (88.center);
		\draw (92) to (93);
		\draw (94.center) to (95.center);
		\draw (97.center) to (96.center);
		\draw (99.center) to (98.center);
		\draw (98.center) to (100.center);
		\draw (104.center) to (103.center);
		\draw (105.center) to (106.center);
	\end{pgfonlayer}
\end{tikzpicture}
}
\end{equation}
Although exactly how to do this gluing process as an action on magnetic quivers is unclear. 

The Hasse diagram is given in Table \ref{tab:Hasse_A1xU1} for the Higgs branch of the $4d$ $\mathcal{N}=2$ rank 1 theory as well as the full moduli space of the $3d$ $\mathcal{N}=4$ theory. 
\begin{table}[t]
    \centering
    \begin{tabular}{c||c} \toprule 
        Hasse diagram for the  & Hasse diagram for the full \\ 
        Higgs branch of the 4d theory  &  moduli space of the 3d theory \\ 
        \midrule 
       \raisebox{-.5\height}{
    \scalebox{0.70}{\begin{tikzpicture}
	\begin{pgfonlayer}{nodelayer}
		\node [style=none] (41) at (-5, 0.25) {};
		\node [style=none] (42) at (-5, -2.75) {};
		\node [style=none] (43) at (-5, -6) {};
		\node [style=bd] (44) at (-5, 0.25) {};
		\node [style=bd] (45) at (-5, -2.75) {};
		\node [style=bd] (46) at (-5, -6) {};
		\node [style=none] (60) at (-11, -1) {$A_1$};
		\node [style=none] (97) at (-10.925, -4.5) {$h_{2,3}$};
		\node [style=gauge3] (436) at (-3.325, 0.25) {};
		\node [style=none] (437) at (-3.325, -0.25) {1};
		\node [style=gauge3] (438) at (-1.825, 0.25) {};
		\node [style=none] (439) at (-1.825, -0.25) {2};
		\node [style=none] (440) at (-3.325, 0.4) {};
		\node [style=none] (441) at (-1.825, 0.4) {};
		\node [style=none] (442) at (-3.325, 0.1) {};
		\node [style=none] (443) at (-1.825, 0.1) {};
		\node [style=none] (444) at (-2.825, 0.25) {};
		\node [style=none] (445) at (-2.325, 0.75) {};
		\node [style=none] (446) at (-2.325, -0.25) {};
		\node [style=gauge3] (447) at (-0.825, 0.25) {};
		\node [style=none] (448) at (-1.825, 0.325) {};
		\node [style=none] (449) at (-1.825, 0.15) {};
		\node [style=none] (450) at (-0.825, 0.325) {};
		\node [style=none] (451) at (-0.825, 0.15) {};
		\node [style=none] (452) at (-0.825, -0.25) {1};
		\node [style=blankflavor] (453) at (-1.8, 0.25) {};
		\node [style=gauge3] (454) at (-3.25, -3) {};
		\node [style=none] (455) at (-3.25, -3.5) {1};
		\node [style=gauge3] (456) at (-1.75, -3) {};
		\node [style=none] (457) at (-1.75, -3.5) {1};
		\node [style=none] (458) at (-3.25, -2.85) {};
		\node [style=none] (459) at (-1.75, -2.85) {};
		\node [style=none] (460) at (-3.25, -3.15) {};
		\node [style=none] (461) at (-1.75, -3.15) {};
		\node [style=none] (462) at (-2.75, -3) {};
		\node [style=none] (463) at (-2.25, -2.5) {};
		\node [style=none] (464) at (-2.25, -3.5) {};
		\node [style=none] (466) at (-1.75, -2.925) {};
		\node [style=none] (467) at (-1.75, -3.1) {};
		\node [style=flavour2] (468) at (-3.275, -2) {};
		\node [style=none] (469) at (-3.275, -1.5) {1};
		\node [style=gauge3] (470) at (-8.325, -1.025) {};
		\node [style=none] (471) at (-8.325, -1.525) {1};
		\node [style=gauge3] (472) at (-6.9, -1.025) {};
		\node [style=none] (473) at (-6.9, -1.525) {1};
		\node [style=none] (474) at (-8.15, -0.95) {};
		\node [style=none] (475) at (-6.9, -0.95) {};
		\node [style=none] (476) at (-8.15, -1.1) {};
		\node [style=none] (477) at (-6.9, -1.1) {};
		\node [style=none] (478) at (-7.9, -1.025) {};
		\node [style=gauge3] (479) at (-8.3, -4.5) {};
		\node [style=none] (480) at (-8.3, -5) {1};
		\node [style=gauge3] (481) at (-6.8, -4.5) {};
		\node [style=none] (482) at (-6.8, -5) {1};
		\node [style=none] (483) at (-8.3, -4.35) {};
		\node [style=none] (484) at (-6.8, -4.35) {};
		\node [style=none] (485) at (-8.3, -4.65) {};
		\node [style=none] (486) at (-6.8, -4.65) {};
		\node [style=none] (487) at (-7.8, -4.5) {};
		\node [style=none] (488) at (-7.3, -4) {};
		\node [style=none] (489) at (-7.3, -5) {};
		\node [style=none] (490) at (-6.8, -4.425) {};
		\node [style=none] (491) at (-6.8, -4.6) {};
		\node [style=flavour2] (492) at (-8.325, -3.5) {};
		\node [style=none] (493) at (-8.325, -3) {1};
		\node [style=gauge3] (494) at (-2.5, -6) {};
		\node [style=none] (495) at (-2.5, -6.5) {1};
	\end{pgfonlayer}
	\begin{pgfonlayer}{edgelayer}
		\draw (44) to (45);
		\draw (45) to (46);
		\draw (436) to (438);
		\draw (440.center) to (441.center);
		\draw (443.center) to (442.center);
		\draw (445.center) to (444.center);
		\draw (444.center) to (446.center);
		\draw (450.center) to (448.center);
		\draw (449.center) to (451.center);
		\draw (454) to (456);
		\draw (458.center) to (459.center);
		\draw (461.center) to (460.center);
		\draw (463.center) to (462.center);
		\draw (462.center) to (464.center);
		\draw (468) to (458.center);
		\draw (474.center) to (475.center);
		\draw (477.center) to (476.center);
		\draw (479) to (481);
		\draw (483.center) to (484.center);
		\draw (486.center) to (485.center);
		\draw (488.center) to (487.center);
		\draw (487.center) to (489.center);
		\draw (492) to (483.center);
	\end{pgfonlayer}
\end{tikzpicture}}
}  & \raisebox{-0.4\height}{\scalebox{.7}{\begin{tikzpicture}
	\begin{pgfonlayer}{nodelayer}
		\node [style=bd] (0) at (-1.5, 2) {};
		\node [style=bd] (1) at (1.5, 2) {};
		\node [style=bd] (2) at (0, 0) {};
		\node [style=none] (3) at (1.25, 1) {$h_{2,3}$};
		\node [style=none] (4) at (-1.25, 0.75) {$a_1$};
		\node [style=bd] (5) at (0, 4) {};
		\node [style=bd] (6) at (3, 4) {};
		\node [style=none] (7) at (2.75, 2.75) {$A_1$};
		\node [style=none] (9) at (-1.25, 3.25) {$h_{2,3}$};
		\node [style=none] (10) at (1.25, 3.25) {$a_1$};
	\end{pgfonlayer}
	\begin{pgfonlayer}{edgelayer}
		\draw [style=blue] (2) to (1);
		\draw [style=red] (2) to (0);
		\draw [style=blue] (0) to (5);
		\draw [style=red] (1) to (5);
		\draw [style=blue] (1) to (6);
	\end{pgfonlayer}
\end{tikzpicture}}} \\ \bottomrule
    \end{tabular}
    \caption{$A_1\times U_1$ theory.  \texttt{Left:} Hasse diagram for the Higgs branch of the $4d$  $\mathcal{N}=2$ rank 1 theory. To the right of the Hasse diagram are the magnetic quivers for the closures of the leaves and to the left are magnetic quivers for the slices. \texttt{Right:} Hasse diagram for the full moduli space of the theory in $3d$ $\mathcal{N}=4$. }
    \label{tab:Hasse_A1xU1}
\end{table}
This Hasse diagram is given in Table \ref{tab:Hasse_A2} for the Higgs branch of the $4d$ $\mathcal{N}=2$ rank 1 theory as well as the full moduli space of the $3d$ $\mathcal{N}=4$ theory. 
\subsection{\texorpdfstring{$A_2$}{A2} rank 1 SCFT}
For this case we repeat the guess which is made for the $A_3$ theory, with adaptation of the details. The twist is $\mathbb{Z}_4$, implying that there should be 4 identical legs, each with an $SU(3)$ global symmetry. Folding this quiver leads to the magnetic quiver in Table \ref{resulttable}.
The Hilbert series results of Table \ref{resultHS} are consistent with the prediction in \cite[Sec.\ 2.3]{Ohmori:2018ona} obtained by compactifying on a torus with a non-trivial flavor background of the $6d$ $\mathcal{N}=(1,0)$ SCFT, which is the UV completion of an $\mathrm{SU}(4)$ gauge theory with 12 flavours in the fundamental representation and one flavour in the antisymmetric representation.  

\begin{table}[t]
    \centering
    \begin{tabular}{c||c} \toprule 
        Hasse diagram for the  & Hasse diagram for the full \\ 
        Higgs branch of the 4d theory  &  moduli space of the 3d theory \\ 
        \midrule 
       \raisebox{-.5\height}{
    \scalebox{0.70}{\begin{tikzpicture}
	\begin{pgfonlayer}{nodelayer}
		\node [style=none] (41) at (-5, 0.25) {};
		\node [style=none] (42) at (-5, -2.75) {};
		\node [style=none] (43) at (-5, -6) {};
		\node [style=bd] (44) at (-5, 0.25) {};
		\node [style=bd] (45) at (-5, -2.75) {};
		\node [style=bd] (46) at (-5, -6) {};
		\node [style=none] (60) at (-10, -1) {$a_2$};
		\node [style=blankflavor] (76) at (-0.5, 0.25) {};
		\node [style=none] (97) at (-10.75, -4.75) {$h_{3,4}$};
		\node [style=gauge3] (98) at (-2, 0.25) {};
		\node [style=none] (99) at (-2, -0.25) {2};
		\node [style=gauge3] (100) at (-0.5, 0.25) {};
		\node [style=none] (101) at (-0.5, -0.25) {3};
		\node [style=none] (102) at (-2, 0.3) {};
		\node [style=none] (103) at (-0.5, 0.3) {};
		\node [style=none] (104) at (-2, 0.175) {};
		\node [style=none] (105) at (-0.5, 0.175) {};
		\node [style=none] (106) at (-1.5, 0.25) {};
		\node [style=none] (107) at (-1, 0.75) {};
		\node [style=none] (108) at (-1, -0.25) {};
		\node [style=none] (109) at (-2, 0.425) {};
		\node [style=none] (110) at (-0.5, 0.425) {};
		\node [style=none] (111) at (-2, 0.05) {};
		\node [style=none] (112) at (-0.5, 0.05) {};
		\node [style=gauge3] (113) at (-3, 0.25) {};
		\node [style=none] (114) at (-3, -0.25) {1};
		\node [style=blankflavor] (115) at (-6, -1) {};
		\node [style=gauge3] (116) at (-7.5, -1) {};
		\node [style=none] (117) at (-7.5, -1.5) {1};
		\node [style=gauge3] (118) at (-6, -1) {};
		\node [style=none] (119) at (-6, -1.5) {2};
		\node [style=none] (120) at (-7.5, -0.95) {};
		\node [style=none] (121) at (-6, -0.95) {};
		\node [style=none] (122) at (-7.5, -1.075) {};
		\node [style=none] (123) at (-6, -1.075) {};
		\node [style=none] (124) at (-7, -1) {};
		\node [style=none] (125) at (-6.5, -0.5) {};
		\node [style=none] (126) at (-6.5, -1.5) {};
		\node [style=none] (127) at (-7.5, -0.825) {};
		\node [style=none] (128) at (-6, -0.825) {};
		\node [style=none] (129) at (-7.5, -1.2) {};
		\node [style=none] (130) at (-6, -1.2) {};
		\node [style=gauge3] (132) at (-2, -3.25) {};
		\node [style=none] (133) at (-2, -3.75) {1};
		\node [style=gauge3] (134) at (-0.5, -3.25) {};
		\node [style=none] (135) at (-0.5, -3.75) {1};
		\node [style=none] (136) at (-2, -3.2) {};
		\node [style=none] (137) at (-0.5, -3.2) {};
		\node [style=none] (138) at (-2, -3.325) {};
		\node [style=none] (139) at (-0.5, -3.325) {};
		\node [style=none] (140) at (-1.5, -3.25) {};
		\node [style=none] (141) at (-1, -2.75) {};
		\node [style=none] (142) at (-1, -3.75) {};
		\node [style=none] (143) at (-2, -3.075) {};
		\node [style=none] (144) at (-0.5, -3.075) {};
		\node [style=none] (145) at (-2, -3.45) {};
		\node [style=none] (146) at (-0.5, -3.45) {};
		\node [style=gauge3] (147) at (-3, -3.25) {};
		\node [style=none] (148) at (-3, -3.75) {1};
		\node [style=flavour2] (149) at (-3, -2.25) {};
		\node [style=none] (150) at (-3, -1.75) {1};
		\node [style=gauge3] (152) at (-7.5, -4.5) {};
		\node [style=none] (153) at (-7.5, -5) {1};
		\node [style=gauge3] (154) at (-6, -4.5) {};
		\node [style=none] (155) at (-6, -5) {1};
		\node [style=none] (156) at (-7.5, -4.45) {};
		\node [style=none] (157) at (-6, -4.45) {};
		\node [style=none] (158) at (-7.5, -4.575) {};
		\node [style=none] (159) at (-6, -4.575) {};
		\node [style=none] (160) at (-7, -4.5) {};
		\node [style=none] (161) at (-6.5, -4) {};
		\node [style=none] (162) at (-6.5, -5) {};
		\node [style=none] (163) at (-7.5, -4.325) {};
		\node [style=none] (164) at (-6, -4.325) {};
		\node [style=none] (165) at (-7.5, -4.7) {};
		\node [style=none] (166) at (-6, -4.7) {};
		\node [style=gauge3] (167) at (-8.5, -4.5) {};
		\node [style=none] (168) at (-8.5, -5) {1};
		\node [style=flavour2] (169) at (-8.5, -3.5) {};
		\node [style=none] (170) at (-8.5, -3) {1};
		\node [style=gauge3] (171) at (-1.75, -6) {};
		\node [style=none] (172) at (-1.75, -6.5) {1};
	\end{pgfonlayer}
	\begin{pgfonlayer}{edgelayer}
		\draw (102.center) to (103.center);
		\draw (105.center) to (104.center);
		\draw (107.center) to (106.center);
		\draw (106.center) to (108.center);
		\draw (109.center) to (110.center);
		\draw (111.center) to (112.center);
		\draw (113) to (98);
		\draw (44) to (45);
		\draw (45) to (46);
		\draw (120.center) to (121.center);
		\draw (123.center) to (122.center);
		\draw (125.center) to (124.center);
		\draw (124.center) to (126.center);
		\draw (127.center) to (128.center);
		\draw (129.center) to (130.center);
		\draw (136.center) to (137.center);
		\draw (139.center) to (138.center);
		\draw (141.center) to (140.center);
		\draw (140.center) to (142.center);
		\draw (143.center) to (144.center);
		\draw (145.center) to (146.center);
		\draw (147) to (132);
		\draw (149) to (147);
		\draw (156.center) to (157.center);
		\draw (159.center) to (158.center);
		\draw (161.center) to (160.center);
		\draw (160.center) to (162.center);
		\draw (163.center) to (164.center);
		\draw (165.center) to (166.center);
		\draw (167) to (152);
		\draw (169) to (167);
	\end{pgfonlayer}
\end{tikzpicture}}
}  & \raisebox{-.5\height}{\scalebox{.7}{ \begin{tikzpicture}
		\begin{pgfonlayer}{nodelayer}
		\node [style=bd] (0) at (-1.5, 2) {};
		\node [style=bd] (1) at (1.5, 2) {};
		\node [style=bd] (2) at (0, 0) {};
		\node [style=none] (3) at (1.25, 1) {$h_{3,4}$};
		\node [style=none] (4) at (-1.25, 0.75) {$A_2$};
		\node [style=bd] (5) at (0, 4) {};
		\node [style=bd] (6) at (3, 4) {};
		\node [style=none] (7) at (2.75, 2.75) {$a_2$};
		\node [style=none] (9) at (-1.25, 3.25) {$h_{3,4}$};
		\node [style=none] (10) at (1.25, 3.25) {$A_2$};
	\end{pgfonlayer}
	\begin{pgfonlayer}{edgelayer}
		\draw [style=blue] (2) to (1);
		\draw [style=red] (2) to (0);
		\draw [style=blue] (0) to (5);
		\draw [style=red] (1) to (5);
		\draw [style=blue] (1) to (6);
	\end{pgfonlayer}
\end{tikzpicture}}} \\ \bottomrule
    \end{tabular}
    \caption{$A_2$ theory.  \texttt{Left:} Hasse diagram for the Higgs branch of the $4d$ $\mathcal{N}=2$ rank 1 theory. To the right of the Hasse diagram are the magnetic quivers for the closures of the leaves and to the left are magnetic quivers for the slices. \texttt{Right:} Hasse diagram for the full moduli space of the theory in $3d$ $\mathcal{N}=4$. }
    \label{tab:Hasse_A2}
\end{table}
\subsection{\texorpdfstring{$\mathbb{H}/\mathbb{Z}_k$}{H mod Zk} rank 1 SCFTs}
For completeness, consider four more rank 1 theories whose Higgs branches are $\mathbb{H}/{\mathbb{Z}_k}$ orbifolds with $k=2,3,4,6$. The magnetic quivers take the unified form:
\begin{equation}
\raisebox{-.5\height}{\scalebox{.8}{
    \begin{tikzpicture}
	\begin{pgfonlayer}{nodelayer}
		\node [style=gauge3] (0) at (-1, 0) {};
		\node [style=gauge3] (1) at (0, 0) {};
		\node [style=none] (2) at (-0.5, 0.25) {$k$};
		\node [style=none] (3) at (-1, -0.5) {1};
		\node [style=none] (4) at (0, -0.5) {1};
	\end{pgfonlayer}
	\begin{pgfonlayer}{edgelayer}
		\draw (1) to (0);
	\end{pgfonlayer}
\end{tikzpicture}}
}
\label{orbifold}
\end{equation}
where $k$ denotes the multiplicity of hypermultiplets. The Hilbert series is well known, see for instance \cite{Cremonesi:2013lqa}:
\begin{equation}
    \mathrm{HS}\eqref{orbifold}=\frac{1-t^{2k}}{(1-t^2)(1-t^kq)(1-t^k/q)} \;.
\end{equation}  
The Coulomb branch global symmetry is $\mathrm{SU(2)}$ for $k=2$ where the generators are all at order $t^2$ transforming as $[2]_{A_1}$ and a singlet relation at order $t^4$. The Coulomb branch global symmetry is $U(1)$ for $k > 2$, with a singlet generator at order $t^2$, and $q+\frac{1}{q}$ generators at order $t^k$ satisfying a singlet relation at order $t^{2k}$. These are consistent with the results in Table \ref{rank1table}. The $U(1)$ global symmetry is a remnant of the accidental enhanced supersymmetry. The moduli space consists of 3 complex scalars, with a starting $SO(6)$ global symmetry. These are the 6 transverse scalars to a D3 brane probe. Complexification breaks the symmetry to $U(3)$, out of which $SU(3)$ is an R symmetry. We are left with a $U(1)$ which is the symmetry that is observed for the orbifold cases $k=3, 4, 6.$ The extra enhancement of symmetry for the case of $k=2$ is a signal of the additional enhancement of supersymmetry to $16$ supercharges.

\section{Higher rank magnetic quivers from 5d \texorpdfstring{$\mathcal{N}=1$}{N=1} theories}\label{generalfamily}
The magnetic quivers for rank 1 theories can be derived by taking magnetic quivers of $5d$ $\mathcal{N}=1$ theories and folding them. The relevant $5d$ theories are summarised in Table \ref{beforefold} with each of them extended to a general family. The folding of $k$ legs of the magnetic quivers of the $5d$ theories is directly related to the compatification of these $5d$ theories with a $\mathbb{Z}_k$ twist discussed in \cite{Zafrir:2016wkk,Ohmori:2018ona}. The folded theories are listed in Table \ref{afterfold}. We also provide the Higgs branch dimension of the magnetic quivers and, where possible, the refined Hilbert series expressed as a highest weight generating function (HWG)\footnote{The HWG here are expressed as a plethystic logarithm (PL) which allows one to express the rational function in an elegant polynomial form. This PL of the HWG is not to be confused with the PL of the Hilbert series where the positive and negative terms encode the generators and relations in the chiral ring.}\cite{Hanany:2016djz}. For a given family the parametrisation is chosen such that \emph{after folding}: 
\begin{itemize}
    \item[--] $n=0$ one obtains a magnetic quiver for flat space $\mathbb{H}^l$ for some $l$. The folded quivers are given in Table \ref{nilpotent}.
    \item[--] $n=1$ one obtains a magnetic quiver for a rank 1 theory without enhanced Coulomb branch, they are closures of minimal nilpotent orbits of some algebra. The folded quivers are given in Table \ref{nilpotent}.
    \item[--] $n=2$ one obtains a magnetic quiver for a rank 1 theory with enhanced Coulomb branch, which partially Higgses to the theory with the $n=1$ magnetic quiver. These are the quivers of main interest in this paper and are given in Table \ref{resulttable}. 
    \item[--] $n>2$ one obtains a magnetic quiver for a higher rank theory, which can be Higgsed to the $n=2$ and $n=1$ case, and possibly other theories. These are tabulated in Table \ref{afterfold}.
\end{itemize}
Before folding, all but one family of magnetic quivers  are either star shaped quivers, such as $T_N$, or those found in \cite{Cabrera:2018jxt}. These are given in Table  \ref{beforefold} and summarised as:
\begin{itemize}
    \item The $C_{n+3}$ family ($n=2$ case gives the $C_5$ rank 1 SCFT) comes from folding the $E_7$ family, which are the magnetic quiver of $5d$ $\mathcal{N}=1$ $SU(n+1)_0$ SQCD  with $2n+4$ flavours. The magnetic quivers here describe the Higgs branch of the $4d$ SCFTs generated by compactifying the $5d$ SCFT lifts of these $\mathcal{N}=1$ SQCD theories to $4d$ with a $\mathbb{Z}_2$ twist. An alternative construction is as the compactification of the $6d$ $\mathcal{N}=(1,0)$ SCFT completion of the $\mathrm{USp}(2n-2)$ gauge theory with $2n+6$ flavours on a torus with a non-trivial flavour background. For $n$ odd, these can be identified as class $\mathcal{S}$ theories associated with the compactification of the $D_{\frac{n+5}{2}}$ $(2,0)$ theory on a sphere with two twisted maximal punctures and one untwisted minimal puncture \cite{Ohmori:2018ona}. For $n$ even, these can also be identified with class $\mathcal{S}$ theories, though the identification is slightly more involved, see \cite[Sec.\ 3.1.2]{Ohmori:2018ona}.
    \item The $C_{n+1}\times A_1$ family ($n=2$ case gives the $C_3\times A_1$ rank 1 SCFT)  comes from folding the two long tails and two short tails of the magnetic quiver for $5d$ $\mathcal{N}=1$ $SU(n+1)_0$ SQCD  with $2n+2$ flavours. The magnetic quivers here describe the Higgs branch of the $4d$ SCFTs generated by compactifying the $5d$ SCFT lifts of these $\mathcal{N}=1$ SQCD theories to $4d$ with a $\mathbb{Z}_2$ twist. For $n$ odd, these can be identified as class $\mathcal{S}$ theories associated with the compactification of the $D_{\frac{n+3}{2}}$ $(2,0)$ theory on a sphere with two twisted maximal punctures and one untwisted puncture of type $[n,1^3]$ \cite{Zafrir:2016wkk}. For $n=4$ the theory seems to match \cite[p.\ 53, \#15]{Chacaltana:2013oka} up to free hypermultiplets.
    \item The $C_{n}\times U_1$ family ($n=2$ case gives the $C_2 \times U_1$ rank 1 SCFT) comes from folding (the two long tails) of the magnetic quiver of one of the two cones of the $5d$ $\mathcal{N}=1$ $SU(n+1)_0$ SQCD theory with $2n$ flavours. The magnetic quivers here describe the Higgs branch of the $4d$ SCFTs generated by compactifying the $5d$ SCFT lifts of these $\mathcal{N}=1$ SQCD theories to $4d$ with a $\mathbb{Z}_2$ twist. For $n$ even, these can be identified as class $\mathcal{S}$ theories named $R_{2,n}$ that were introduced in \cite{Chacaltana:2014nya}. These can be constructed by the compactification of the $A_{n}$ $(2,0)$ theory on a sphere with two twisted maximal punctures and one untwisted minimal puncture. Alternatively, they can also be constructed by the compactification of the $A_{2n}$ $(2,0)$ theory on a sphere with one twisted maximal puncture and one twisted irregular puncture \cite{Tachikawa:2018rgw,Wang:2018gvb}.
    \item The $A_{n+1}$ family ($n=2$ case gives the $A_3$ rank 1 SCFT)  comes from folding the three legs of a $T_{n+2}$ theory. The magnetic quivers here describe the Higgs branch of the $4d$ SCFTs generated by compactifying the $5d$ $T_{n+2}$ theories to $4d$ with a $\mathbb{Z}_3$ twist. There is also an alternative construction involving the compactification of a family of $6d$ $\mathcal{N}=(1,0)$ SCFTs on a torus with a non-trivial flavour background. The exact description of these SCFTs, in terms of, for instance, the low-energy gauge theory on the tensor branch, is quite involved and can be found in \cite[Sec.\ 3.2.2]{Ohmori:2018ona}.
    \item The $A_{n-1}\times U_1$ family ($n=2$ case gives the $A_1\times U_1$ rank 1 SCFT)  comes from taking the extension of the magnetic quiver of the $T_n$ theory with a $U(1)$ connected by a multiplicity 2 edge to the central $U(n)$ node and folding the three long legs. The magnetic quivers here describe the Higgs branch of the $4d$ SCFTs generated by compactifying particular $5d$ SCFTs to $4d$ with a $\mathbb{Z}_3$ twist. The $5d$ SCFTs in question can be conveniently described as the result of a $\mathbb{Z}_3$ symmetric mass deformation of the $5d$ $T_{n+2}$ SCFTs. 
    \item The $A_{n}{}'$ family ($n=2$ case gives the $A_2$ rank 1 SCFT) comes from folding all four legs of the magnetic quiver for the Higgs branch of a class $\mathcal{S}$ theory defined by a sphere with four maximal punctures. The magnetic quivers here describe the Higgs branch of the $4d$ SCFTs generated by compactifying particular $5d$ SCFTs to $4d$ with a $\mathbb{Z}_4$ twist. The $5d$ SCFTs in question are the UV completions of the $5d$ gauge theories made from a linear quiver of $n$ $SU(n+1)$ groups, connected via bifundamental hypers, without Chern-Simons terms and with $n+1$ fundamental hypers for both edge groups. These $5d$ SCFTs can be engineered in string theory through the intersection of $n+1$ D$5$-branes and $n+1$ NS$5$-branes. In order to read the magnetic quiver associated to this brane web, one should impose the $\mathbb{Z}_4$ invariance when decomposing the web into subwebs. 
\end{itemize}
The Higgs branch dimension of the magnetic quivers 	$\text{dim}_{\mathbb{H}}(\mathcal{H}(\mathsf{Q}'))$ listed in Table \ref{beforefold} gives indication of the complexity of the moduli space. For dimensions linear in $n$, the HWG of the Coulomb branch has a simple expression. For those quadratic in $n$, the HWG is complicated. This is also reflected in the simplicity of the Hasse diagrams as demonstrated in Section \ref{generalfamily}.

\begin{table}[t]
\small
\centering
\begin{adjustbox}{center}
	\begin{tabular}{ccccc}
\toprule
Family & Unfolded Magnetic quiver $\mathsf{Q}'$ &
	$\text{dim}_{\mathbb{H}}(\mathcal{H}(\mathsf{Q}'))$ &
	$\text{dim}_{\mathbb{H}}(\mathcal{C}(\mathsf{Q}'))$ &$\text{PL}(\mathrm{HWG}(\mathcal{C}(\mathsf{Q}')))$ \\ 
\midrule
      $C_{n+3}$ &		
    \begin{tabular}{c}
    \scalebox{0.70}{
    \begin{tikzpicture}
	\begin{pgfonlayer}{nodelayer}
		\node [style=gauge3] (0) at (0, 0) {};
		\node [style=gauge3] (1) at (1, 0) {};
		\node [style=gauge3] (2) at (-1, 0) {};
		\node [style=none] (4) at (0, -0.5) {$n{+}3$};
		\node [style=none] (5) at (1, -0.5) {$n{+}2$};
		\node [style=none] (6) at (-1, -0.5) {$n{+}2$};
		\node [style=none] (8) at (-1.75, 0) {\ldots};
		\node [style=none] (10) at (1.75, 0) {\ldots};
		\node [style=gauge3] (11) at (2.5, 0) {};
		\node [style=gauge3] (12) at (-2.5, 0) {};
		\node [style=none] (14) at (2.5, -0.5) {1};
		\node [style=none] (15) at (-2.5, -0.5) {1};
		\node [style=none] (22) at (-2.125, 0) {};
		\node [style=none] (23) at (-1.35, 0) {};
		\node [style=none] (24) at (1.375, 0) {};
		\node [style=none] (25) at (2.15, 0) {};
		\node [style=gauge3] (28) at (0, 1) {};
		\node [style=none] (33) at (0, 1.5) {2};
	\end{pgfonlayer}
	\begin{pgfonlayer}{edgelayer}
		\draw (0) to (1);
		\draw (0) to (2);
		\draw (24.center) to (1);
		\draw (25.center) to (11);
		\draw (23.center) to (2);
		\draw (22.center) to (12);
		\draw (28) to (0);
	\end{pgfonlayer}
\end{tikzpicture}
}
    \end{tabular}	& 
    $n$ & $n^2 + 6n +10$&
    \begin{tabular}{c}
        \parbox{4cm}{$\sum\limits_{i=1}^{n+2}\mu_i\mu_{2n+6-i}t^{2i}+t^4\\+\mu_{2n+6}(t^{n+1}+t^{n+3})$}
    \end{tabular}\\ 
    \midrule
    
     $C_{n+1} \times A_1$ &		
    \begin{tabular}{c}
    \scalebox{0.70}{
    \begin{tikzpicture}
	\begin{pgfonlayer}{nodelayer}
		\node [style=gauge3] (0) at (0, 0) {};
		\node [style=gauge3] (1) at (1, 0) {};
		\node [style=gauge3] (2) at (-1, 0) {};
		\node [style=none] (4) at (0, -0.5) {$n{+}1$};
		\node [style=none] (5) at (1, -0.5) {$n$};
		\node [style=none] (6) at (-1, -0.5) {$n$};
		\node [style=none] (8) at (-1.75, 0) {\ldots};
		\node [style=none] (10) at (1.75, 0) {\ldots};
		\node [style=gauge3] (11) at (2.5, 0) {};
		\node [style=gauge3] (12) at (-2.5, 0) {};
		\node [style=none] (14) at (2.5, -0.5) {1};
		\node [style=none] (15) at (-2.5, -0.5) {1};
		\node [style=none] (22) at (-2.125, 0) {};
		\node [style=none] (23) at (-1.35, 0) {};
		\node [style=none] (24) at (1.375, 0) {};
		\node [style=none] (25) at (2.15, 0) {};
		\node [style=gauge3] (26) at (0, 1) {};
		\node [style=gauge3] (27) at (1, 1) {};
		\node [style=gauge3] (28) at (-1, 1) {};
		\node [style=none] (29) at (-1, 1.5) {1};
		\node [style=none] (30) at (0, 1.5) {2};
		\node [style=none] (31) at (1, 1.5) {1};
	\end{pgfonlayer}
	\begin{pgfonlayer}{edgelayer}
		\draw (0) to (1);
		\draw (0) to (2);
		\draw (24.center) to (1);
		\draw (25.center) to (11);
		\draw (23.center) to (2);
		\draw (22.center) to (12);
		\draw (26) to (0);
		\draw (26) to (27);
		\draw (26) to (28);
	\end{pgfonlayer}
\end{tikzpicture}

}
    \end{tabular}	& 
    $n $ & $n(n+2)+4$ &
    \begin{tabular}{c}
        \parbox{5cm}{   $\sum\limits_{i=1}^{n+1}\mu_i\mu_{2n+2-i}t^{2i}+(\nu_1^2+\nu_2^2)t^2\\+t^4+\nu_1\nu_2\mu_{2n+2}(t^{n+1}+t^{n+3})\\-\nu_1^2\nu_2^2\mu_{n+1}^2t^{2n+6}$}
    \end{tabular}\\ 
    \midrule
      $C_n \times U_1$ &		
    \begin{tabular}{c}
    \scalebox{0.70}{
\begin{tikzpicture}
	\begin{pgfonlayer}{nodelayer}
		\node [style=gauge3] (0) at (0, 0) {};
		\node [style=gauge3] (1) at (1, 0) {};
		\node [style=gauge3] (2) at (-1, 0) {};
		\node [style=none] (4) at (0, -0.5) {$n$};
		\node [style=none] (5) at (1, -0.5) {$n{-}1$};
		\node [style=none] (6) at (-1, -0.5) {$n{-}1$};
		\node [style=none] (8) at (-1.75, 0) {\ldots};
		\node [style=none] (10) at (1.75, 0) {\ldots};
		\node [style=gauge3] (11) at (2.5, 0) {};
		\node [style=gauge3] (12) at (-2.5, 0) {};
		\node [style=none] (14) at (2.5, -0.5) {1};
		\node [style=none] (15) at (-2.5, -0.5) {1};
		\node [style=none] (22) at (-2.125, 0) {};
		\node [style=none] (23) at (-1.35, 0) {};
		\node [style=none] (24) at (1.375, 0) {};
		\node [style=none] (25) at (2.15, 0) {};
		\node [style=gauge3] (26) at (-0.75, 1) {};
		\node [style=gauge3] (27) at (0.75, 1) {};
		\node [style=none] (28) at (-0.75, 1.5) {1};
		\node [style=none] (29) at (0.75, 1.5) {1};
	\end{pgfonlayer}
	\begin{pgfonlayer}{edgelayer}
		\draw (0) to (1);
		\draw (0) to (2);
		\draw (24.center) to (1);
		\draw (25.center) to (11);
		\draw (23.center) to (2);
		\draw (22.center) to (12);
		\draw (26) to (0);
		\draw (0) to (27);
		\draw (27) to (26);
	\end{pgfonlayer}
\end{tikzpicture}
}
    \end{tabular}	& 
    $n$ & $n^2+1$&
    \begin{tabular}{c}
        \parbox{4.5cm}{ $\sum\limits_{i=1}^n\mu_i\mu_{2n-i}t^{2i} + t^2 + \mu_n (q\\+\frac{1}{q}) t^{n+1}-\mu_n^2t^{2(n+1)}$}
    \end{tabular}\\ 
    \midrule
   
  $A_{n+1}$ &		
    \begin{tabular}{c}
    \scalebox{0.70}{
\begin{tikzpicture}
	\begin{pgfonlayer}{nodelayer}
		\node [style=gauge3] (0) at (0, 0) {};
		\node [style=gauge3] (1) at (1, 0) {};
		\node [style=gauge3] (2) at (-1, 0) {};
		\node [style=gauge3] (3) at (0, 1) {};
		\node [style=none] (4) at (0, -0.5) {$n{+}2$};
		\node [style=none] (5) at (1, -0.5) {$n{+}1$};
		\node [style=none] (6) at (-1, -0.5) {$n{+}1$};
		\node [style=none] (7) at (0.75, 1) {$n{+}1$};
		\node [style=none] (8) at (-1.75, 0) {\ldots};
		\node [style=none] (9) at (0, 1.75) {\vdots};
		\node [style=none] (10) at (1.75, 0) {\ldots};
		\node [style=gauge3] (11) at (2.5, 0) {};
		\node [style=gauge3] (12) at (-2.5, 0) {};
		\node [style=gauge3] (13) at (0, 2.45) {};
		\node [style=none] (14) at (2.5, -0.5) {1};
		\node [style=none] (15) at (-2.5, -0.5) {1};
		\node [style=none] (16) at (0.5, 2.45) {1};
		\node [style=none] (22) at (-2.125, 0) {};
		\node [style=none] (23) at (-1.35, 0) {};
		\node [style=none] (24) at (1.375, 0) {};
		\node [style=none] (25) at (2.15, 0) {};
		\node [style=none] (28) at (0, 2.025) {};
		\node [style=none] (29) at (0, 1.4) {};
	\end{pgfonlayer}
	\begin{pgfonlayer}{edgelayer}
		\draw (3) to (0);
		\draw (0) to (1);
		\draw (0) to (2);
		\draw (28.center) to (13);
		\draw (28.center) to (13);
		\draw (29.center) to (3);
		\draw (24.center) to (1);
		\draw (25.center) to (11);
		\draw (23.center) to (2);
		\draw (22.center) to (12);
	\end{pgfonlayer}
\end{tikzpicture}

}
    \end{tabular}	& 
    $\frac{n(n+1)}{2}$ & $\frac{1}{2} (1 + n) (8 + 3 n)$&
    \begin{tabular}{c}
        \parbox{4cm}{ Complicated}
    \end{tabular}\\ 
    \midrule
     
      $A_{n-1} \times U_1$ &		
    \begin{tabular}{c}
    \scalebox{0.70}{\begin{tikzpicture}
	\begin{pgfonlayer}{nodelayer}
		\node [style=gauge3] (0) at (0, 0) {};
		\node [style=gauge3] (1) at (1, 0) {};
		\node [style=gauge3] (2) at (-1, 0) {};
		\node [style=none] (4) at (-0.25, -0.425) {$n$};
		\node [style=none] (5) at (1, -0.5) {$n{-}1$};
		\node [style=none] (6) at (-1, -0.5) {$n{-}1$};
		\node [style=none] (8) at (-1.75, 0) {\ldots};
		\node [style=none] (10) at (1.75, 0) {\ldots};
		\node [style=gauge3] (11) at (2.5, 0) {};
		\node [style=gauge3] (12) at (-2.5, 0) {};
		\node [style=none] (14) at (2.5, -0.5) {1};
		\node [style=none] (15) at (-2.5, -0.5) {1};
		\node [style=gauge3] (17) at (0, -1) {};
		\node [style=none] (18) at (0, -1.65) {\vdots};
		\node [style=gauge3] (19) at (0, -2.4) {};
		\node [style=none] (20) at (0.5, -2.4) {1};
		\node [style=none] (21) at (0.75, -1) {$n{-}1$};
		\node [style=none] (22) at (-2.125, 0) {};
		\node [style=none] (23) at (-1.35, 0) {};
		\node [style=none] (24) at (1.375, 0) {};
		\node [style=none] (25) at (2.15, 0) {};
		\node [style=none] (26) at (0, -1.375) {};
		\node [style=none] (27) at (0, -2.05) {};
		\node [style=gauge3] (28) at (0, 1) {};
		\node [style=none] (29) at (-0.125, 1) {};
		\node [style=none] (30) at (0.125, 1) {};
		\node [style=none] (31) at (-0.125, 0) {};
		\node [style=none] (32) at (0.125, 0) {};
		\node [style=none] (33) at (0.5, 1) {1};
	\end{pgfonlayer}
	\begin{pgfonlayer}{edgelayer}
		\draw (0) to (1);
		\draw (0) to (2);
		\draw (0) to (17);
		\draw (24.center) to (1);
		\draw (25.center) to (11);
		\draw (26.center) to (17);
		\draw (23.center) to (2);
		\draw (22.center) to (12);
		\draw (27.center) to (19);
		\draw (29.center) to (31.center);
		\draw (30.center) to (32.center);
	\end{pgfonlayer}
\end{tikzpicture}

}
    \end{tabular}	& 
     $\frac{n(n+1)}{2}$  & $\frac{n(3n-1)}{2}$&
    \begin{tabular}{c}
        \parbox{4cm}{Complicated}
    \end{tabular}\\ 
    \midrule
    
       $A_{n}{}'$ &		
    \begin{tabular}{c}
    \scalebox{0.70}{
\begin{tikzpicture}
	\begin{pgfonlayer}{nodelayer}
		\node [style=gauge3] (0) at (0, 0) {};
		\node [style=gauge3] (1) at (1, 0) {};
		\node [style=gauge3] (2) at (-1, 0) {};
		\node [style=gauge3] (3) at (0, 1) {};
		\node [style=none] (4) at (-0.375, 0.325) {$n{+}1$};
		\node [style=none] (5) at (1, -0.5) {$n$};
		\node [style=none] (6) at (-1, -0.5) {$n$};
		\node [style=none] (7) at (0.75, 1) {$n$};
		\node [style=none] (8) at (-1.75, 0) {\ldots};
		\node [style=none] (9) at (0, 1.75) {\vdots};
		\node [style=none] (10) at (1.75, 0) {\ldots};
		\node [style=gauge3] (11) at (2.5, 0) {};
		\node [style=gauge3] (12) at (-2.5, 0) {};
		\node [style=gauge3] (13) at (0, 2.45) {};
		\node [style=none] (14) at (2.5, -0.5) {1};
		\node [style=none] (15) at (-2.5, -0.5) {1};
		\node [style=none] (16) at (0.5, 2.45) {1};
		\node [style=gauge3] (17) at (0, -1) {};
		\node [style=none] (18) at (0, -1.65) {\vdots};
		\node [style=gauge3] (19) at (0, -2.4) {};
		\node [style=none] (20) at (0.5, -2.4) {1};
		\node [style=none] (21) at (0.75, -1) {$n$};
		\node [style=none] (22) at (-2.125, 0) {};
		\node [style=none] (23) at (-1.35, 0) {};
		\node [style=none] (24) at (1.375, 0) {};
		\node [style=none] (25) at (2.15, 0) {};
		\node [style=none] (26) at (0, -1.375) {};
		\node [style=none] (27) at (0, -2.05) {};
		\node [style=none] (28) at (0, 2.025) {};
		\node [style=none] (29) at (0, 1.4) {};
	\end{pgfonlayer}
	\begin{pgfonlayer}{edgelayer}
		\draw (3) to (0);
		\draw (0) to (1);
		\draw (0) to (2);
		\draw (0) to (17);
		\draw (28.center) to (13);
		\draw (28.center) to (13);
		\draw (29.center) to (3);
		\draw (24.center) to (1);
		\draw (25.center) to (11);
		\draw (26.center) to (17);
		\draw (23.center) to (2);
		\draw (22.center) to (12);
		\draw (27.center) to (19);
	\end{pgfonlayer}
\end{tikzpicture}
}
    \end{tabular}	& 
    $n^2$& $n(2n+3)$ &
    \begin{tabular}{c}
        \parbox{4cm}{Complicated}
    \end{tabular}\\ 
    \bottomrule
	\end{tabular}
\end{adjustbox}
\caption{Magnetic quivers of $5d$ $\mathcal{N}=1$ theories. In the case of $n=2$, folding these theories reproduces the magnetic quivers of $4d$ $\mathcal{N}=2$ theories of Table \ref{resulttable}. We provide the dimension of both the Higgs branch $\mathcal{H}(\mathsf{Q}')$ and the Coulomb branch $\mathcal{C}(\mathsf{Q}')$ of the unfolded magnetic quivers. The HWGs are given in \cite{Ferlito:2017xdq,TropicalHWG}. The prime in the label of the last family is to distinguish it from the fourth family. }
\label{beforefold}
\end{table}
\begin{table}[t]
\small
\centering
\begin{adjustbox}{center}
	\begin{tabular}{ccccc}
\toprule
		Family & Magnetic quiver $\mathsf{Q}$ & $\mathrm{dim}_{\mathbb{H}}(\mathcal{C}(\mathsf{Q}))$ &$\text{PL}(\mathrm{HWG}(\mathcal{C}(\mathsf{Q})))$ \\ 
\midrule
      $C_{n+3}$ &		
    \begin{tabular}{c}
    \scalebox{0.70}{\begin{tikzpicture}
	\begin{pgfonlayer}{nodelayer}
		\node [style=gauge3] (2) at (-1, 0) {};
		\node [style=none] (6) at (-1, -0.5) {$n{+}2$};
		\node [style=none] (8) at (-1.75, 0) {\ldots};
		\node [style=gauge3] (12) at (-2.5, 0) {};
		\node [style=none] (15) at (-2.5, -0.5) {1};
		\node [style=none] (22) at (-2.125, 0) {};
		\node [style=none] (23) at (-1.35, 0) {};
		\node [style=gauge3] (24) at (0.25, 0) {};
		\node [style=none] (29) at (0.25, -0.5) {$n{+}3$};
		\node [style=none] (30) at (-1, 0.075) {};
		\node [style=none] (31) at (0.25, 0.075) {};
		\node [style=none] (32) at (-1, -0.075) {};
		\node [style=none] (33) at (0.25, -0.075) {};
		\node [style=none] (34) at (-0.5, 0) {};
		\node [style=none] (35) at (-0.125, 0.375) {};
		\node [style=none] (36) at (-0.125, -0.375) {};
		\node [style=gauge3] (37) at (1, 0) {};
		\node [style=none] (38) at (1, -0.5) {2};
		\node [style=blankflavor] (39) at (0.25, 0) {};
	\end{pgfonlayer}
	\begin{pgfonlayer}{edgelayer}
		\draw (23.center) to (2);
		\draw (22.center) to (12);
		\draw (30.center) to (31.center);
		\draw (33.center) to (32.center);
		\draw (35.center) to (34.center);
		\draw (34.center) to (36.center);
		\draw (37) to (24);
	\end{pgfonlayer}
\end{tikzpicture}

} 
    \end{tabular}	& $\frac{(n+3)(n+4)}{2}+1$&
    \begin{tabular}{c}
  $\sum\limits_{i=1}^{n+2}\mu_i^2t^{2i}+t^4+\mu_{n+3}(t^{n+1}+t^{n+3})$
    \end{tabular}\\ 
    \midrule
        $C_{n+1} \times A_1$ &		
    \begin{tabular}{c}
    \scalebox{0.70}{\begin{tikzpicture}
	\begin{pgfonlayer}{nodelayer}
		\node [style=gauge3] (2) at (-1, 0) {};
		\node [style=none] (6) at (-1, -0.5) {$n$};
		\node [style=none] (8) at (-1.75, 0) {\ldots};
		\node [style=gauge3] (12) at (-2.5, 0) {};
		\node [style=none] (15) at (-2.5, -0.5) {1};
		\node [style=none] (22) at (-2.125, 0) {};
		\node [style=none] (23) at (-1.35, 0) {};
		\node [style=gauge3] (24) at (0.25, 0) {};
		\node [style=none] (29) at (0.25, -0.5) {$n{+}1$};
		\node [style=none] (30) at (-1, 0.075) {};
		\node [style=none] (31) at (0.25, 0.075) {};
		\node [style=none] (32) at (-1, -0.075) {};
		\node [style=none] (33) at (0.25, -0.075) {};
		\node [style=none] (34) at (-0.5, 0) {};
		\node [style=none] (35) at (-0.125, 0.375) {};
		\node [style=none] (36) at (-0.125, -0.375) {};
		\node [style=gauge3] (37) at (1, 0) {};
		\node [style=none] (38) at (1, -0.5) {2};
		\node [style=gauge3] (39) at (2.25, 0) {};
		\node [style=none] (40) at (1, 0.075) {};
		\node [style=none] (41) at (2.25, 0.075) {};
		\node [style=none] (42) at (1, -0.075) {};
		\node [style=none] (43) at (2.25, -0.075) {};
		\node [style=none] (44) at (1.75, 0) {};
		\node [style=none] (45) at (1.375, 0.375) {};
		\node [style=none] (46) at (1.375, -0.375) {};
		\node [style=none] (47) at (2.25, -0.5) {1};
		\node [style=blankflavor] (48) at (0.25, 0) {};
	\end{pgfonlayer}
	\begin{pgfonlayer}{edgelayer}
		\draw (23.center) to (2);
		\draw (22.center) to (12);
		\draw (30.center) to (31.center);
		\draw (33.center) to (32.center);
		\draw (35.center) to (34.center);
		\draw (34.center) to (36.center);
		\draw (37) to (24);
		\draw (40.center) to (41.center);
		\draw (43.center) to (42.center);
		\draw (45.center) to (44.center);
		\draw (44.center) to (46.center);
	\end{pgfonlayer}
\end{tikzpicture}

}
    \end{tabular}& $\frac{(n+1)(n+2)}{2}+2$	& 
    \begin{tabular}{c}
        \parbox{5.5cm}{   $\sum\limits_{i=1}^{n+1}\mu_i^2t^{2i}+\nu^2t^2+t^4+\nu^2\mu_{n+1}(t^{n+1}\\+t^{n+3})-\nu^4\mu_{n+1}^2t^{2n+6}$}
    \end{tabular}\\ 
    \midrule
 
      $C_n \times U_1$ &		
    \begin{tabular}{c}
    \scalebox{0.70}{\begin{tikzpicture}
	\begin{pgfonlayer}{nodelayer}
		\node [style=gauge3] (2) at (-1, 0) {};
		\node [style=none] (6) at (-1, -0.5) {$n{-}1$};
		\node [style=none] (8) at (-1.75, 0) {\ldots};
		\node [style=gauge3] (12) at (-2.5, 0) {};
		\node [style=none] (15) at (-2.5, -0.5) {1};
		\node [style=none] (22) at (-2.125, 0) {};
		\node [style=none] (23) at (-1.35, 0) {};
		\node [style=gauge3] (24) at (0.25, 0) {};
		\node [style=none] (29) at (0.25, -0.5) {$n$};
		\node [style=none] (30) at (-1, 0.075) {};
		\node [style=none] (31) at (0.25, 0.075) {};
		\node [style=none] (32) at (-1, -0.075) {};
		\node [style=none] (33) at (0.25, -0.075) {};
		\node [style=none] (34) at (-0.5, 0) {};
		\node [style=none] (35) at (-0.125, 0.375) {};
		\node [style=none] (36) at (-0.125, -0.375) {};
		\node [style=gauge3] (37) at (1.25, 0.5) {};
		\node [style=gauge3] (38) at (1.25, -0.5) {};
		\node [style=none] (39) at (1.75, 0.5) {1};
		\node [style=none] (40) at (1.75, -0.5) {1};
		\node [style=blankflavor] (41) at (0.25, 0) {};
	\end{pgfonlayer}
	\begin{pgfonlayer}{edgelayer}
		\draw (23.center) to (2);
		\draw (22.center) to (12);
		\draw (30.center) to (31.center);
		\draw (33.center) to (32.center);
		\draw (35.center) to (34.center);
		\draw (34.center) to (36.center);
		\draw (37) to (33.center);
		\draw (38) to (31.center);
		\draw (37) to (38);
	\end{pgfonlayer}
\end{tikzpicture}

}
    \end{tabular}& $\frac{n(n+1)}{2}+1$	& 
    \begin{tabular}{c}
        \parbox{5cm}{ $\sum\limits_{i=1}^n\mu_i^2t^{2i} + t^2 + \mu_n (q+\frac{1}{q}) t^{n+1}\\-\mu_n^2t^{2(n+1)}$}
    \end{tabular}\\ 
    \midrule
 
    $A_{n+1}$ &		
    \begin{tabular}{c}
    \scalebox{0.70}{
\begin{tikzpicture}
	\begin{pgfonlayer}{nodelayer}
		\node [style=gauge3] (2) at (-1, 0) {};
		\node [style=none] (6) at (-1, -0.5) {$n{+}1$};
		\node [style=none] (8) at (-1.75, 0) {\ldots};
		\node [style=gauge3] (12) at (-2.5, 0) {};
		\node [style=none] (15) at (-2.5, -0.5) {1};
		\node [style=none] (22) at (-2.125, 0) {};
		\node [style=none] (23) at (-1.35, 0) {};
		\node [style=gauge3] (24) at (0.5, 0) {};
		\node [style=none] (29) at (0.5, -0.5) {$n{+}2$};
		\node [style=none] (30) at (-1, 0.15) {};
		\node [style=none] (31) at (0.5, 0.15) {};
		\node [style=none] (32) at (-1, -0.15) {};
		\node [style=none] (33) at (0.5, -0.15) {};
		\node [style=none] (34) at (-0.5, 0) {};
		\node [style=none] (35) at (0, 0.5) {};
		\node [style=none] (36) at (0, -0.5) {};
		\node [style=blankflavor] (37) at (0.5, 0) {};
	\end{pgfonlayer}
	\begin{pgfonlayer}{edgelayer}
		\draw (23.center) to (2);
		\draw (22.center) to (12);
		\draw (2) to (24);
		\draw (30.center) to (31.center);
		\draw (33.center) to (32.center);
		\draw (35.center) to (34.center);
		\draw (34.center) to (36.center);
	\end{pgfonlayer}
\end{tikzpicture}
}
    \end{tabular}& $\frac{(n+2)(n+3)}{2}-1$	& 
    \begin{tabular}{c}
        \parbox{4cm}{ Complicated}
    \end{tabular}\\ 
    \midrule
   
      $A_{n-1} \times U_1$ &		
    \begin{tabular}{c}
    \scalebox{0.70}{\begin{tikzpicture}
	\begin{pgfonlayer}{nodelayer}
		\node [style=gauge3] (2) at (-1, 0) {};
		\node [style=none] (6) at (-1, -0.5) {$n{-}1$};
		\node [style=none] (8) at (-1.75, 0) {\ldots};
		\node [style=gauge3] (12) at (-2.5, 0) {};
		\node [style=none] (15) at (-2.5, -0.5) {1};
		\node [style=none] (22) at (-2.125, 0) {};
		\node [style=none] (23) at (-1.35, 0) {};
		\node [style=gauge3] (24) at (0.5, 0) {};
		\node [style=none] (29) at (0.5, -0.5) {$n$};
		\node [style=none] (30) at (-1, 0.15) {};
		\node [style=none] (31) at (0.5, 0.15) {};
		\node [style=none] (32) at (-1, -0.15) {};
		\node [style=none] (33) at (0.5, -0.15) {};
		\node [style=none] (34) at (-0.5, 0) {};
		\node [style=none] (35) at (0, 0.5) {};
		\node [style=none] (36) at (0, -0.5) {};
		\node [style=gauge3] (37) at (1.5, 0) {};
		\node [style=none] (38) at (0.5, 0.075) {};
		\node [style=none] (39) at (0.5, -0.1) {};
		\node [style=none] (40) at (1.5, 0.075) {};
		\node [style=none] (41) at (1.5, -0.1) {};
		\node [style=none] (42) at (1.5, -0.5) {1};
		\node [style=blankflavor] (43) at (0.5, 0) {};
	\end{pgfonlayer}
	\begin{pgfonlayer}{edgelayer}
		\draw (23.center) to (2);
		\draw (22.center) to (12);
		\draw (2) to (24);
		\draw (30.center) to (31.center);
		\draw (33.center) to (32.center);
		\draw (35.center) to (34.center);
		\draw (34.center) to (36.center);
		\draw (40.center) to (38.center);
		\draw (39.center) to (41.center);
	\end{pgfonlayer}
\end{tikzpicture}

}
    \end{tabular}& $\frac{n(n+1)}{2}$	& 
    \begin{tabular}{c}
        \parbox{4cm}{ Complicated}
    \end{tabular}\\ 
    \midrule
    $A_{n}{}'$ &		
    \begin{tabular}{c}
    \scalebox{0.70}{
\begin{tikzpicture}
	\begin{pgfonlayer}{nodelayer}
		\node [style=gauge3] (2) at (-1, 0) {};
		\node [style=none] (6) at (-1, -0.5) {$n$};
		\node [style=none] (8) at (-1.75, 0) {\ldots};
		\node [style=gauge3] (12) at (-2.5, 0) {};
		\node [style=none] (15) at (-2.5, -0.5) {1};
		\node [style=none] (22) at (-2.125, 0) {};
		\node [style=none] (23) at (-1.35, 0) {};
		\node [style=gauge3] (24) at (0.5, 0) {};
		\node [style=none] (29) at (0.5, -0.5) {$n{+}1$};
		\node [style=none] (30) at (-1, 0.05) {};
		\node [style=none] (31) at (0.5, 0.05) {};
		\node [style=none] (32) at (-1, -0.075) {};
		\node [style=none] (33) at (0.5, -0.075) {};
		\node [style=none] (34) at (-0.5, 0) {};
		\node [style=none] (35) at (0, 0.5) {};
		\node [style=none] (36) at (0, -0.5) {};
		\node [style=none] (37) at (-1, 0.15) {};
		\node [style=none] (38) at (0.5, 0.15) {};
		\node [style=none] (39) at (-1, -0.175) {};
		\node [style=none] (40) at (0.5, -0.175) {};
		\node [style=blankflavor] (41) at (0.5, 0) {};
	\end{pgfonlayer}
	\begin{pgfonlayer}{edgelayer}
		\draw (23.center) to (2);
		\draw (22.center) to (12);
		\draw (30.center) to (31.center);
		\draw (33.center) to (32.center);
		\draw (35.center) to (34.center);
		\draw (34.center) to (36.center);
		\draw (37.center) to (38.center);
		\draw (39.center) to (40.center);
	\end{pgfonlayer}
\end{tikzpicture}

}
    \end{tabular}& $\frac{(n+1)(n+2)}{2}-1$	& 
    \begin{tabular}{c}
        \parbox{4cm}{Complicated}
    \end{tabular}\\ 
    \bottomrule

	\end{tabular}
\end{adjustbox}
\caption{General quiver families obtained by folding the legs in the magnetic quivers $\mathsf{Q}'$ in Table \ref{beforefold}. In the case of $n=2$ these families correspond to the magnetic quivers of $4d$ $\mathcal{N}=2$ rank 1 theories with enhanced Coulomb branch. For $n>1$ the families are labelled by their global symmetry. For $n=1$ the magnetic quivers describe rank 1 theories without enhanced Coulomb branch, and for $n=0$ each of the moduli spaces is some $\mathbb{H}^l$ for a suitable $l$. The dimensions and the HWGs of the Coulomb branches of the magnetic quivers $\mathcal{C}(\mathsf{Q})$ are provided.}
\label{afterfold}
\end{table}
\begin{table}[t]
\small
\centering
\begin{adjustbox}{center}
	\begin{tabular}{c|cc|cc}
\toprule
\multirow{2}{*}{Family} & \multicolumn{2}{c|}{Case $n=1$} & \multicolumn{2}{c}{Case $n=0$} \\
		 & Magnetic quiver&Moduli space & Magnetic quiver&Moduli space \\ 
\midrule
     $C_{n+3}$ &		
    \begin{tabular}{c}
    \scalebox{0.70}{\begin{tikzpicture}
	\begin{pgfonlayer}{nodelayer}
		\node [style=gauge3] (2) at (-0.925, 0) {};
		\node [style=none] (6) at (-0.925, -0.5) {3};
		\node [style=gauge3] (12) at (-1.925, 0) {};
		\node [style=none] (15) at (-1.925, -0.5) {2};
		\node [style=gauge3] (24) at (0.25, 0) {};
		\node [style=none] (29) at (0.25, -0.5) {4};
		\node [style=none] (30) at (-0.75, 0.075) {};
		\node [style=none] (31) at (0.25, 0.075) {};
		\node [style=none] (32) at (-0.75, -0.075) {};
		\node [style=none] (33) at (0.25, -0.075) {};
		\node [style=none] (34) at (-0.5, 0) {};
		\node [style=none] (35) at (-0.125, 0.375) {};
		\node [style=none] (36) at (-0.125, -0.375) {};
		\node [style=gauge3] (37) at (1, 0) {};
		\node [style=none] (38) at (1, -0.5) {2};
		\node [style=gauge3] (39) at (-2.925, 0) {};
		\node [style=none] (42) at (-2.925, -0.5) {1};
		\node [style=blankflavor] (43) at (0.25, 0) {};
	\end{pgfonlayer}
	\begin{pgfonlayer}{edgelayer}
		\draw (30.center) to (31.center);
		\draw (33.center) to (32.center);
		\draw (35.center) to (34.center);
		\draw (34.center) to (36.center);
		\draw (37) to (24);
		\draw (12) to (2);
		\draw (39) to (12);
	\end{pgfonlayer}
\end{tikzpicture}
}
    \end{tabular}	& $\overline{\mathcal{O}}^{\text{min}}_{E_6}$ &
    \begin{tabular}{c}
    \scalebox{0.70}{\begin{tikzpicture}
	\begin{pgfonlayer}{nodelayer}
		\node [style=gauge3] (2) at (-0.925, 0) {};
		\node [style=none] (6) at (-0.925, -0.5) {2};
		\node [style=gauge3] (12) at (-1.925, 0) {};
		\node [style=none] (15) at (-1.925, -0.5) {1};
		\node [style=gauge3] (24) at (0.25, 0) {};
		\node [style=none] (29) at (0.25, -0.5) {3};
		\node [style=none] (30) at (-0.75, 0.075) {};
		\node [style=none] (31) at (0.25, 0.075) {};
		\node [style=none] (32) at (-0.75, -0.075) {};
		\node [style=none] (33) at (0.25, -0.075) {};
		\node [style=none] (34) at (-0.5, 0) {};
		\node [style=none] (35) at (-0.125, 0.375) {};
		\node [style=none] (36) at (-0.125, -0.375) {};
		\node [style=gauge3] (37) at (1, 0) {};
		\node [style=none] (38) at (1, -0.5) {2};
		\node [style=blankflavor] (39) at (0.25, 0) {};
	\end{pgfonlayer}
	\begin{pgfonlayer}{edgelayer}
		\draw (30.center) to (31.center);
		\draw (33.center) to (32.center);
		\draw (35.center) to (34.center);
		\draw (34.center) to (36.center);
		\draw (37) to (24);
		\draw (12) to (2);
	\end{pgfonlayer}
\end{tikzpicture}
}
    \end{tabular}	&$\mathbb{H}^7$
    \\ 
    \midrule
       $C_{n+1} \times A_1$ &		
    \begin{tabular}{c}
    \scalebox{0.70}{\begin{tikzpicture}
	\begin{pgfonlayer}{nodelayer}
		\node [style=gauge3] (2) at (-1, 0) {};
		\node [style=none] (6) at (-1, -0.5) {1};
		\node [style=gauge3] (24) at (0.25, 0) {};
		\node [style=none] (29) at (0.25, -0.5) {2};
		\node [style=none] (30) at (-0.75, 0.075) {};
		\node [style=none] (31) at (0.25, 0.075) {};
		\node [style=none] (32) at (-0.75, -0.075) {};
		\node [style=none] (33) at (0.25, -0.075) {};
		\node [style=none] (34) at (-0.5, 0) {};
		\node [style=none] (35) at (-0.125, 0.375) {};
		\node [style=none] (36) at (-0.125, -0.375) {};
		\node [style=gauge3] (37) at (1, 0) {};
		\node [style=none] (38) at (1, -0.5) {2};
		\node [style=gauge3] (39) at (2.25, 0) {};
		\node [style=none] (40) at (1, 0.075) {};
		\node [style=none] (41) at (2.25, 0.075) {};
		\node [style=none] (42) at (1, -0.075) {};
		\node [style=none] (43) at (2.25, -0.075) {};
		\node [style=none] (44) at (1.75, 0) {};
		\node [style=none] (45) at (1.375, 0.375) {};
		\node [style=none] (46) at (1.375, -0.375) {};
		\node [style=none] (49) at (2.25, -0.5) {1};
		\node [style=blankflavor] (50) at (0.25, 0) {};
	\end{pgfonlayer}
	\begin{pgfonlayer}{edgelayer}
		\draw (30.center) to (31.center);
		\draw (33.center) to (32.center);
		\draw (35.center) to (34.center);
		\draw (34.center) to (36.center);
		\draw (37) to (24);
		\draw (40.center) to (41.center);
		\draw (43.center) to (42.center);
		\draw (45.center) to (44.center);
		\draw (44.center) to (46.center);
	\end{pgfonlayer}
\end{tikzpicture}
}
    \end{tabular}	& $\overline{\mathcal{O}}^{\text{min}}_{D_4}$ &
    \begin{tabular}{c}
    \scalebox{0.70}{\begin{tikzpicture}
	\begin{pgfonlayer}{nodelayer}
		\node [style=gauge3] (24) at (0.25, 0) {};
		\node [style=none] (29) at (0.25, -0.5) {1};
		\node [style=none] (31) at (0.25, 0.075) {};
		\node [style=none] (33) at (0.25, -0.075) {};
		\node [style=gauge3] (37) at (1, 0) {};
		\node [style=none] (38) at (1, -0.5) {2};
		\node [style=gauge3] (39) at (2.25, 0) {};
		\node [style=none] (40) at (1, 0.075) {};
		\node [style=none] (41) at (2.25, 0.075) {};
		\node [style=none] (42) at (1, -0.075) {};
		\node [style=none] (43) at (2.25, -0.075) {};
		\node [style=none] (44) at (1.75, 0) {};
		\node [style=none] (45) at (1.375, 0.375) {};
		\node [style=none] (46) at (1.375, -0.375) {};
		\node [style=none] (49) at (2.25, -0.5) {1};
		\node [style=blankflavor] (50) at (1, 0) {};
	\end{pgfonlayer}
	\begin{pgfonlayer}{edgelayer}
		\draw (37) to (24);
		\draw (40.center) to (41.center);
		\draw (43.center) to (42.center);
		\draw (45.center) to (44.center);
		\draw (44.center) to (46.center);
	\end{pgfonlayer}
\end{tikzpicture}
}
    \end{tabular}	&$\mathbb{H}^3$
    \\
    \midrule
    
   $C_n \times U_1$ &		
    \begin{tabular}{c}
    \scalebox{0.70}{\begin{tikzpicture}
	\begin{pgfonlayer}{nodelayer}
		\node [style=gauge3] (24) at (0.25, 0) {};
		\node [style=none] (29) at (0.25, -0.5) {1};
		\node [style=none] (31) at (0.25, 0.075) {};
		\node [style=none] (33) at (0.25, -0.075) {};
		\node [style=gauge3] (37) at (1.25, 0.5) {};
		\node [style=gauge3] (38) at (1.25, -0.5) {};
		\node [style=none] (39) at (1.75, 0.5) {1};
		\node [style=none] (40) at (1.75, -0.5) {1};
	\end{pgfonlayer}
	\begin{pgfonlayer}{edgelayer}
		\draw (37) to (33.center);
		\draw (38) to (31.center);
		\draw (37) to (38);
	\end{pgfonlayer}
\end{tikzpicture}
}
    \end{tabular}	& $\overline{\mathcal{O}}^{\text{min}}_{A_2}$
    &		
    \begin{tabular}{c}
    \scalebox{0.70}{\begin{tikzpicture}
	\begin{pgfonlayer}{nodelayer}
		\node [style=gauge3] (37) at (1.25, 0.5) {};
		\node [style=gauge3] (38) at (1.25, -0.5) {};
		\node [style=none] (39) at (1.75, 0.5) {1};
		\node [style=none] (40) at (1.75, -0.5) {1};
	\end{pgfonlayer}
	\begin{pgfonlayer}{edgelayer}
		\draw (37) to (38);
	\end{pgfonlayer}
\end{tikzpicture}
}
    \end{tabular}	& $\mathbb{H}$\\ 
    \midrule
    
    $A_{n+1}$ &		
    \begin{tabular}{c}
    \scalebox{0.70}{
\begin{tikzpicture}
	\begin{pgfonlayer}{nodelayer}
		\node [style=gauge3] (2) at (-1, 0) {};
		\node [style=none] (6) at (-1, -0.5) {2};
		\node [style=gauge3] (24) at (0.5, 0) {};
		\node [style=none] (29) at (0.5, -0.5) {3};
		\node [style=none] (30) at (-1, 0.15) {};
		\node [style=none] (31) at (0.5, 0.15) {};
		\node [style=none] (32) at (-1, -0.15) {};
		\node [style=none] (33) at (0.5, -0.15) {};
		\node [style=none] (34) at (-0.5, 0) {};
		\node [style=none] (35) at (0, 0.5) {};
		\node [style=none] (36) at (0, -0.5) {};
		\node [style=gauge3] (37) at (-2, 0) {};
		\node [style=none] (39) at (-2, -0.5) {1};
		\node [style=blankflavor] (40) at (0.5, 0) {};
	\end{pgfonlayer}
	\begin{pgfonlayer}{edgelayer}
		\draw (2) to (24);
		\draw (30.center) to (31.center);
		\draw (33.center) to (32.center);
		\draw (35.center) to (34.center);
		\draw (34.center) to (36.center);
		\draw (37) to (2);
	\end{pgfonlayer}
\end{tikzpicture}
}
    \end{tabular}	& $\overline{\mathcal{O}}^{\text{min}}_{D_4}$
    &		
    \begin{tabular}{c}
    \scalebox{0.70}{
\begin{tikzpicture}
	\begin{pgfonlayer}{nodelayer}
		\node [style=gauge3] (2) at (-1, 0) {};
		\node [style=none] (6) at (-1, -0.5) {1};
		\node [style=gauge3] (24) at (0.5, 0) {};
		\node [style=none] (29) at (0.5, -0.5) {2};
		\node [style=none] (30) at (-1, 0.15) {};
		\node [style=none] (31) at (0.5, 0.15) {};
		\node [style=none] (32) at (-1, -0.15) {};
		\node [style=none] (33) at (0.5, -0.15) {};
		\node [style=none] (34) at (-0.5, 0) {};
		\node [style=none] (35) at (0, 0.5) {};
		\node [style=none] (36) at (0, -0.5) {};
		\node [style=blankflavor] (37) at (0.5, 0) {};
	\end{pgfonlayer}
	\begin{pgfonlayer}{edgelayer}
		\draw (2) to (24);
		\draw (30.center) to (31.center);
		\draw (33.center) to (32.center);
		\draw (35.center) to (34.center);
		\draw (34.center) to (36.center);
	\end{pgfonlayer}
\end{tikzpicture}
}
    \end{tabular}	& $\mathbb{H}^2$\\ 
    \midrule

      $A_{n-1} \times U_1$ &		
    \begin{tabular}{c}
    \scalebox{0.70}{\begin{tikzpicture}
	\begin{pgfonlayer}{nodelayer}
		\node [style=gauge3] (24) at (0.5, 0) {};
		\node [style=none] (29) at (0.5, -0.5) {1};
		\node [style=none] (31) at (0.5, 0.15) {};
		\node [style=none] (33) at (0.5, -0.15) {};
		\node [style=none] (35) at (0, 0.5) {};
		\node [style=none] (36) at (0, -0.5) {};
		\node [style=gauge3] (37) at (1.5, 0) {};
		\node [style=none] (38) at (0.5, 0.075) {};
		\node [style=none] (39) at (0.5, -0.1) {};
		\node [style=none] (40) at (1.5, 0.075) {};
		\node [style=none] (41) at (1.5, -0.1) {};
		\node [style=none] (42) at (1.5, -0.5) {1};
	\end{pgfonlayer}
	\begin{pgfonlayer}{edgelayer}
		\draw (40.center) to (38.center);
		\draw (39.center) to (41.center);
	\end{pgfonlayer}
\end{tikzpicture}
}
    \end{tabular}	& $\overline{\mathcal{O}}^{\text{min}}_{A_1}$
    &		
    \begin{tabular}{c}
    \scalebox{0.70}{\begin{tikzpicture}
	\begin{pgfonlayer}{nodelayer}
		\node [style=gauge3] (37) at (1.5, 0) {};
		\node [style=none] (40) at (1.5, 0.075) {};
		\node [style=none] (41) at (1.5, -0.1) {};
		\node [style=none] (42) at (1.5, -0.5) {1};
	\end{pgfonlayer}
	\begin{pgfonlayer}{edgelayer}
	\end{pgfonlayer}
\end{tikzpicture}
}
    \end{tabular}	&Trivial\\ 
    \midrule
    $A_n{}'$ &		
    \begin{tabular}{c}
    \scalebox{0.70}{
\begin{tikzpicture}
	\begin{pgfonlayer}{nodelayer}
		\node [style=gauge3] (2) at (-1, 0) {};
		\node [style=none] (6) at (-1, -0.5) {1};
		\node [style=gauge3] (24) at (0.5, 0) {};
		\node [style=none] (29) at (0.5, -0.5) {2};
		\node [style=none] (30) at (-1, 0.05) {};
		\node [style=none] (31) at (0.5, 0.05) {};
		\node [style=none] (32) at (-1, -0.075) {};
		\node [style=none] (33) at (0.5, -0.075) {};
		\node [style=none] (34) at (-0.5, 0) {};
		\node [style=none] (35) at (0, 0.5) {};
		\node [style=none] (36) at (0, -0.5) {};
		\node [style=none] (37) at (-1, 0.15) {};
		\node [style=none] (38) at (0.5, 0.15) {};
		\node [style=none] (39) at (-1, -0.175) {};
		\node [style=none] (40) at (0.5, -0.175) {};
		\node [style=blankflavor] (41) at (0.5, 0) {};
	\end{pgfonlayer}
	\begin{pgfonlayer}{edgelayer}
		\draw (30.center) to (31.center);
		\draw (33.center) to (32.center);
		\draw (35.center) to (34.center);
		\draw (34.center) to (36.center);
		\draw (37.center) to (38.center);
		\draw (39.center) to (40.center);
	\end{pgfonlayer}
\end{tikzpicture}

}
    \end{tabular}	&$\overline{\mathcal{O}}^{\text{min}}_{A_2}$ 
    &		
    \begin{tabular}{c}
    \scalebox{0.70}{
\begin{tikzpicture}
	\begin{pgfonlayer}{nodelayer}
		\node [style=gauge3] (24) at (0.5, 0) {};
		\node [style=none] (29) at (0.5, -0.5) {1};
		\node [style=none] (31) at (0.5, 0.05) {};
		\node [style=none] (33) at (0.5, -0.075) {};
		\node [style=none] (38) at (0.5, 0.175) {};
		\node [style=none] (40) at (0.5, -0.2) {};
	\end{pgfonlayer}
	\begin{pgfonlayer}{edgelayer}   
	\end{pgfonlayer}
\end{tikzpicture}

}
    \end{tabular}	&Trivial\\
    \bottomrule

	\end{tabular}
\end{adjustbox}
\caption{The $n=1$ and $n=0$ members of the general Families of Table \ref{afterfold}, where the Coulomb branches are closures of minimal nilpotent orbits and freely generated theories respectively. The $n=1$ cases correspond to rank 1 theories without enhanced Coulomb branch. Notice that the global symmetry here does not match the labelling of the family.}
\label{nilpotent}
\end{table}

\paragraph{Orbifolds $\mathbb{H}/\mathbb{Z}_k$  for $k=2,3,4,6$.}
Another pattern that emerges is the relation between magnetic quivers before folding and $5d$ $\mathcal{N}=1$ SQCD theories of $SU(n+1)_{0}$ with $N_f$ flavours. In the $\mathbb{Z}_2$ column in Table \ref{resulttable}, we start with the $C_5$ theory with $N_f=2n+4$ and the flavor reduces by 2 when we go to the next row for $C_3\times A_1$ and another 2 for $C_2 \times U_1$. Following this, the $C_1\times \chi_0$ (which is the $\mathbb{H}/\mathbb{Z}_2$ orbifold) family should come as the magnetic quiver of the $n=1$ member of the family $SU(n+1)_0$ with $2n-2$ flavours.   The magnetic quiver of the  $5d$ $\mathcal{N}=1$ theory takes the form:
\begin{equation}\scalebox{.8}{
\raisebox{-.5\height}{
\begin{tikzpicture}
	\begin{pgfonlayer}{nodelayer}
		\node [style=gauge3] (0) at (0, 0) {};
		\node [style=gauge3] (1) at (1, 0) {};
		\node [style=gauge3] (2) at (-1, 0) {};
		\node [style=none] (4) at (0, -0.5) {$n{-}1$};
		\node [style=none] (5) at (1, -0.5) {$n{-}2$};
		\node [style=none] (6) at (-1, -0.5) {$n{-}2$};
		\node [style=none] (8) at (-1.75, 0) {\ldots};
		\node [style=none] (10) at (1.75, 0) {\ldots};
		\node [style=gauge3] (11) at (2.5, 0) {};
		\node [style=gauge3] (12) at (-2.5, 0) {};
		\node [style=none] (14) at (2.5, -0.5) {1};
		\node [style=none] (15) at (-2.5, -0.5) {1};
		\node [style=none] (22) at (-2.125, 0) {};
		\node [style=none] (23) at (-1.35, 0) {};
		\node [style=none] (24) at (1.375, 0) {};
		\node [style=none] (25) at (2.15, 0) {};
		\node [style=gauge3] (26) at (-0.75, 1) {};
		\node [style=gauge3] (27) at (0.75, 1) {};
		\node [style=none] (28) at (-0.75, 1.5) {1};
		\node [style=none] (29) at (0.75, 1.5) {1};
		\node [style=none] (30) at (-0.7, 1.1) {};
		\node [style=none] (31) at (-0.7, 0.925) {};
		\node [style=none] (32) at (0.8, 1.1) {};
		\node [style=none] (33) at (0.8, 0.925) {};
	\end{pgfonlayer}
	\begin{pgfonlayer}{edgelayer}
		\draw (0) to (1);
		\draw (0) to (2);
		\draw (24.center) to (1);
		\draw (25.center) to (11);
		\draw (23.center) to (2);
		\draw (22.center) to (12);
		\draw (26) to (0);
		\draw (0) to (27);
		\draw (30.center) to (32.center);
		\draw (33.center) to (31.center);
	\end{pgfonlayer}
\end{tikzpicture}
}}
\label{orbifoldfamily}
\end{equation}
The HWG reads
\begin{equation}
\mathrm{HWG}(\mu_i,t)=\mathrm{PE}\left[\sum\limits_{i=1}^{n+1}\mu_i\mu_{2n-2-i}t^{2i} + t^2 + \mu_{n-1} \left(q+\frac{1}{q}\right) t^{n+1}-\mu_{n-1}^2t^{2n+2} \right]
 \end{equation}
 and 
 \begin{equation}
 \mathrm{dim}_{\mathbb{H}}\mathcal{H}\eqref{orbifoldfamily}=n-1 \,.
 \end{equation}
 Folding the quiver \eqref{orbifoldfamily} yields the general family of the $\mathbb{H}/\mathbb{Z}_2$ rank 1 theories is tabulated in Table \ref{orbifolds}. Since the orbifold itself is a minimal nilpotent orbit, the $n=1$ case gives our desired $\mathbb{H}/\mathbb{Z}_2$ theory. For the remaining orbifolds $\mathbb{H}/\mathbb{Z}_k$ with $k=2,3,4,6$, the general family of quiver before folding will be \eqref{orbifoldfamily} but with $k$ multiplicity of hypermultiplets between the two $U(1)$ nodes and $k$ long legs from 1 to $n-1$. The folded quivers are listed in Table \ref{orbifolds}. The $n=1$ cases reduces to the $\mathbb{H}\slash \mathbb{Z}_k$ orbifolds. 
\begin{table}[t]
\small
\centering
\begin{adjustbox}{center}
	\begin{tabular}{cccc}
\toprule
	Family & Magnetic quiver&$G_{\mathrm{global}}$  &$\text{PL}(\mathrm{HWG})$ \\ 
\midrule
  $\mathbb{H}/\mathbb{Z}_2$ &		
    \begin{tabular}{c}
    \scalebox{0.70}{\begin{tikzpicture}
	\begin{pgfonlayer}{nodelayer}
		\node [style=gauge3] (2) at (-1, 0) {};
		\node [style=none] (6) at (-1, -0.5) {$n{-}2$};
		\node [style=none] (8) at (-1.75, 0) {\ldots};
		\node [style=gauge3] (12) at (-2.5, 0) {};
		\node [style=none] (15) at (-2.5, -0.5) {1};
		\node [style=none] (22) at (-2.125, 0) {};
		\node [style=none] (23) at (-1.35, 0) {};
		\node [style=gauge3] (24) at (0.25, 0) {};
		\node [style=none] (29) at (0.25, -0.5) {$n{-}1$};
		\node [style=none] (30) at (-1, 0.075) {};
		\node [style=none] (31) at (0.25, 0.075) {};
		\node [style=none] (32) at (-1, -0.075) {};
		\node [style=none] (33) at (0.25, -0.075) {};
		\node [style=none] (34) at (-0.5, 0) {};
		\node [style=none] (35) at (-0.125, 0.375) {};
		\node [style=none] (36) at (-0.125, -0.375) {};
		\node [style=gauge3] (37) at (1.25, 0.5) {};
		\node [style=gauge3] (38) at (1.25, -0.5) {};
		\node [style=none] (39) at (1.75, 0.5) {1};
		\node [style=none] (40) at (1.75, -0.5) {1};
		\node [style=none] (41) at (1.15, 0.425) {};
		\node [style=none] (42) at (1.35, 0.425) {};
		\node [style=none] (43) at (1.15, -0.5) {};
		\node [style=none] (44) at (1.35, -0.5) {};
		\node [style=blankflavor] (45) at (0.25, 0) {};
	\end{pgfonlayer}
	\begin{pgfonlayer}{edgelayer}
		\draw (23.center) to (2);
		\draw (22.center) to (12);
		\draw (30.center) to (31.center);
		\draw (33.center) to (32.center);
		\draw (35.center) to (34.center);
		\draw (34.center) to (36.center);
		\draw (37) to (33.center);
		\draw (38) to (31.center);
		\draw (41.center) to (43.center);
		\draw (42.center) to (44.center);
	\end{pgfonlayer}
\end{tikzpicture}
}
    \end{tabular}	& 
$C_{n-1}\times U_1$ &
    \begin{tabular}{c}
        \parbox{4.3cm}{\footnotesize $\sum\limits_{i=1}^{n-1}\mu_i^2t^{2i} + t^2 + \mu_{n-1} (q+\frac{1}{q}) t^{n+1}\\-\mu_{n-1}^2t^{2n+2}$}
    \end{tabular}\\ 
      $\mathbb{H}/\mathbb{Z}_3$ &		
    \begin{tabular}{c}
    \scalebox{0.70}{\begin{tikzpicture}
	\begin{pgfonlayer}{nodelayer}
		\node [style=gauge3] (2) at (-1, 0) {};
		\node [style=none] (6) at (-1, -0.5) {$n{-}2$};
		\node [style=none] (8) at (-1.75, 0) {\ldots};
		\node [style=gauge3] (12) at (-2.5, 0) {};
		\node [style=none] (15) at (-2.5, -0.5) {1};
		\node [style=none] (22) at (-2.125, 0) {};
		\node [style=none] (23) at (-1.35, 0) {};
		\node [style=gauge3] (24) at (0.25, 0) {};
		\node [style=none] (29) at (0.25, -0.5) {$n{-}1$};
		\node [style=none] (34) at (-0.5, 0) {};
		\node [style=none] (35) at (-0.125, 0.375) {};
		\node [style=none] (36) at (-0.125, -0.375) {};
		\node [style=gauge3] (37) at (1.25, 0.5) {};
		\node [style=gauge3] (38) at (1.25, -0.5) {};
		\node [style=none] (39) at (1.75, 0.5) {1};
		\node [style=none] (40) at (1.75, -0.5) {1};
		\node [style=none] (41) at (-1, 0.125) {};
		\node [style=none] (42) at (-1, 0) {};
		\node [style=none] (43) at (-1, -0.125) {};
		\node [style=none] (44) at (0.25, 0.125) {};
		\node [style=none] (45) at (0.25, 0) {};
		\node [style=none] (46) at (0.25, -0.125) {};
		\node [style=none] (47) at (1.1, -0.6) {};
		\node [style=none] (48) at (1.4, -0.6) {};
		\node [style=none] (49) at (1.25, -0.625) {};
		\node [style=none] (50) at (1.1, 0.4) {};
		\node [style=none] (51) at (1.4, 0.4) {};
		\node [style=none] (52) at (1.25, 0.375) {};
		\node [style=blankflavor] (53) at (0.25, 0) {};
	\end{pgfonlayer}
	\begin{pgfonlayer}{edgelayer}
		\draw (23.center) to (2);
		\draw (22.center) to (12);
		\draw (35.center) to (34.center);
		\draw (34.center) to (36.center);
		\draw (44.center) to (41.center);
		\draw (42.center) to (45.center);
		\draw (46.center) to (43.center);
		\draw (50.center) to (47.center);
		\draw (48.center) to (51.center);
		\draw (52.center) to (49.center);
		\draw (24) to (37);
		\draw (38) to (24);
	\end{pgfonlayer}
\end{tikzpicture}}
    \end{tabular}	& 
    $A_{n-2}\times U_1$ &
    \begin{tabular}{c}
Complicated
    \end{tabular}\\ 
      $\mathbb{H}/\mathbb{Z}_4$ &		
    \begin{tabular}{c}
    \scalebox{0.70}{\begin{tikzpicture}
	\begin{pgfonlayer}{nodelayer}
		\node [style=gauge3] (2) at (-1, 0) {};
		\node [style=none] (6) at (-1, -0.5) {$n{-}2$};
		\node [style=none] (8) at (-1.75, 0) {\ldots};
		\node [style=gauge3] (12) at (-2.5, 0) {};
		\node [style=none] (15) at (-2.5, -0.5) {1};
		\node [style=none] (22) at (-2.125, 0) {};
		\node [style=none] (23) at (-1.35, 0) {};
		\node [style=gauge3] (24) at (0.25, 0) {};
		\node [style=none] (29) at (0.25, -0.5) {$n{-}1$};
		\node [style=none] (34) at (-0.5, 0) {};
		\node [style=none] (35) at (-0.125, 0.375) {};
		\node [style=none] (36) at (-0.125, -0.375) {};
		\node [style=gauge3] (37) at (1.25, 0.5) {};
		\node [style=gauge3] (38) at (1.275, -0.5) {};
		\node [style=none] (39) at (1.75, 0.5) {1};
		\node [style=none] (40) at (1.75, -0.5) {1};
		\node [style=none] (41) at (-1, 0.075) {};
		\node [style=none] (42) at (-1, -0.05) {};
		\node [style=none] (43) at (-1, -0.15) {};
		\node [style=none] (44) at (0.25, 0.075) {};
		\node [style=none] (45) at (0.25, -0.05) {};
		\node [style=none] (46) at (0.25, -0.15) {};
		\node [style=none] (47) at (1.075, -0.45) {};
		\node [style=none] (48) at (1.45, -0.45) {};
		\node [style=none] (49) at (1.2, -0.475) {};
		\node [style=none] (50) at (1.075, 0.4) {};
		\node [style=none] (51) at (1.45, 0.4) {};
		\node [style=none] (52) at (1.2, 0.375) {};
		\node [style=none] (53) at (-1, 0.175) {};
		\node [style=none] (54) at (0.25, 0.175) {};
		\node [style=none] (55) at (1.325, -0.45) {};
		\node [style=none] (56) at (1.325, 0.4) {};
		\node [style=blankflavor] (57) at (0.25, 0) {};
	\end{pgfonlayer}
	\begin{pgfonlayer}{edgelayer}
		\draw (23.center) to (2);
		\draw (22.center) to (12);
		\draw (35.center) to (34.center);
		\draw (34.center) to (36.center);
		\draw (44.center) to (41.center);
		\draw (42.center) to (45.center);
		\draw (46.center) to (43.center);
		\draw (50.center) to (47.center);
		\draw (48.center) to (51.center);
		\draw (52.center) to (49.center);
		\draw (24) to (37);
		\draw (38) to (24);
		\draw (54.center) to (53.center);
		\draw (55.center) to (56.center);
	\end{pgfonlayer}
\end{tikzpicture}}
    \end{tabular}	& 
    $A_{n-2}\times U_1$ &
    \begin{tabular}{c}
Complicated
    \end{tabular}\\ 
      $\mathbb{H}/\mathbb{Z}_6$ &		
    \begin{tabular}{c}
    \scalebox{0.70}{\begin{tikzpicture}
	\begin{pgfonlayer}{nodelayer}
		\node [style=gauge3] (2) at (-1, 0) {};
		\node [style=none] (6) at (-1, -0.5) {$n{-}2$};
		\node [style=gauge3] (12) at (-2.5, 0) {};
		\node [style=none] (15) at (-2.5, -0.5) {1};
		\node [style=none] (22) at (-2.125, 0) {};
		\node [style=none] (23) at (-1.35, 0) {};
		\node [style=gauge3] (24) at (0.25, 0) {};
		\node [style=none] (29) at (0.25, -0.5) {$n{-}1$};
		\node [style=none] (34) at (-0.5, 0) {};
		\node [style=none] (35) at (-0.125, 0.375) {};
		\node [style=none] (36) at (-0.125, -0.375) {};
		\node [style=blankflavor] (57) at (0.25, 0) {};
		\node [style=gauge3] (37) at (1.25, 0.5) {};
		\node [style=gauge3] (38) at (1.25, -0.5) {};
		\node [style=none] (39) at (1.75, 0.5) {1};
		\node [style=none] (40) at (1.75, -0.5) {1};
		\node [style=none] (41) at (-1, 0.075) {};
		\node [style=none] (42) at (-1, -0.05) {};
		\node [style=none] (43) at (-1, -0.15) {};
		\node [style=none] (44) at (0.25, 0.075) {};
		\node [style=none] (45) at (0.25, -0.05) {};
		\node [style=none] (46) at (0.25, -0.15) {};
		\node [style=none] (47) at (1.075, -0.35) {};
		\node [style=none] (48) at (1.325, -0.35) {};
		\node [style=none] (49) at (1.175, -0.375) {};
		\node [style=none] (50) at (1.075, 0.4) {};
		\node [style=none] (51) at (1.325, 0.4) {};
		\node [style=none] (52) at (1.175, 0.375) {};
		\node [style=none] (53) at (-1, 0.175) {};
		\node [style=none] (54) at (0.25, 0.175) {};
		\node [style=none] (55) at (1.425, -0.35) {};
		\node [style=none] (56) at (1.425, 0.4) {};
		\node [style=none] (8) at (-1.75, 0) {\ldots};
	\end{pgfonlayer}
	\begin{pgfonlayer}{edgelayer}
		\draw (23.center) to (2);
		\draw (22.center) to (12);
		\draw (35.center) to (34.center);
		\draw (34.center) to (36.center);
		\draw (24) to (37);
		\draw (38) to (24);
		\draw (-1,.03) to (.25,.03);
		\draw (-1,.09) to (.25,.09);
		\draw (-1,.15) to (.25,.15);
		\draw (-1,-.03) to (.25,-.03);
		\draw (-1,-.09) to (.25,-.09);
		\draw (-1,-.15) to (.25,-.15);
		\draw (1.10,-.5) to (1.10,.5);
		\draw (1.16,-.5) to (1.16,.5);
		\draw (1.22,-.5) to (1.22,.5);
		\draw (1.28,-.5) to (1.28,.5);
		\draw (1.34,-.5) to (1.34,.5);
		\draw (1.40,-.5) to (1.40,.5);
	\end{pgfonlayer}
\end{tikzpicture}}
    \end{tabular}	& 
    $A_{n-2}\times U_1$ &
    \begin{tabular}{c}
      Complicated
    \end{tabular}\\ 
    \bottomrule
	\end{tabular}
\end{adjustbox}
\caption{The $n=1$ case of these families correspond to the magnetic quivers of $4d$ $\mathcal{N}=2$ rank 1 theories whose Higgs branch are $\mathbb{H}/\mathbb{Z}_k$ orbifolds for $k=2,3,4,6$. }
\label{orbifolds}
\end{table}

\paragraph{Hasse diagram for general families.}
Some of the generalised families have simple and linear Hasse diagrams, which are presented in Table \ref{tabHasse}. For the remaining three families $A_3$, $A_1 \times U_1$, and $A_2$ (which do not come from folding a magnetic quiver of infinite coupling SQCD in 5d) the Hasse diagrams are elaborate and are not detailed here. It can be noted, that these are also the theories with complicated HWGs.

\begin{table}[t]
    \centering
\begin{tabular}{ccc||c} \toprule 
    $C_{n+3}$ & $C_{n+1}\times A_1$ & $C_n \times U_1$ & $\mathbb{H}/\mathbb{Z}_2$ \\  \midrule  
\raisebox{-.5\height}{\begin{tikzpicture}[scale=0.7]
	\begin{pgfonlayer}{nodelayer}
	\node at (-1, 7) {};\node at (1, -6) {};
		\node [style=bd] (0) at (0, 6) {};
		\node [style=bd] (1) at (0, 4) {};
		\node [style=bd] (2) at (0, 2) {};
		\node [style=bd] (3) at (0, 0) {};
		\node [style=none] (4) at (0.5, 5) { $e_6$};
		\node [style=none] (5) at (0.5, 3) {$c_5$};
		\node [style=none] (6) at (0.5, 1) {$c_6$};
		\node [style=none] (7) at (0, -0.75) {};
		\node [style=none] (8) at (0, -2.25) {};
		\node [style=bd] (9) at (0, -3) {};
		\node [style=none] (10) at (0, -1.5) {\vdots};
		\node [style=bd] (11) at (0, -5) {};
		\node [style=none] (12) at (0.8, -4) {$c_{n+3}$ };
	\end{pgfonlayer}
	\begin{pgfonlayer}{edgelayer}
		\draw (0) to (1);
		\draw (1) to (2);
		\draw (2) to (3);
		\draw (3) to (7.center);
		\draw (8.center) to (9);
		\draw (9) to (11);
	\end{pgfonlayer}
\end{tikzpicture}}
 & \raisebox{-.5\height}{\begin{tikzpicture}[scale=0.7]
	\begin{pgfonlayer}{nodelayer}
		\node at (-1, 7) {};
		\node at (2.5, -8) {};
		\node [style=bd] (0) at (0, 6) {};
		\node [style=bd] (1) at (1.25, 4) {};
		\node [style=bd] (2) at (1.25, 2) {};
		\node [style=bd] (3) at (1.25, 0) {};
		\node [style=none] (4) at (1, 5.25) {$d_4$};
		\node [style=none] (5) at (1.75, 3) {$c_3$};
		\node [style=none] (6) at (1.75, 1) {$c_4$};
		\node [style=none] (7) at (1.25, -2.75) {};
		\node [style=none] (8) at (1.25, -4.5) {};
		\node [style=bd] (9) at (1.25, -5.5) {};
		\node [style=none] (10) at (1.25, -3.75) {\vdots};
		\node [style=bd] (11) at (1.25, -7.5) {};
		\node [style=none] (12) at (2.05, -6.5) {$c_{n+1}$};
		\node [style=bd] (13) at (0, 1.5) {};
		\node [style=bd] (14) at (0, -0.5) {};
		\node [style=none] (15) at (0, -1.25) {};
		\node [style=none] (16) at (0, -3.25) {};
		\node [style=none] (17) at (0, -2.5) {\vdots};
		\node [style=bd] (18) at (0, -4) {};
		\node [style=bd] (19) at (0, -6) {};
		\node [style=none] (20) at (-0.5, 3.5) {$e_6$};
		\node [style=none] (22) at (0.775, 1.1) {$A_1$};
		\node [style=bd] (23) at (1.25, -2) {};
		\node [style=none] (24) at (0.75, -0.8) {$A_1$};
		\node [style=none] (25) at (1.8, -1) {$c_5$};
		\node [style=none] (26) at (-0.45, 0.5) {$c_5$};
		\node [style=none] (27) at (0.775, -4.3) {$A_1$};
		\node [style=none] (28) at (0.8, -6.325) {$A_1$};
		\node [style=none] (29) at (-0.65, -5) {$c_{n+1}$};
	\end{pgfonlayer}
	\begin{pgfonlayer}{edgelayer}
		\draw (0) to (1);
		\draw (1) to (2);
		\draw (2) to (3);
		\draw (3) to (7.center);
		\draw (8.center) to (9);
		\draw (9) to (11);
		\draw (0) to (13);
		\draw (13) to (14);
		\draw (14) to (15.center);
		\draw (16.center) to (18);
		\draw (18) to (19);
		\draw (19) to (11);
		\draw (13) to (3);
		\draw (14) to (23);
		\draw (18) to (9);
	\end{pgfonlayer}
\end{tikzpicture}}
 & \raisebox{-.5\height}{\begin{tikzpicture}[scale=0.7]
	\begin{pgfonlayer}{nodelayer}
	\node at (-1, 7) {};\node at (1, -6) {};
		\node [style=bd] (0) at (0, 6) {};
		\node [style=bd] (1) at (0, 4) {};
		\node [style=bd] (2) at (0, 2) {};
		\node [style=bd] (3) at (0, 0) {};
		\node [style=none] (4) at (0.5, 5) {$a_2$};
		\node [style=none] (5) at (0.5, 3) {$c_2$};
		\node [style=none] (6) at (0.5, 1) { $c_3$};
		\node [style=none] (7) at (0, -0.75) {};
		\node [style=none] (8) at (0, -2.25) {};
		\node [style=bd] (9) at (0, -3) {};
		\node [style=none] (10) at (0, -1.5) {\vdots};
		\node [style=bd] (11) at (0, -5) {};
		\node [style=none] (12) at (0.6, -4) {$c_n$};
	\end{pgfonlayer}
	\begin{pgfonlayer}{edgelayer}
		\draw (0) to (1);
		\draw (1) to (2);
		\draw (2) to (3);
		\draw (3) to (7.center);
		\draw (8.center) to (9);
		\draw (9) to (11);
	\end{pgfonlayer}
\end{tikzpicture}} & \raisebox{-.5\height}{\begin{tikzpicture}[scale=0.7]
	\begin{pgfonlayer}{nodelayer}
		\node at (-1, 7) {};
		\node at (1, -6) {};
		\node [style=bd] (0) at (0, 6) {};
		\node [style=bd] (1) at (0, 4) {};
		\node [style=bd] (2) at (0, 2) {};
		\node [style=bd] (3) at (0, 0) {};
		\node [style=none] (4) at (0.5, 5) {$A_1$};
		\node [style=none] (5) at (0.5, 3) {$A_1$};
		\node [style=none] (6) at (0.5, 1) {$c_2$};
		\node [style=none] (7) at (0, -0.75) {};
		\node [style=none] (8) at (0, -2.25) {};
		\node [style=bd] (9) at (0, -3) {};
		\node [style=none] (10) at (0, -1.5) {\vdots};
		\node [style=bd] (11) at (0, -5) {};
		\node [style=none] (12) at (0.7, -4) {$c_{n-1}$};
	\end{pgfonlayer}
	\begin{pgfonlayer}{edgelayer}
		\draw (0) to (1);
		\draw (1) to (2);
		\draw (2) to (3);
		\draw (3) to (7.center);
		\draw (8.center) to (9);
		\draw (9) to (11);
	\end{pgfonlayer}
\end{tikzpicture}}
 \\  \bottomrule 
\end{tabular}
    \caption{Hasse diagrams for the first three families of quivers in Table \ref{afterfold}, and for the generalised $\mathbb{H}/\mathbb{Z}_2$ family of quivers in Table \ref{orbifolds}.}
    \label{tabHasse}
\end{table}
Similarly, we provide the Hasse diagram for the 
generalised family of $\mathbb{H}/\mathbb{Z}_2$ in Table \ref{tabHasse} as the other orbifold families have elaborate Hasse diagrams. 

\section{Outlook}
In this note we study Higgs branches of $4d$ $\mathcal{N}=2$ SCFTs through the use of magnetic quivers. Due to recent work on understanding $3d$ $\mathcal{N}=4$ Coulomb branches, in all cases these magnetic quivers can be guessed based on little information, such as global symmetry, dimension, etc.\ of the Higgs branch of the SCFT. It is possible to compare these results with previous work on compactifying higher dimensional theories to obtain said SCFTs. Through quiver subtraction the Hasse diagrams of the Higgs branches are obtained. A goal for the future would be a systematic study of magnetic quivers for higher rank theories other than the families discussed here. Current work on classifying these theories may provide a Cockaigne for the dedicated magnetic quiverist.

\section*{Acknowledgements}
We are grateful to Philip Argyres, Simone Giacomelli, Rudolph Kalveks, Mario Martone, Noppadol Mekareeya, Dominik Miketa, Travis Schedler, Alex Weekes and Anton Zajac for helpful discussions. The work of AB, JFG, AH and ZZ is supported by the STFC Consolidated Grant  ST/P000762/1.  The work of MS is supported by the National Thousand-Young-Talents Program of China, the National Natural Science Foundation of China (grant no.\ 11950410497), and the China Postdoctoral Science Foundation (grant no.\ 2019M650616). The work of GZ is supported in part by World Premier International Research Center Initiative (WPI), MEXT, Japan, by the ERC-STG grant 637844-HBQFTNCER and by the INFN.

\appendix 

\section{Two parameter family of orbifold minimal slices \texorpdfstring{$h_{d,k}$}{h dk}}
\label{orbifoldappendix}
  The $k=2$ case for the $h_{d,k}$ orbifolds gives the $c_d$ minimal nilpotent orbits and $k=3$ gives the  $cg_2$ slices \cite{stembridge1998partial,2003math......5095M} generalised to $d$ nodes. 
Consider the following quiver, with a non-simply laced edge of multiplicity $k$ (in the drawing, $k=3$): 
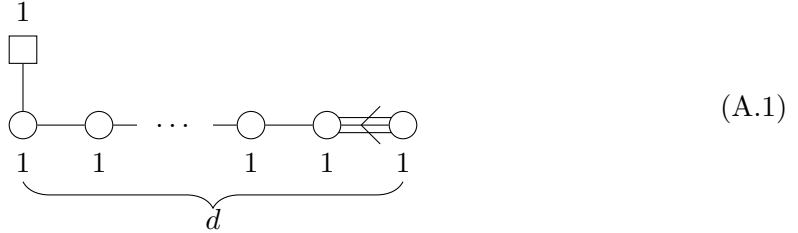
\begin{equation}
\raisebox{-.5\height}{
\begin{tikzpicture}
	\begin{pgfonlayer}{nodelayer}
		\node [style=gauge3] (0) at (0, 0) {};
		\node [style=gauge3] (1) at (1, 0) {};
		\node [style=gauge3] (3) at (-1, 0) {};
		\node [style=none] (4) at (-2, 0) {$\dots$};
		\node [style=gauge3] (5) at (-3, 0) {};
		\node [style=gauge3] (6) at (-4, 0) {};
		\node [style=none] (9) at (-2.5, 0) {};
		\node [style=none] (10) at (-1.5, 0) {};
		\node [style=none] (11) at (0, 0.1) {};
		\node [style=none] (12) at (1, 0.1) {};
		\node [style=none] (13) at (1, -0.1) {};
		\node [style=none] (14) at (0, -0.1) {};
		\node [style=none] (15) at (0.7, 0.25) {};
		\node [style=none] (16) at (0.7, -0.25) {};
		\node [style=none] (17) at (0.45, 0) {};
		\node [style=none] (18) at (-4, 1.5) {1};
		\node [style=none] (20) at (-4, -0.5) {1};
		\node [style=none] (21) at (-3, -0.5) {1};
		\node [style=none] (22) at (-1, -0.5) {1};
		\node [style=none] (23) at (0, -0.5) {1};
		\node [style=none] (24) at (1, -0.5) {1};
		\node [style=none] (25) at (-4, -0.75) {};
		\node [style=none] (27) at (1, -0.75) {};
		\node [style=flavor1] (28) at (-4, 1) {};
		\node [style=none] (29) at (-1.5, -1.25) {$d$};
	\end{pgfonlayer}
	\begin{pgfonlayer}{edgelayer}
		\draw (6) to (5);
		\draw (5) to (9.center);
		\draw (10.center) to (3);
		\draw (3) to (0);
		\draw (0) to (1);
		\draw (11.center) to (12.center);
		\draw (13.center) to (14.center);
		\draw (15.center) to (17.center);
		\draw (17.center) to (16.center);
		\draw (28) to (6);
		\draw [style=brace] (27.center) to (25.center);
	\end{pgfonlayer}
\end{tikzpicture}
}
\end{equation}
The Coulomb branch is $\mathbb{H}^d / \mathbb{Z}_k = (\mathbb{C}^d \oplus \mathbb{C}^d) / \mathbb{Z}_k$ where the action of $\mathbb{Z}_k$ on the coordinates $(z_1 , \dots , z_d , z_1', \dots , z_d') \in \mathbb{C}^d \oplus \mathbb{C}^d$ is given by 
\begin{equation}
    \zeta \cdot (z_1 , \dots , z_d , z_1', \dots , z_d') = (\zeta z_1 , \dots , \zeta z_d , \zeta^{-1} z_1', \dots , \zeta^{-1} z_d')
\end{equation}
for $\zeta = e^{2 i \pi /k} $.\footnote{This family as a set of elementary slices was proposed to us via email correspondence by Mario Martone, Caro Meneghelli, Wolfger Peelaers and Leonardo Rastelli.} The global symmetry is $\mathrm{U}(d)$ (or $\mathrm{Sp}(d)$ for $k=2$) and the Hilbert series can be computed as a Molien sum. Hence, the HWG reads
\begin{equation}
\mathrm{HWG}(\mu_i,q,t)= \mathrm{PE}\left[t^2 + \mu_1\mu_{d-1}t^2 + \frac{\mu_1^k}{q^k}t^k + \mu_{d-1}^kq^kt^k - \mu_1^k\mu_{d-1}^kt^{2k}\right] \,.
\end{equation}

\bibliographystyle{JHEP2}
\bibliography{bibli.bib}

\end{document}